\documentclass[12pt]{article}
\usepackage{graphicx,amssymb,bbold}
\usepackage{amsmath,amsfonts,epsfig}
\usepackage{chessfss}
\usepackage{braket}
\usepackage{tikz}

\newcommand{\tm}{\textrm}
\newcommand{\vp}{\varphi}
\newcommand{\blue}{\textcolor{blue}}

\newcommand{\pslash}{p\kern-1ex /}
\newcommand{\qslash}{q\kern-1ex /}
\newcommand{\lslash}{l\kern-1ex /}
\newcommand{\sslash}{s\kern-1ex /}
\newcommand{\kaslash}{k_a\kern-2ex /}
\newcommand{\kbslash}{k_b\kern-2ex /}
\newcommand{\Dslash}{{\cal D}\kern-1.5ex /}

\newcommand{\bc}{\overline{c}}
\newcommand{\tr}{{\rm tr}}

\newcommand{\beqa}{\begin{eqnarray}}
\newcommand{\eeqa}{\end{eqnarray}}

\newcommand{\bpm}{\begin{pmatrix}}
\newcommand{\epm}{\end{pmatrix}}
\newcommand{\bbm}{\begin{bmatrix}}
\newcommand{\ebm}{\end{bmatrix}}

\newcommand{\plaq}[1]{U_{{\rm P},#1}}  
\makeatletter

\@addtoreset{equation}{section}
\makeatother
\usepackage{subfig}
\begin{document}


\voffset -0.7 true cm
\hoffset 1.5 true cm
\topmargin 0.0in
\evensidemargin 0.0in
\oddsidemargin 0.0in
\textheight 8.6in
\textwidth 5.4in
\parskip 9 pt
 
\def\Tr{\hbox{Tr}}
\newcommand{\be}{\begin{equation}}
\newcommand{\ee}{\end{equation}}
\newcommand{\bea}{\begin{eqnarray}}
\newcommand{\eea}{\end{eqnarray}}
\newcommand{\beas}{\begin{eqnarray*}}
\newcommand{\eeas}{\end{eqnarray*}}
\newcommand{\nn}{\nonumber}
\font\cmsss=cmss8
\def\C{{\hbox{\cmsss C}}}
\font\cmss=cmss10
\def\bigC{{\hbox{\cmss C}}}
\def\scriptlap{{\kern1pt\vbox{\hrule height 0.8pt\hbox{\vrule width 0.8pt
  \hskip2pt\vbox{\vskip 4pt}\hskip 2pt\vrule width 0.4pt}\hrule height 0.4pt}
  \kern1pt}}
\def\ba{{\bar{a}}}
\def\bb{{\bar{b}}}
\def\bc{{\bar{c}}}
\def\bphi{{\Phi}}
\def\Bigggl{\mathopen\Biggg}
\def\Bigggr{\mathclose\Biggg}
\def\Biggg#1{{\hbox{$\left#1\vbox to 25pt{}\right.\n@space$}}}
\def\n@space{\nulldelimiterspace=0pt \m@th}
\def\m@th{\mathsurround = 0pt}

\begin{titlepage}
\begin{flushright}
{\small OU-HET-931}, {\small YITP-17-27} 
 \\
\end{flushright}

\begin{center}

\vspace{5mm}

{\Large \bf Entanglement Entropy} \\[3pt] 
\vspace{1mm}
{\Large \bf  for 2D Gauge Theories with Matters}

\vspace{6mm}

\renewcommand\thefootnote{\mbox{$\fnsymbol{footnote}$}}
Sinya Aoki${}^{\textsymbishop}$, 
Norihiro Iizuka${}^{\textsymking}$, 
Kotaro Tamaoka${}^{\textsymking}$ and 
Tsuyoshi Yokoya${}^{\textsymking}$


\vspace{3mm}

${}^\textsymbishop$
{\small \sl Center for Gravitational Physics,} \\  
{\small \sl Yukawa Institute for Theoretical Physics, Kyoto 606-8502, JAPAN,} 

${}^\textsymking$
{\small \sl Department of Physics, Osaka University} \\ 
{\small \sl Toyonaka, Osaka 560-0043, JAPAN}

\vspace{4mm}

{\small \tt 
{saoki at yukawa.kyoto-u.ac.jp}, {iizuka at phys.sci.osaka-u.ac.jp},   \\
{k-tamaoka, yokoya at het.phys.sci.osaka-u.ac.jp}
}

\end{center}

\vspace{3mm}

\noindent

We investigate the entanglement entropy in 1+1-dimensional $SU(N)$ gauge theories with various matter fields using the lattice regularization. Here we use extended Hilbert space definition for entanglement entropy, which contains three contributions; (1) classical Shannon entropy associated with superselection sector distribution, where sectors are labelled by irreducible representations of boundary penetrating fluxes, (2) logarithm of the dimensions of their representations, which is associated with ``color entanglement'', and (3) EPR Bell pairs, which give ``genuine'' entanglement. We explicitly show that entanglement entropies (1) and (2) above indeed appear for various multiple ``meson'' states in gauge theories with matter fields. Furthermore, we employ transfer matrix formalism for gauge theory with fundamental matter field and analyze its ground state using hopping parameter expansion (HPE), where the hopping parameter $K$ is roughly the inverse square of the mass for the matter. We evaluate the entanglement entropy for the ground state and show that all (1), (2), (3) above appear in the HPE, though the Bell pair part (3) appears in higher order than (1) and (2) do. With these results, we discuss how the  ground state entanglement entropy in the continuum limit can be understood from the lattice ground state obtained in the HPE.

\end{titlepage}

\setcounter{footnote}{0}
\renewcommand\thefootnote{\mbox{\arabic{footnote}}}

\tableofcontents
\section{Introduction\label{sect:intro}}

Entanglement is an key feature, which distinguishes quantum worlds from classical worlds. Simply saying, entanglement allows us to know detailed information about subsystem A once we measure the other subsystem B, even though we  know nothing about each subsystem A \& B separately before we make a measurement. Recently these entanglement were caught attention since it becomes more and more clear that the notion of entanglement is one of the key feature to understand the gauge/gravity duality \cite{Maldacena:1997re} and emerging smooth space-time (see for example, \cite{VanRaamsdonk:2010pw}). 
Needless to say, all of the forces except for gravity in Nature are described by gauge theories, 
and furthermore due to the gauge/gravity duality, quantum gravity in asymptotic anti-de Sitter space is also equivalent to certain gauge theory non-perturbatively. 
In order to understand how the space-time emerges through the idea of entanglement and gauge/gravity duality, 
deepening our understanding of entanglement in gauge theory must be crucial.

Entanglement in spin system is well-defined and there is no ambiguity for its definition. Decomposing the Hilbert space into ``inside'' and ``outside'', and by tracing out the ``outside'' Hilbert space, we obtain the density matrix of the ``inside'' states.  Its von Neumann entropy is the entanglement entropy between ``inside'' and ``outside''.  
However the situation is a bit more subtle in gauge theories. 
In gauge theories, Hilbert space {\it cannot} be decomposed into two gauge invariant subsystem properly, 
due to the local gauge invariance condition, which gives non-local constraints for the allowed states.  
As a result, there exists non-local operators such as 
Wilson loops which spread both ``inside'' and ``outside'', and thus restrict Hilbert spaces of ``inside'' and ``outside'' through Gauss's law
constraints. 
The absence of the gauge invariant decomposition brought  
some confusions for how to define the entanglement entropy in gauge theories.

The main problem of how to define the entanglement entropy associated with the non-product nature of the Hilbert space in gauge theories is now solved through recent works \cite{Casini:2013rba, Ghosh:2015iwa, Aoki:2015bsa, Soni:2015yga}.  For Abelian gauge theory, 
Casini {\it et al.} in \cite{Casini:2013rba} pointed out that the presence of a non-trivial center, 
which commute with all the operators on the ``inside'' (Hilbert space), characterizes 
the ambiguity of the entanglement entropy in gauge theories.  
Clearly this center corresponds physically to gauge invariant Wilson loop operators penetrating the boundary. They connect ``inside'' and ``outside'' Hilbert spaces, and also split the ``inside'' Hilbert space into several different superselection sectors labeled by fluxes of the penetrating loop . 
In each superselection sector, the Hilbert space {\it can} now be written as a tensor product of ``inside'' and ``outside'' Hilbert spaces ${\hat {\cal H}}^{\bf{k}}_{in}$, ${\hat {\cal H}}^{\bf{k}}_{out}$. 
They allow us to define reduced density matrix ${\rho}_{in}^{\bf{k}}$ such that $\Tr {\rho}_{in}^{\bf{k}}  = 1$, 
where $\bf{k}$ is the label for different superselection sectors,  specifying the penetrating gauge flux `representations' at all boundaries.  
Then the definition of the entanglement entropy is given as \cite{Casini:2013rba}
\be
S_{EE}=-\sum_{\bf{k}} p_{\bf{k}} \log (p_{\bf{k}}) 
- \sum_{\bf{k}} p_{\bf{k}}\Tr_{{\hat {\cal H}}^{\bf{k}}_{in}}{ \rho}_{in}^{\bf{k}} \log {\rho}_{in}^{\bf{k}} \,,
\label{abelianEE}
\ee
where the second term is
the weighted average of the ``genuine'' entanglement on each sector $\bf{k}$ with the probability $p_{\bf{k}}$, which we mean EPR Bell pairs obtained in entanglement distillation, 
\be
S({\rho}_{in}^{\bf{k}} ) = - \Tr_{{\hat {\cal H}}^{\bf{k}}_{in}}{ \rho}_{in}^{\bf{k}} \log {\rho}_{in}^{\bf{k}} 
 \,,
\ee 
while the first term is
the classical Shannon entropy for the probability distribution of the variables on the center\footnote{In \cite{Casini:2013rba}, it is also shown that different choices of the ``inside'' operators give different centers and then it is possible to take a trivial center such that the classical entropy for the sector distribution part vanishes.}. 
Note that this classical entropy is different from the ``genuine'' entanglement entropy.

The ``extended Hilbert space'' definition of the entanglement entropy is given in \cite{Ghosh:2015iwa,Aoki:2015bsa,Soni:2015yga}. In these, we literary extend the Hilbert space in such a way that the Hilbert space is {\it no more} restricted to gauge invariant state only 
\be
Q_{\rm BRST} \ket{{\rm phys}}  =  0 \,.
\ee 
As a result of this extension, the Hilbert space 
can now be decomposed as a tensor products of two (gauge non-invariant) subsystems without ambiguity. In the lattice formulation of gauge theories, the extended Hilbert space can be identified to the Hilbert space of a spin system, so that one can define the entanglement entropy unambiguously. 
For example in $U(1)$ case, the explicit calculation becomes possible \cite{Casini:2015dsg, Soni:2016ogt}, 
and it has been  shown in \cite{Ghosh:2015iwa} that this definition agrees with \eqref{abelianEE}.

In non-Abelian gauge theory, however, the extended Hilbert space entanglement entropy definition needs an extension of  \eqref{abelianEE},  which consists of three terms as \cite{Soni:2016ogt} 
\be
\label{colorEE}
S_{EE}=-\sum_{\bf{k}} p_{\bf{k}} \log p_{\bf{k}} + \sum_{\bf{k}} p_{\bf{k}} \left(\sum_i \log d_{{\bf k}^i} \right)- \sum_{\bf{k}} p_{\bf{k}}\Tr_{{\hat {\cal H}}^{\bf{k}}_{in}}{ \rho}_{in}^{\bf{k}} \log {\rho}_{in}^{\bf{k}} .
\ee
The first and third term are essentially the same as  \eqref{abelianEE}, 
while the peculiarity of the non-Abelian gauge theory appears in the second term, 
which contains
the sum over boundary vertices index $i$, where $i$ runs all boundary vertices and ${\bf{k}}^i$ is the irreducible representation of the penetrating gauge loop at that boundary with  $d_{{\bf k}^i} $ being the dimension of the representation ${\bf{k}}^i$.\footnote{Note that ${\bf k} =\{{\bf k}^1, {\bf k}^2, \cdots, {\bf k}^{n_b}\}$, where $n_b$ is a total number of boundary vertices.} 
Thus the second term vanishes for the abelian case since all representations are 1-dimensional. 
Here the superselection sector is labelled by ${\bf{k}} = \{ {\bf{k}}^i \ \vert \ i \in \mbox{all boundary vertices in }  {\hat {\cal H}}^{\bf{k}}_{in} \}$. 
Since the representations in non-Abelian gauge theory are no more one dimensional, 
the requirement of wave function being gauge invariant (singlet) at the boundaries 
 generates a new type of ``entanglement" between  ``inside'' and ``outside'' states. 
In other words, the non-Abelian gauge theory has a new term in (\ref{colorEE}), which is the entanglement entropy 
associated with ``color" at each boundary.

Although the appropriate definition is given, definitely more detailed aspects of the entanglement entropy, especially for non-Abelian gauge theories,  need to be better understood both qualitatively and quantitatively. 
A purpose of this paper is twofold: one is to deepen our understanding of the formula \eqref{colorEE} in non-Abelian gauge theories with various matter fields, by explicitly evaluating the contributions to each of the three terms in \eqref{colorEE}. 
This is because the non-Bell pair contributions, {\it i.e.,} the first and second terms of \eqref{colorEE} are less familiar.  
The other is to study the vacuum entanglement entropy of non-abelian gauge theories through the lattice formulation. 
Gauge theories are well-defined on the lattice, and moreover,
once we employ the extended Hilbert space definition, the gauge theory on the lattice effectively 
reduces to the one essentially equivalent to the usual spin system. 

The entanglement entropy for the ground state in non-Abelian gauge theories is especially interesting and 
it is well studied 
by the strong coupling expansion in the lattice formulation\cite{Donnelly:2011hn, Chen:2015kfa, Radicevic:2015sza, Soni:2015yga}. In the formulation by Kogut-Susskind \cite{Kogut:1974ag}, the Hamiltonian for pure gauge theories (without matter fields) in lattice regularization is given by \cite{Creutz:1983}  
\bea
H = \frac{g_{YM}^2}{2 a} \sum_{{\rm link} \, (ij)} \hat{J}_{ij}^2 + \frac{1}{ g_{YM}^{2} a} \sum (\mbox{plaquette terms}) 
\label{eq:H_KS}
\eea
where $a$ is the lattice spacing, $g_{YM}$ is the bare gauge coupling on the lattice, and $\hat{J}_{ij}$ is the generator of the gauge transformation at the vertex $i$ for the link $(ij)$, which satisfies  $\hat{J}_{ij}^2 = \hat{J}_{ji}^2$. 
In the strong coupling limit that $g_{YM} \to \infty$, 
the ground state, which we call the {\it strong coupling ground state} $\ket{0}_{\rm strong}$, is given by the {\it tensor product} of the  ground state $\ket{0}_{ij}$ of each link as
\be
\label{zerostrong}
\ket{0}_{\rm strong} = \bigotimes_{ij} \ket{0}_{ij} \,,
\ee
where $\ket{0}_{ij}$ satisfies $\hat{J}_{ij}^2\ket{0}_{ij} =0$. 
Therefore  
there is {\it no entanglement} for the strong coupling ground state. Note that plaquette terms disappear in 2 dimension, so that  one can always obtain this $\ket{0}_{\rm strong}$ as a ground state in 
2-dimensional pure gauge theories at an arbitrary value  of the coupling constant. 
In other words,  not only the ground state obtained in higher dimensional ($d \ge 3$) pure gauge theories at strong coupling limit but also that of  2-dimensional gauge theories at an arbitrary coupling ground state 
are given by $\ket{0}_{\rm strong}$ on the lattice.

On the other hand, the vacuum in continuum gauge theories, which we call the {\it continuum ground state},  
is manifestly entangled:
tracing out the subsystem makes the rest subsystem into mixed states like the Bogoliubov transformation. 
This is not a contradiction, however, 
since the lattice gauge theories at the strong coupling limit is far from the continuum limit.
Due to the asymptotic freedom of gauge theories, 
the continuum gauge theory with non-zero renormalized coupling  (the IR theory) is obtained from
the lattice gauge theory in the limit of zero bare gauge coupling  (the UV theory).

Therefore, it is important to understand how the strong coupling ground state 
approaches the entangled continuum ground state in the process of the continuum limit. 
In generic dimensions, however, solving the gauge theory on the lattice analytically is very hard exercise, unless we take the strong coupling limit or the expansion around it. That is  why people use numerical simulations in lattice gauge theories, which are shown to be very successful. 
This situation is a little different in 2-dimensions,  since a 2-dimensional pure gauge theory is in some sense ``trivial'' due to the absence of  local physical degrees of freedom.  As a result, we can calculate entanglement entropy for any states at an arbitrary coupling constant \cite{Aoki:2016lma}, so that we can take the continuum limit analytically.
Unfortunately, ``genuine'' entanglement, {\it i.e.,} the third term in \eqref{colorEE}, vanishes  
in 2-dimensional pure gauge theories even in the continuum limit \cite{Aoki:2016lma} as is expected.

Once we add matter fields to pure gauge theories in 2-dimensions, ``genuine'' entangled states emerge due to the existence of local degrees of freedom.
We thus take these gauge plus matter theories as toy models of pure gauge theories in higher dimensions,
since gauge plus adjoint matters in 2-dimensions, for example,  are expected to have analogous behaviors  as  higher dimensional pure Yang-Mills theories with  compactified extra ($d-2$) dimensions. 
While pure gauge theories plus matters can not be solved analytically even in 2-dimensions,\footnote{Unless we take large $N$ limit \cite{tHooft:1974pnl}.}
we can include effects of matter fields order by order in the hopping parameter expansion (HPE) for the small  hopping parameter $K \equiv 1/(2+ (ma)^2)$, where $m$ is the bare mass of matter field and $ma$ must be large for the HPE to work.\footnote{The massless theory or the continuum limit with the finite mass corresponds to $K=1/2$, its maximum value.}

In this paper, using the HPE but at an arbitrary gauge coupling, we demonstrate how the ``genuine" entanglement  entropy emerges for the ground state of gauge plus matter fields in 2-dimensions. 
We mainly consider matter fields in the fundamental representation, but an essential idea works similarly for adjoint matters and other representations. Adding adjoint matters is an interesting set-up, since it resembles the large $N$ D1-brane gauge theory, which is dual to the string theory in the curved space-time \cite{Itzhaki:1998dd}.

The organization of this paper is as follows. In \S \ref{puregaugeEE}, we review the lattice study in \cite{Soni:2015yga}
for the pure gauge theory in 2-dimensions, which has no local physical degrees of freedom.
Therefore, there is no ``genuine'' entanglement in 2-dimensional pure gauge theory. 
Then in \S \ref{singlemesonEE} and \ref{multimesonEE}, we add matter fields, and study entanglement of various meson excited states. \S \ref{commentsection} gives a short summary of the first part. 
Then in \S \ref{TmatrixandHPE}, we show at the leading order of HPE that these mesons states appear in the ground  state of this theory,
which is the eigenstates of the ``transfer matrix'' $\hat{T}$ with the largest eigenvalues. 
The transfer matrix is the time translation operator on the lattice with one time unit and is related to    
the Hamiltonian $H$ as  $\hat{T} = e^{- a H}$.
Then later in \S \ref{Bellsection}, we consider the higher order corrections of HPE
and show that the strong coupling ground state and lattice meson states mix to form the true ground state, and 
at the $K^6$ order, the ground state of the transfer matrix shows nonzero ``genuine'' entanglement, and we end with discussion in 
\S \ref{discussionsession} on our picture of how the strong coupling ground state, which has no entanglement, is connected to the continuum entangled ground state. 

Throughout this paper\footnote{For $N=2$ case, the analysis, especially in \S \ref{multimesonEE}, is slightly modified since meson is un-oriented due to the fact that fundamental = anti-fundamental for $SU(2)$.}, we consider $SU(N)$ gauge theory with $N \ge 3$.

\section{Entanglement entropy for pure gauge theory in lattice formulation}
\label{puregaugeEE}
In this section, 
we briefy illustrate  how the second terms of the entanglement entropy in eq.~(\ref{colorEE}) 
appear in the 2-dimensional pure gauge theory on the lattice formulation \cite{Aoki:2016lma},
using explicit examples.  

\begin{figure}[tbp]
\begin{center}
\includegraphics[height=4.7cm, width=5.5cm]{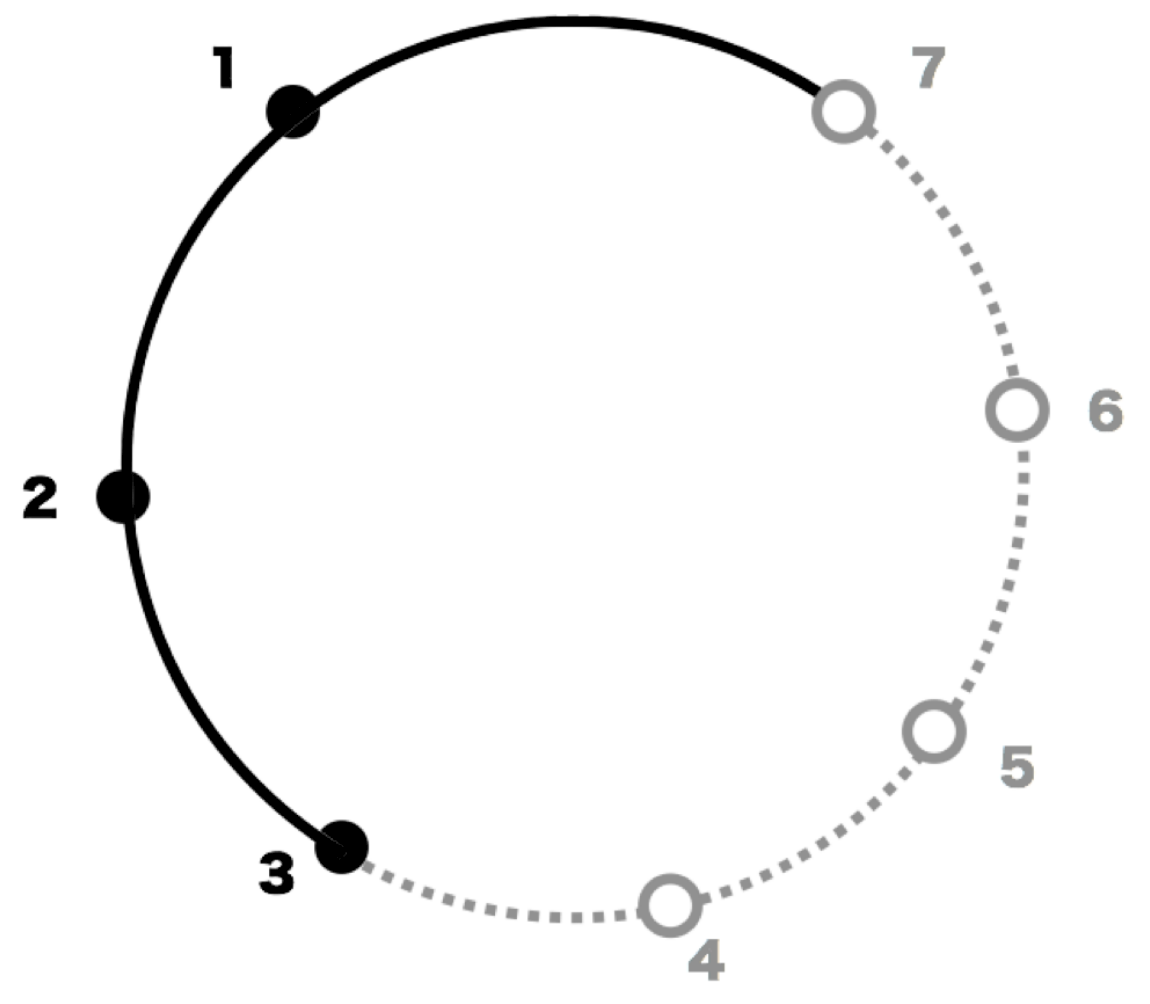}
\caption{Toy seven vertex lattice setup. 
Black vertices and solid lines belongs to ``inside" and white vertices and dotted lines to ``outside".\label{sevenvertexmodel}}  
\end{center}
\end{figure}

We will consider the 7 vertex spatial lattice given in Fig. \ref{sevenvertexmodel} as a 
simple example, which is good enough to see the essential points, and   
one can easily generalize the results in this section to more general cases. 

Consider following wave function
\be
{\bf R}(U)\equiv \chi_{\bf F}(U) = \underset{\bf F}{\Tr}(U) \quad (U \equiv U_{12}U_{23}U_{34}U_{45}U_{56}U_{67}U_{71}) \,,\label{eq:wf1}
\ee
where $U_{ij} \in SU(N)$ is the spatial gauge link variable between the vertices  $i$ and $j$, 
which satisfies $U_{ji} \equiv U_{ij}^\dagger$, and 
$\chi_{\bf F}(U)$ is the character for the `fundamental representation' ${\bf F}$.\footnote{We take the temporal gauge $A_0 =0$ throughout this paper.   As will be seen later, this ${\bf R}(U)$ is the eigenfunction of the transfer matrix \cite{Aoki:2016lma}.}

Straightforward calculation shows that the reduced density matrix becomes
\begin{align}
& \bra{U_{12},U_{23}, U_{71}}\rho\ket{V_{12},V_{23}, V_{71}} \nn \\
&\quad =\, \int dW_{34}dW_{45}dW_{56}dW_{67}\chi_{\bf F}(U_{71}U_{12}U_{23}W_{34}W_{45}W_{56}W_{67})\nonumber \\
& \qquad \qquad \times\chi_{\bf F}(W_{67}^\dagger W_{56}^\dagger W_{45}^\dagger W_{34}^\dagger V_{23}^\dagger V_{12}^\dagger V_{71}^\dagger)\nonumber \\
& \quad =\, \frac{1}{N}\chi_{\bf F}(U_{71}U_{12}U_{23}V_{23}^\dagger V_{12}^\dagger V_{71}^\dagger) \,,
\end{align}
where we used \eqref{formula1} and integrated out  ``outside''-link variables $W_{34},  W_{45}, W_{56}, W_{67}$. 
Therefore the square of the reduced density matrix is 
\begin{align}
\bra{U_{12},U_{23}, U_{71}}\rho^2\ket{V_{12},V_{23}, V_{71}}= &\, \frac{1}{N^2}\int dW_{12}dW_{23}dW_{71}\chi_{\bf F}(U_{71}U_{12}U_{23}W_{23}^\dagger W_{12}^\dagger W_{71}^\dagger)\nonumber \\
&\qquad \times\chi_{\bf F}(W_{71}W_{12}W_{23}V_{23}^\dagger V_{12}^\dagger V_{71}^\dagger)\nonumber \\
=&\,\frac{1}{N^3}\chi_{\bf F}(U_{71}U_{12}U_{23}V_{23}^\dagger V_{12}^\dagger V_{71}^\dagger)\nonumber \\
=&\,\frac{1}{N^2}\bra{U_{12},U_{23}, U_{71}}\rho\ket{V_{12},V_{23}, V_{71}},
\end{align}
where again we used \eqref{formula1}.  This implies
\be
\Tr \rho^n = \frac{1}{N^{2(n-1)}} \,.
\ee
As a result, we obtain an entanglement entropy $S_{EE}$ as 
\be
S_{EE}\equiv -\Tr \rho \log \rho = - \lim_{n\to1} \frac{\partial}{\partial n} \Tr \rho^n 
=2 \log N = n_b \log N \,.
\label{eq:EE_puregauge}
\ee
This is consistent with the ``area-law'' of the entanglement entropy \cite{Srednicki:1993im}, where the boundary is consists of two sites, {\it i.e.,} site 3 and 7, so the ``boundary site number'' $n_b = 2$.  
To see this further,  as an example of $n_b = 4$, 
we consider a different separation of in and out regions 
in such a way that link 2-3 and 5-6 are outside and others are inside. 
Then using \eqref{formula1} and \eqref{formula2}, it is straightforward to check 
the reduced density matrix and its square become 
\begin{align}
\bra{U_{in}}\rho\ket{V_{in}}=& \, \int dW_{23}dW_{56}\chi_{\bf F}(U_{71}U_{12}W_{23}U_{34}U_{45}W_{56}U_{67})\nonumber \\
&\qquad \times\chi_{\bf F}(V_{67}^\dagger W_{56}^\dagger V_{45}^\dagger V_{34}^\dagger W_{23}^\dagger V_{12}^\dagger V_{71}^\dagger)\nonumber \\
=&\, \frac{1}{N^2}\chi_{\bf F}(U_{67}U_{71}U_{12}V_{12}^\dagger V_{71}^\dagger V_{67}^\dagger )\chi_{\bf F}(U_{34}U_{45}V_{45}^\dagger V_{34}^\dagger ) \,,\\ 
\bra{U_{in}}\rho^2\ket{V_{in}}=&\, \frac{1}{N^4}\int dW_{12}dW_{71}dW_{67}\chi_{\bf F}(U_{67}U_{71}U_{12}W_{12}^\dagger W_{71}^\dagger W_{67}^\dagger ) \nn \\
& \qquad\qquad\qquad\qquad\quad  \quad \times  \chi_{\bf F}(W_{67}W_{71}W_{12}V_{12}^\dagger V_{71}^\dagger V_{67}^\dagger )\nonumber \\
& \times\int dW_{34}dW_{45}\chi_{\bf F}(U_{34}U_{45}W_{45}^\dagger W_{34}^\dagger )\chi_{\bf F}(W_{34}W_{45}V_{45}^\dagger V_{34}^\dagger )\nonumber \\
=&\, \frac{1}{N^4}\bra{U_{in}}\rho\ket{V_{in}} \,,
\end{align}
so that we obtain 
\be
S_{EE}= 4\log N = n_{b} \log N \,,
\ee
for $n_b = 4$. It is easy to see in general that 
\be
\label{Aokietalresult}
S_{EE} = n_b \log d_{\bf R} \,,  
\ee
where $d_{\bf R}$ is the dimension of the irreducible representation {\bf R}. 
This is the essential results of \cite{Aoki:2015bsa, Aoki:2016lma}. 
Before we end this section, 
we have several comments. 

Since there is no physical degrees of freedoms in the 2-dimensional pure gauge theory, the result \eqref{Aokietalresult} cannot represent the ``genuine'' entanglement in the sprint of the information theory,  
which is equivalent to the number of Bell pairs obtained in the entanglement distillation.  See \S4 of \cite{Soni:2015yga}, for example.

All calculations in the above are done in the extended Hilbert space definition \cite{Ghosh:2015iwa, Aoki:2015bsa, Soni:2015yga}. 
The Hilbert space in the gauge theory cannot be written as a tensor product of ``inside'' Hilbert space and ``outside'' Hilbert space. 
In above calculations, however, we trace over all of the out states without worrying about the gauge constraint.
This is possible {\it only}  in the extended Hilbert space.

In the extended Hilbert space, we can define the entanglement entropy, which consists of three contributions as is given \eqref{colorEE}. 
Different superselection sectors are distinguished by the electric flux for the Abelian gauge theory and 
by the quadratic Casimir for the non-Abelian gauge theory at each boundary, and the different Casimir corresponds to the different ``spin'', or representation. Due to the Gauss's law in 1+1 dimension, we have only one sector, $p_{\bf F} = 1$,
in our wave function \eqref{eq:wf1}, restricted in  the fundamental representation. 
Therefore \eqref{Aokietalresult} gives only the second term in \eqref{colorEE}, as the first and the third term in \eqref{colorEE} vanish.

Clearly this entanglement entropy \eqref{Aokietalresult} is associated with the fact that  
in and out link variables connected with each other  at the boundary vertex cannot take values freely due to the gauge invariance constraint, and this gauge invariance correlates the two link variables. As a result, 
this correlation produces the entanglement obtained in \eqref{Aokietalresult}, which is the {``color entanglement"}.

\section{Entanglement entropy for single meson states}
\label{singlemesonEE}

\subsection{2d gauge theory with the fundamental scalar field}  

Now we consider the 2-dimensional gauge theory with the fundamental scalar field. Again we consider the Fig. \ref{sevenvertexmodel} lattice setup. For each vertex $n$, there is a scalar field $\vp_n$, in addition to the link variable $U_{ij} \equiv U_{ji}^\dagger$ on  
 each link $(ij)$.

Let us consider the following wave function, 
\begin{align}
\label{fixedgaugeone}
\Psi(\vp_i,U_{ij})& \equiv\dfrac{1}{\mathcal{N}}\;\left[\vp_1^\dagger U_{12}U_{23}U_{34}U_{45}\vp_{5}\right]\;\prod^7_{m=1}e^{-\frac{\gamma}{2}\vp_n^\dagger\vp_n} \,, \\
|\mathcal{N}|^2 &= \dfrac{N}{\gamma^2}\left(\dfrac{\pi}{\gamma}\right)^{7N}\label{eq:norm} \,, 
\end{align}
where  $\mathcal{N}$ is the normalization constant. 
This is a single ``meson'' state composed by a scalar ``quark'' (at site $n=1$) and ``anti-quark'' (at site $n=5$) pair. 
For the wave function of the scalar field to be normalizable, we have introduced 
the Gaussian suppression factor $\propto e^{-\frac{\gamma}{2} \vp^\dagger \vp}$ with the 
Gaussian parameter $\gamma$.
The normalization constant  $\mathcal{N}$ is obtained from the condition 
\begin{align}
1 & = 
\int [d\vp_1d\vp_2\cdots d\vp_7]\int[dU_{12}dU_{23}\cdots dU_{71}]\Psi^\ast(\vp_i,U_{ij})\Psi(\vp_i,U_{ij}) \,,\nn
\end{align}
where we use \eqref{eq:gauss0} and \eqref{eq:UUd}.  
%
Similarly, using \eqref{eq:gauss0}, \eqref{eq:gauss2} and \eqref{eq:norm},  
the reduced density matrix $\rho(\vp_{in},U_{in};\phi_{in},V_{in})$ becomes 
\begin{align}
\rho(\vp_{in},U_{in};\phi_{in},V_{in})
&=\int[d\tilde{\vp}_4\cdots d\tilde{\vp}_7]\int[dW_{34}\cdots dW_{67}]\Phi(\vp_{in},\tilde{\vp}_{out};U_{in},W_{out}) \nn \\
& \qquad \times \Phi^\ast(\phi_{in},\tilde{\vp}_{out};V_{in},W_{out})\nn\\
&=\dfrac{\gamma}{N}\left(\dfrac{\pi}{\gamma}\right)^{-3N}\left[(\vp_1^\dagger U_{12}V^\dagger_{12}\phi_1)\prod_{n=1}^3\,e^{-\frac{\gamma}{2}\vp_n^\dagger\vp_n-\frac{\gamma}{2}\phi_n^\dagger\phi_n}\right] \,,
\end{align}
and a square of the reduced density matrix thus is given by 
\begin{align}
\rho^{2}(\vp,U;\phi,V)&=\int[d \tilde{\vp} dW]\;\rho(\vp,U;\tilde{\vp},W)\rho(\tilde{\vp},W;\phi,V) \nn\\
&=\left[\dfrac{\gamma}{N}\left(\dfrac{\pi}{\gamma}\right)^{-3N}\right]^2\prod_{n=1}^3e^{-\frac{\gamma}{2}\vp^\dagger_n\vp_n-\frac{\gamma}{2}\phi^\dagger_n\phi_n}\nn\\
&\quad \times\int[d \tilde{\vp}{dW}]\left(\vp^\dagger_1U_{12}{W^\dagger_{12}}\tilde{\vp}_1\right)\,\left(\tilde{\vp}_1^\dagger {W_{12}}V^\dagger_{12}\phi_1\right)\,e^{-\gamma(\tilde{\vp}_1^\dagger \tilde{\vp}_1+\tilde{\vp}_2^\dagger \tilde{\vp}_2+\tilde{\vp}_3^\dagger \tilde{\vp}_3)}\nn\\
&=\left[\dfrac{\gamma}{N}\left(\dfrac{\pi}{\gamma}\right)^{-3N}\right]^2\prod_{n=1}^3e^{-\frac{\gamma}{2}\vp^\dagger_n\vp_n-\frac{\gamma}{2}\phi^\dagger_n\phi_n}\left(\vp^\dagger_1U_{12}\right)_c\left(V^\dagger_{12}\phi_1\right)^b\nn\\
&\quad \times{\int[d\tilde{\vp}_2d\tilde{\vp}_3]\;e^{-\gamma(\tilde{\vp}_2^\dagger \tilde{\vp}_2+\tilde{\vp}_3^\dagger \tilde{\vp}_3)}}{\dfrac{1}{N}\delta^a{}_d\delta^c{}_b}\;{\int [d\tilde{\vp}_1]\;\tilde{\vp}_1^d\tilde{\vp}_{1a}^{\dagger}\,e^{-\gamma(\tilde{\vp}_1^\dagger \tilde{\vp}_1)}}\nn\\
%
%
&= \dfrac{1}{N} \, \rho(\vp_{in},U_{in};\phi_{in},V_{in}) \,, 
\label{eq:rho2}
\end{align}
where we have performed the $W_{12}$ integral using the formula \eqref{eq:UUd} in the third equality, and then
the $\vp$ integral using \eqref{eq:gauss2} in the fourth equality. 
From eq.~(\ref{eq:rho2}),  the entanglement entropy is obtained as 
\be
S^{\tm{Fund.}}_{EE}=-\Tr\rho\log\rho=\log N  \,.
\label{singleFundEE}
\ee
Here $\log N$  simply represents the color charge entanglement between scalar quark and anti-quark in the fundamental representation. 

A few comments are in order.
\begin{itemize}
\item This $\log N$ term corresponds to the 2nd term of \eqref{colorEE}. 
First of all, since a color is neither physical nor observable, 
this term cannot be the ``genuine'' entanglement related to the Bell pair, {\it i.e.,} the 3rd term in \eqref{colorEE}.  
A reason why eq.~(\ref{singleFundEE}) does not satisfy the area-law of the entanglement is simply because  the flux takes 
the fundamental representation at the ``boundary vertex'' 3 only but the trivial representation at the ``boundary vertex'' 7.
Furthermore, since we have already fixed the representation in this setup, the 1st term of \eqref{colorEE} can not appear in eq.~(\ref{singleFundEE}).
\item We can easily show the followings. 
The entanglement entropy is given again by \eqref{singleFundEE}
for the  wave functions 
\bea
\vp_{5}^\dagger  U_{56} U_{67} U_{71} \vp_1  
\eea
instead of \eqref{fixedgaugeone}, 
while it vanishes if all fields (quark, anti-quark and all link variables) belong to either ``inside'' or ``outside'' such that 
\bea
\vp_1^\dagger \vp_{1} \,, \quad \vp_1^\dagger U_{12} U_{23}  \vp_{3} \,, \quad  \vp_4^\dagger 
U_{45} U_{56} U_{67} \vp_{7} \,,
\eea
as expected.  
\item The situation is very similar to the pure gauge theory in \S \ref{puregaugeEE}. Regarding that the link variable $U_{56}{}^c{}_d$ made up of two scalar fields $\vp_5{}^c$ and $\vp_{6 \, d}^\dagger$ as $U_{56}{}^c{}_d \approx \vp_5{}^c \vp_{6 \, d}^\dagger$,
the result in eq.~(\ref{eq:EE_puregauge}) can be understood as follows. 
The argument of $\log$ for the entanglement entropy is the dimensions of the representation, {\it i.e.,} the entanglement associated with color numbers.  
The coefficient in front of $\log N$ counts a number of boundary vertices in which the gauge flux penetrates. 
As we will see in the next subsection,
the adjoint matter field gives the $\log \left( N^2 -1 \right)$ instead of $\log N$  contribution to the entanglement entropy. 
\end{itemize}

\subsection{2d gauge theory with the adjoint scalar field}  
For completeness, we show the result with the adjoint matter field $\Phi$.     
We take 
\be
\label{singleadjointmesonwavefn}
\Psi (\Phi_i,U_{ij})=\dfrac{1}{\mathcal{N}}\left[\chi(\Phi_1 U_{12} U_{23}U_{34}U_{45}\Phi_5U_{45}^{\dagger}U_{34}^{\dagger}U_{23}^{\dagger} U_{12}^{\dagger})\right]\prod_{i=1}^7e^{-\beta\Tr \Phi^2_i}
\ee
for the wave function 
with the adjoint scalar field $\Phi$ at the vertex 1 and 5, where $\beta$ is the Gaussian suppression factor. The lattice setup is same as Fig. \ref{sevenvertexmodel}.  

Applying \eqref{eq:sunadjg} and \eqref{eq:U4d} to the condition 
\begin{align}
1=&\frac{1}{|\mathcal{N}|^2}\int [d\Phi][dU] \, \chi (\Phi_1 U_{12}U_{23}U_{34}U_{45}\Phi_5U^{\dagger}_{45}U^{\dagger}_{34}U^{\dagger}_{23}U^{\dagger}_{12}) \nn \\
& \qquad \times \chi (\Phi_1 U_{12}U_{23}U_{34}U_{45}\Phi_5U^{\dagger}_{45}U^{\dagger}_{34}U^{\dagger}_{23}U^{\dagger}_{12})\prod_{i=1}^7e^{-2\beta\Tr\Phi^2_i} ,
%
\end{align}
the normalization constant is determined as
\be
\dfrac{1}{|\mathcal{N}|^2}=\frac{16\beta^2}{N^2-1}\left(\sqrt{\frac{2\beta}{\pi}}\right)^{7(N^2-1)} \,.
\ee
Then, the reduced density matrix is given by
\begin{align}
&\bra{\tilde{\Phi}_{in},V_{in}}\rho\ket{\Phi_{in}, U_{in}}\nn\\
=&\, \frac{1}{|\mathcal{N}|^2}\prod_{i=1}^{3}e^{-\beta\Tr\Phi^2_i-\beta\Tr\tilde{\Phi}^2_i}{\int[d X_{4,6,7}]\prod_{i=4,6,7}e^{-2\beta\Tr X^2_i}} \int [dW] [{dX_5}] \nn\\
&\qquad\qquad \times{\chi (\Phi_1 U_{12} U_{23}W_{34}W_{45}X_5W^{\dagger}_{45}W^{\dagger}_{34}U^{\dagger}_{23} U^{\dagger}_{12})} \nn\\
& \qquad \qquad\quad \times {\chi (\tilde{\Phi}_1 V_{12} V_{23}W_{34}W_{45}X_5W^{\dagger}_{45}W^{\dagger}_{34}V^{\dagger}_{23} V^{\dagger}_{12} )e^{-2\beta\Tr X^2_5}}\nn\\
=&\, \underbrace{\frac{4\beta}{N^2-1}\left(\sqrt{\frac{2\beta}{\pi}}\right)^{3(N^2-1)}}_{:=A}
\chi(U^{\dagger}_{23} U^{\dagger}_{12} \Phi_1 U_{12} U_{23}V^{\dagger}_{23}  V^{\dagger}_{12}  \tilde{\Phi}_1 V_{12}  V_{23})\prod_{i=1}^{3}e^{-\beta\Tr\Phi^2_i-\beta\Tr\tilde{\Phi}^2_i},
\label{adjoint_reduced}
\end{align}
and its square becomes 
\begin{align}
&\bra{\tilde{\Phi}_{in},V_{in}}\rho^2\ket{\Phi_{in}, U_{in}}
=\, A^2{\int[dX_2][dX_3]\prod_{i=2,3}e^{-2\beta\Tr X^2_i}} \nn \\
& \qquad \times \int[{dX_1}][dW] 
{\chi(U^{\dagger}_{23}   U^{\dagger}_{12} \Phi_1 U_{12}  U_{23}W^{\dagger}_{23} W^{\dagger}_{12}  X_1 W_{12}   W_{23})} \nn\\
& \qquad \times {\chi(W^{\dagger}_{23} W^{\dagger}_{12} X_1 W_{12} W_{23} V^{\dagger}_{23} V^{\dagger}_{12} \tilde{\Phi}_1 V_{12} V_{23})e^{-2\beta\Tr X^2_1}}\prod_{i=1}^{3}e^{-\beta\Tr\Phi^2_i-\beta\Tr\tilde{\Phi}^2_i}\nn\\
=&\, A \, {\frac{4\beta}{N^2-1}\left(\sqrt{\frac{2\beta}{\pi}}\right)^{3(N^2-1)}}\frac{1}{4\beta}\left(\sqrt{\frac{\pi}{2\beta}}\right)^{3(N^2-1)}  
\nn\\
& \qquad \times 
\chi(U^{\dagger}_{23}  U^{\dagger}_{12} \Phi_1 U_{12}  U_{23}V^{\dagger}_{23} V^{\dagger}_{12} \tilde{\Phi}_1 V_{12}V_{23})
\prod_{i=1}^{3}e^{-\beta\Tr\Phi^2_i-\beta\Tr\tilde{\Phi}^2_i}\nn\\
=&\, \frac{1}{N^2-1}\bra{\tilde{\Phi}_{in},V_{in}}\rho\ket{\Phi_{in}, U_{in}},
\label{adjredsquare}
\end{align}
Therefore, the entanglement entropy is obtained as
\be
S_{EE}=\log (N^2-1) \,,
\label{logNsquare}
\ee
which confirms  that the argument of $\log$ counts a dimension of the representation for the flux at the boundary vertex.

\subsection{Entanglement entropy for a single meson with the multiple splitting}
\begin{figure}[htbp]
\begin{center}
\includegraphics[height=5.6cm, width=12.4cm]{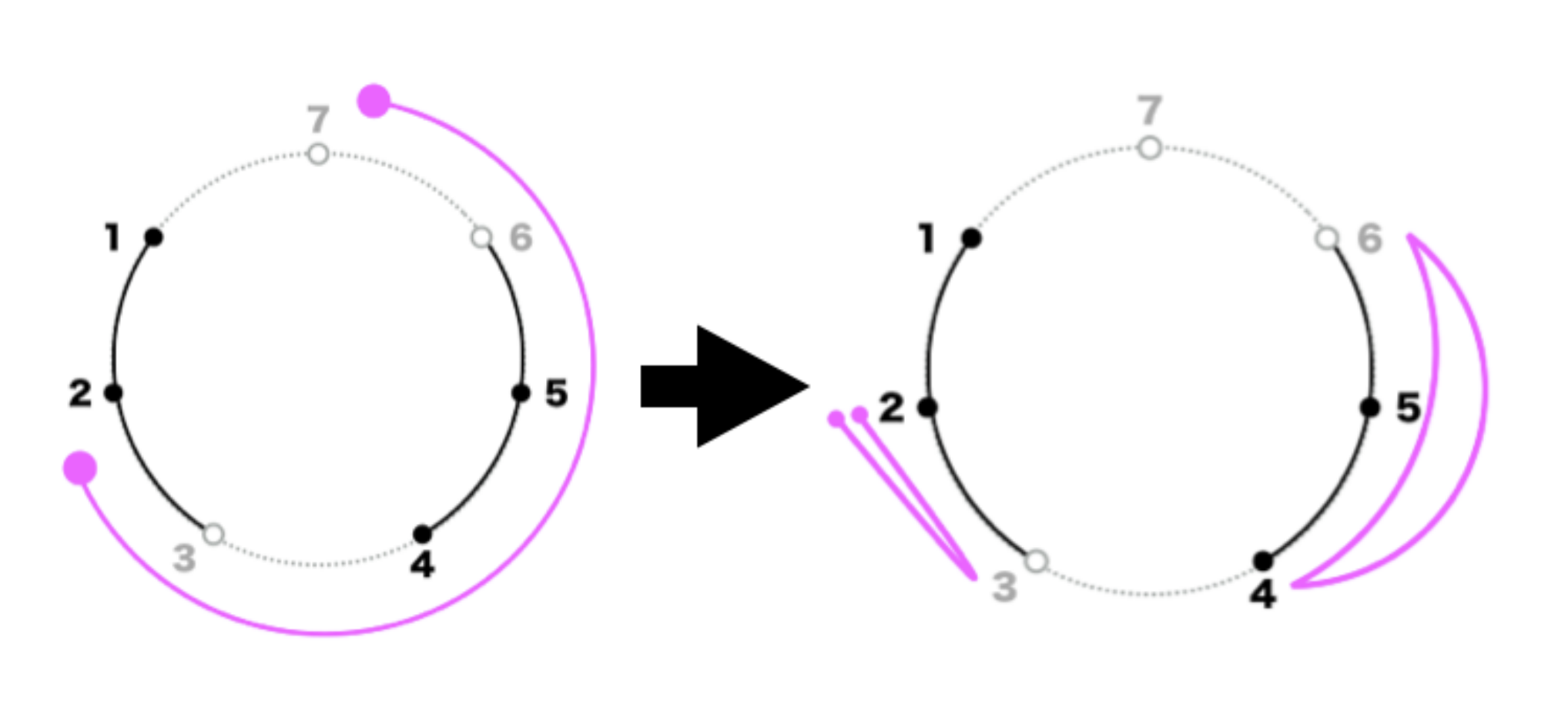}
\caption{Black vertices and solid lines belong to ``inside" and white vertices and dotted lines to ``outside" as before. Scalar quark/anti-quark are at vertices 2 and 7 and gluon is penetrating at the boundary vertices 3, 4, and 6 but not 1. The color indices for the reduced density matrix can be seen pictorially in the right figure.  
}\label{fig:red}
\end{center}
\end{figure}

Let us consider the situation where vertices $1, 2, 4, 5$ and links $(12), (23), (45), (56)$ belong to ``inside'' and the rest belong to ``outside''. See Fig. \ref{fig:red}. 

Let us consider the following wave function   
\begin{align}
\Psi(\vp_i,U_{ij})&\equiv\dfrac{1}{\mathcal{N}}\;\left[\vp_2^\dagger U_{23}{U_{34}}U_{45}U_{56}{U_{67}\vp_{7}}\right]\;\prod_{m=1,2,4,5}e^{-\frac{a}{2}\vp_m^\dagger\vp_m}{\prod_{n=3,6,7}e^{-\frac{a}{2}\vp_n^\dagger\vp_n}} \,, \nn \\ \\
\dfrac{1}{|\mathcal{N}|^{2}}&=\dfrac{a^2}{N}\left(\dfrac{\pi}{a}\right)^{-7N} \,.
\end{align}
It is straightforward to show 
\begin{align}
\rho_{in}&=\dfrac{1}{|\mathcal{N}|^2}\int [{d\tilde{\vp}dW}]\;\left(\vp_2^\dagger U_{23}{W_{34}}U_{45}U_{56}{W_{67}\tilde{\vp}_{7}}\right)\left(
{\tilde{\vp}_{7}^\dagger W_{67}^\dagger}V^\dagger_{56}V^\dagger_{45}{W_{34}^\dagger}V^\dagger_{23}\phi_2\right)\; \nn \\
& \qquad \times
e^{-a({\tilde{\vp}_{3}^\dagger\tilde{\vp}_{3}+\tilde{\vp}_{6}^\dagger\tilde{\vp}_{6}+\tilde{\vp}_{7}^\dagger\tilde{\vp}_{7}})}
e^{-\frac{a}{2}(\vp_1^\dagger\vp_1+\vp_2^\dagger\vp_2+\vp_4^\dagger\vp_4+\vp_5^\dagger\vp_5)-\frac{a}{2}(\phi_1^\dagger\phi_1+\phi_2^\dagger\phi_2+\phi_4^\dagger\phi_4+\phi_5^\dagger\phi_5)}\nn\\
&=\dfrac{a}{N^2}\left(\dfrac{\pi}{a}\right)^{-4N}\,\left( {\vp^\dagger_2U_{23}V^\dagger_{23}\phi_2}\right)\,\left[ {\chi_{\bf F}\left(U_{45}U_{56}V^\dagger_{56}V^\dagger_{45}\right)}\right]\, \nn \\
& \qquad \times e^{-\frac{a}{2}(\vp_1^\dagger\vp_1+\vp_2^\dagger\vp_2+\vp_4^\dagger\vp_4+\vp_5^\dagger\vp_5)-\frac{a}{2}(\vp\leftrightarrow\phi)} \,. \label{eq:reduced4}
\end{align}
This reduced density matrix can be shown pictorially in Fig. \ref{fig:red}. We thus obtain 
\begin{align}
\rho_{in}^2 &=\dfrac{1}{N^3}\rho_{in} \,, \\   
S_{EE}
&= 3\log N \,,
\end{align}
which is again consistent with the second term in \eqref{colorEE}, since a number of boundaries on which  the penetrating flux of the fundamental representation exists is $n_b = 3$ (at vertices 3, 4, and 6).  The boundary 1 does not contribute since there is no penetrating flux there. 

So far, we obtain the entanglement entropy 
\be
\label{numbernontrivialboundary}
S_{EE} = n_b \log d_{\bf R} \,,
\ee
where $d_{\bf R}$ is the dimension of the representation {\bf R}, and $n_b$ is the number of boundaries on which there is nontrivial flux in the representation {\bf R} of the gauge group. 


\section{Entanglement for multiple meson states}
\label{multimesonEE}
We next consider multiple meson states and evaluate their entanglement entropy. 
In \S \ref{twowithoutoverlap}, we first consider a  two meson state where two meson excitations do not overlap each other.  Next in \S \ref{quarkantiquarkpariinoppositedirections}, we consider a various types of overlapped two meson states whose  excited fluxes go through the same boundary. We classify these states in Fig. \ref{fig:2meson},  
and consider the entanglement entropy for all of these possibilities. 
In \S \ref{subsec:4m}, 
we finally consider a four meson state where all excited fluxes  penetrate the same boundary. 

One of the main differences between  these multiple meson excitations and single meson excitations in the previous section is that
we need to decompose  the product of the same link variables of  
multiple meson excitations at the same boundary into a sum of  irreducible representations. 
As a results of this decomposition, we have several different superselection sectors,
labeled by the irreducible representation {\bf R} of the penetrating flux. 
This results in nonzero contribution to the 1st term of the entanglement entropy in \eqref{colorEE}, 
which is the Shannon entropy associated with the superselection sector distribution. 

In this section, we again use the lattice setup in Fig. \ref{sevenvertexmodel}. 

\subsection{Two mesons without overlapping}
\label{twowithoutoverlap}
We first  consider a two meson state without overlap. Explicitly, let us consider the following wave function,  
\begin{align}
\Psi(\vp_i,U_{ij})&=\dfrac{1}{\mathcal{N}}\left(\vp_5^\dagger U_{56}U_{67}U_{71}\vp_1\right)\left(\vp_2^\dagger U_{23}U_{34}\vp_4\right)\prod^7_{i=1}e^{-\frac{\gamma}{2}\vp^\dagger_i\vp_i} \,, \\
{|\mathcal{N}|^{-2}}&=\left( \dfrac{\gamma^2}{N} \right)^{2}\left(\dfrac{\pi}{\gamma}\right)^{-7N} \,. \label{eq:norm2}
\end{align}
A straightforward calculation shows that the reduced density matrix and its square are given by
\begin{align}
\rho_{in}(\vp,U;\phi,V) &= \left( \dfrac{\gamma}{N} \right)^{2}\left(\dfrac{\pi}{\gamma}\right)^{-3N} 
\left(\phi^\dagger_1V^\dagger_{71}U_{71}\vp_1\right)\left(\vp^\dagger_2U_{23}V^\dagger_{23}\phi_2\right) \nn \\
& \qquad \qquad  \qquad \qquad \times \prod^3_{i=1}e^{-\frac{a}{2}\vp^\dagger_i\vp_i-\frac{\gamma}{2}\phi^\dagger_i\phi_i} \,. 
\label{eq:redu2}\\
\rho_{in}^{2}(\vp,U;\phi,V)&=\dfrac{1}{N^{2}}\, \rho_{in}(\vp,U;\phi,V) \,, \label{eq:sqrho}
\end{align}
Thus the entanglement entropy is
\be
S_{EE}=\log N^2=2\log N, 
\ee
which is simply  the twice of the single meson result \eqref{singleFundEE}, and  can be understood from \eqref{numbernontrivialboundary}. 

\subsection{Two mesons sharing the same boundary}
\label{quarkantiquarkpariinoppositedirections}

We next consider several types of two overlapping meson states whose excited fluxes penetrate the same boundary, as shown in Fig. \ref{fig:2meson}. 
\subsubsection{Case (a): Opposite meson direction with 4 (anti)quarks at different positions} 
\begin{figure}[tbp]
\begin{center}
\includegraphics[height=7cm, width=14cm]{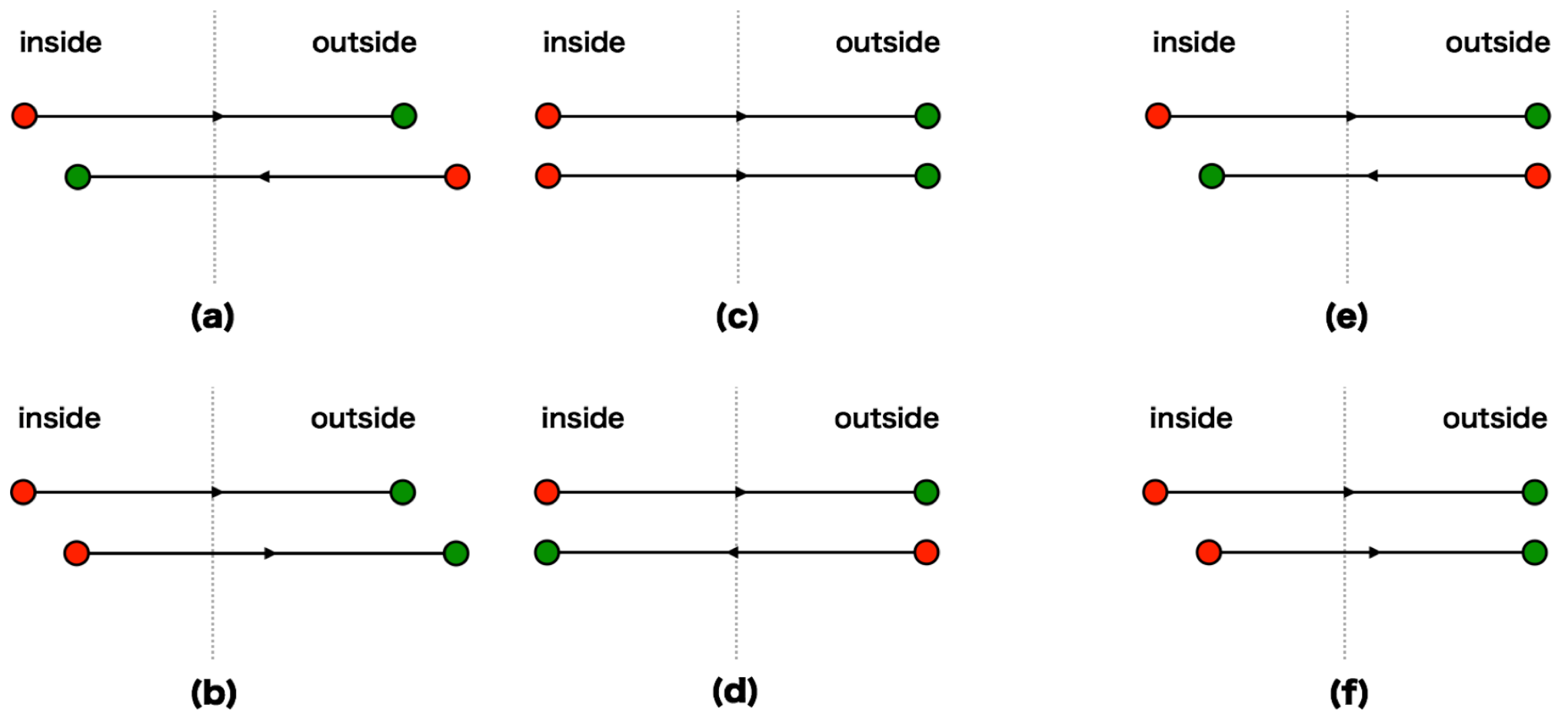}
\caption{Two meson configurations which we consider in this subsection. }\label{fig:2meson}
\end{center}
\end{figure} 
Let us consider the following state corresponding to Fig. \ref{fig:2meson} (a), 
\begin{align}
\Psi_a(\vp_i,U_{ij})&=\dfrac{1}{\mathcal{N}_a}\left(\vp_2^\dagger U_{23}U_{34}U_{45}\vp_5\right)\left(\vp_6^\dagger U^{\dagger}_{56}U^{\dagger}_{45}U^{\dagger}_{34}\vp_3\right)\prod^{7}_{n=1}e^{-\frac{\gamma}{2}\vp^\dagger_n\vp_n} \,, 
\label{wftmdpo} \\
|\mathcal{N}_a|^{-2}&= \left( \dfrac{\gamma^2}{N} \right)^2 \left(\dfrac{\pi}{\gamma}\right)^{-7N} \,.
\end{align}
Overlapping links need to be decomposed into a sum of  irreducible representations. 
Explicitly,  let us consider the link variable between 3 and 4 vertices.  Since there are one fundamental ($U_{34}$) and one anti-fundamental ($U^\dagger_{34}$) links,  
this state split into a sum of ``singlet'' and ``adjoint'' states as follows. 
Let us first rewrite our state as
\be
\Psi_a(\vp_i,U_{ij})=\dfrac{1}{\mathcal{N}_a}\left(\vp_{2\rightarrow3}^\dagger U_{34} \vp_{4\rightarrow5}\right)\left(\vp_{6\rightarrow4}^\dagger U_{34}^\dagger \vp_{3}\right)\prod^{7}_{n=1}e^{-\frac{\gamma}{2}\vp^\dagger_n\vp_n} \,, 
\ee
where $\vp_{i\rightarrow j}$'s are defined by
\begin{align}
\left(\vp^\dagger_{2\rightarrow3}\right)_a&\equiv\left(\vp_2^\dagger U_{23}\right)_a \,, \label{eq:vp2to4}\\
(\vp_{6\rightarrow4}^\dagger)_a&\equiv (\vp_6^\dagger U_{56}^\dagger U_{45}^\dagger)_a \,,  \\
(\vp_{4\rightarrow5})^a&\equiv (U_{45}\vp_5)^a \,.
\end{align}
We then decompose this state as 
\begin{align}
&\hspace{-1cm}\left(\vp_{2\rightarrow3}^\dagger U_{34} \vp_{4\rightarrow5}\right)\left(\vp_{6\rightarrow4}^\dagger U_{34}^\dagger \vp_{3}\right) 
=(\Phi^\dagger_{23})^j{}_a   (\Phi_{46})^b{}_i   \left(  (U_{34})^a{}_b   (U^\dagger_{34})^i{}_j   - \frac{1}{N}  \delta^a{}_j  \delta^i{}_b \right) \nn \\
& \qquad  \qquad  \qquad  \qquad \qquad  \qquad \quad +\dfrac{1}{N}(\vp_{2\rightarrow3}^\dagger\vp_{3})(\vp_{6\rightarrow4}^\dagger\vp_{4\rightarrow5})\,, \label{eq:dec_adj0}
\end{align}
where
\begin{align}
(\Phi^\dagger_{23})^j{}_a&\equiv\left[(\vp_{3})^j(\vp_{2\rightarrow3}^\dagger)_a-\dfrac{1}{N}\delta^j{}_{a}(\vp_{2\rightarrow3}^\dagger\vp_{3})\right],\nn\\
(\Phi_{46})^b{}_i&\equiv\left[\vp_{4\rightarrow5}^b(\vp_{6\rightarrow4}^\dagger)_i-\dfrac{1}{N}\delta^b{}_{i}\vp_{6\rightarrow4}^\dagger\vp_{4\rightarrow5}\right]. 
\end{align}
As mentioned, the first and the second terms in the r.h.s. of \eqref{eq:dec_adj0} represent  the adjoint and the singlet states, respectively. 

The reduced density matrix for this state becomes
\begin{align}
\rho_{in}(\vp_{in},U_{in};\phi_{in},V_{in}) &= \dfrac{1}{N^2} \, \rho_{(1)}(\vp_{in},U_{in};\phi_{in},V_{in}) \nn \\
&\quad \, +\left(1-\dfrac{1}{N^2}\right) \, \rho_{(adj)}(\vp_{in},U_{in};\phi_{in},V_{in}) \,,\quad
\end{align}
where 
\begin{align}
&\rho_{(1)}(\vp_{in},U_{in};\phi_{in},V_{in})\nn\\
& \qquad =\dfrac{\gamma^2}{N}\left(\dfrac{\pi}{\gamma}\right)^{-4N}\left(\vp_2^\dagger U_{23}\vp_3\right)\left(\phi_3^\dagger V_{23}^\dagger\phi_2\right)\prod^{4}_{n=1}e^{-\frac{\gamma}{2}\vp^\dagger_n\vp_n-\frac{\gamma}{2}\phi^\dagger_n\phi_n} \,, \qquad \\ 
&\rho_{(adj)}(\vp_{in},U_{in};\phi_{in},V_{in})  \nn \\
& \qquad=\dfrac{\gamma^2}{N^2-1}\left(\dfrac{\pi}{\gamma}\right)^{-4N}\left(\vp_2^\dagger U_{23}V_{23}^\dagger \phi_2\right)\left(\phi^\dagger_3\vp_3\right)\prod^{4}_{n=1}e^{-\frac{\gamma}{2}\vp^\dagger_n\vp_n-\frac{\gamma}{2}\phi^\dagger_n\phi_n}\nn\\
&\hspace{0.5cm}   \qquad-\dfrac{1}{N^2-1} \, \rho_{(1)}(\vp_{in},U_{in};\phi_{in},V_{in}) \,,
\end{align}
and these matrices satisfy
\begin{subequations}\label{subeq:algrho}
\begin{align}
\Tr\rho_{(1)}&=\Tr\rho_{(adj)}=1 \,, \qquad  
\rho^2_{(1)}  
=\rho_{(1)} \,, \\ 
\rho^2_{(adj)}   
&=\dfrac{1}{N^2-1} \, \rho_{(adj)} \,, \qquad    
\rho_{(1)}\rho_{(adj)}  
=\rho_{(adj)}\rho_{(1)} = 0 \,.  
\end{align}
\end{subequations}
Using these, the entanglement entropy for this state $\Psi_a$ is given by\footnote{We here use $\lim_{n\to1} \frac{\partial A^n}{\partial n}  = A \log A$ and $\lim_{n\to1} \frac{\partial A^{n-1}}{\partial n}  = \log A$.}  
\begin{align}
S_{EE}
&= - \lim_{n\to1} \frac{\partial }{\partial n} \Tr \rho^n \nn \\
&= - \lim_{n \to 1}\sum_{\bold{R}}  \frac{\partial }{\partial n} 
\left(  
\dfrac{1}{N^{2n}} + \left(1-\dfrac{1}{N^2}\right)^n \left( \dfrac{1}{N^2-1} \right)^{n-1}
\right) \nn \\
&=-\left\{\dfrac{1}{N^2}\log\dfrac{1}{N^2}+\left(1-\dfrac{1}{N^2}\right)\log\left(1-\dfrac{1}{N^2}\right)\right\} 
+\left(1-\dfrac{1}{N^2}\right)\log(N^2-1) \nn\\
&=\log N^2 \,.
\end{align}
In the third line, the first two terms correspond to the Shannon entropy for the superselection sector distribution ($p_{(1)}=1/N^2$ and $p_{(adj)}=1-1/N^2$), {\it i.e.}, the first 
term in \eqref{colorEE}, while the third  
term corresponds to the dimension of the adjoint representation, {\it i.e.,} the second term in \eqref{colorEE}.
On the other hand, since the genuine entanglement, the third term in \eqref{colorEE}, is absent here,    
we cannot extract any Bell pairs from this state. 

\subsubsection{Case (b): Two excited mesons in the same direction with 4 (anti)quarks at different positions} 
Instead of the wave function (\ref{wftmdpo}), we next consider the state 
\begin{align}
\Psi_b (\vp_i,U_{ij})&=\dfrac{1}{\mathcal{N}_b}\left(\vp_2^\dagger U_{23}U_{34}U_{45}\vp_5\right)\left(\vp_3^\dagger U_{34}U_{45}U_{56}\vp_6\right)\prod^{7}_{n=1}e^{-\frac{\gamma}{2}\vp^\dagger_n\vp_n}, \\
{|\mathcal{N}_b|^{-2}} &= \left(\dfrac{\gamma^2}{N}\right)^2\left(\dfrac{\pi}{\gamma}\right)^{-7N} \,,
\end{align}
where quark-anti-quark pairs lie in the same direction as Fig. \ref{fig:2meson} (b). In this case, we can decompose the state 
into ``symmetric'' and ``anti-symmetric'' states.  Similarly to the previous case, the reduced density matrix becomes  
%
\begin{align}
\rho_{in}(\vp_{in},U_{in};\phi_{in},V_{in})&=\dfrac{N+1}{2N}\rho_{(sym)}(\vp_{in},U_{in};\phi_{in},V_{in})\nn\\
&\,\quad + \, \dfrac{N-1}{2N}\rho_{(asym)}(\vp_{in},U_{in};\phi_{in},V_{in}) \,, 
\end{align}
where these matrices satisfy
\begin{subequations}\label{subeq:algrho2}
\bea
&& \Tr\rho_{(sym)} = \Tr\rho_{(asym)}=1 \,, \qquad  
\rho^2_{(sym)} =\dfrac{2}{N(N+1)}\rho_{(sym)} \,, \qquad \,
\label{subeqalgrho2a}
\\ 
&&\rho^2_{(asym)}
=\dfrac{2}{N(N-1)}\rho_{(asym)} \,, \quad 
\rho_{(sym)} \rho_{(asym)}
=\rho_{(asym)}\rho_{(sym)}  
=0 \,.\qquad \qquad
\eea
\end{subequations}
Therefore the entanglement entropy for $\Psi_b$ is evaluated as
\begin{align}
S_{EE}
%
&= - \lim_{n \to 1}\sum_{\bold{R}}  \frac{\partial }{\partial n} 
\left\{  
\left( \dfrac{N+1}{2N} \right)^n \left( \dfrac{2}{N(N+1)} \right)^{n-1} + \left( \dfrac{N-1}{2N}
\right)^n \left( \dfrac{2}{N(N-1)}  \right)^{n-1}
\right\}  \nn \\
&= -\left\{\dfrac{N+1}{2N}\log\dfrac{N+1}{2N}+\dfrac{N-1}{2N}\log\dfrac{N-1}{2N}\right\}\nn\\
&\quad +\left\{\dfrac{N+1}{2N}\log\dfrac{N(N+1)}{2}+\dfrac{N-1}{2N}\log\dfrac{N(N-1)}{2}\right\}\nn\\
&=\log N^2 \,. 
\end{align}
The result is very similar to the previous case: 
The first two terms in the second equality correspond to the Shannon entropy for the superselection sector distribution 
with $p_{sym}=(N+1)/(2N)$ and $p_{asym}=(N-1)/(2N)$,  
and the next two terms correspond to the color entanglement with $d_{sym}=N(N+1)/2$ and $d_{asym}=N(N-1)/2$, 
while there is no Bell pairs in this state.  

\subsubsection{Case (c) and (d): 4 (anti)quarks at the same positions} 
In the previous two examples, that entanglement entropy for two mesons in different quark-antiquark positions
is  $\log N^2$, which however does not contain any Bell pairs.
We here  consider two meson states in the same (anti)quark positions, which are shown to have
the different entanglement entropy. However,  again all contributions come from non-Bell pair parts. 

Let us consider the following two wave functions,
\begin{align}
\Psi_c(\vp_i,U_{ij})&=\dfrac{1}{\mathcal{N}_c}\left(\vp_{2\rightarrow3}^\dagger U_{34} \vp_{4\rightarrow5}\right)^2\prod^{7}_{n=1}e^{-\frac{\gamma}{2}\vp^\dagger_n\vp_n} \,, \\
\Psi_d(\vp_i,U_{ij})&=\dfrac{1}{\mathcal{N}_d}\left(\vp_{2\rightarrow3}^\dagger U_{34} \vp_{4\rightarrow5}\right)\left(\vp_{4\rightarrow5}^\dagger U^\dagger_{34} \vp_{2\rightarrow3}\right)\prod^{7}_{n=1}e^{-\frac{\gamma}{2}\vp^\dagger_n\vp_n} \,, \\
{|\mathcal{N}_c|^{-2}}&= {|\mathcal{N}_d|^{-2}}=\dfrac{\gamma^4}{2N(N+1)}\left(\dfrac{\pi}{\gamma}\right)^{-7N} \,,
\end{align} 
which correspond to two meson excitations in the same and opposite directions at the same (anti)quark positions, in Fig. \ref{fig:2meson} (c) and (d), respectively.

The reduced density matrices for these states become
\begin{align}
\rho_{c,in}(\vp_{in},U_{in};\phi_{in},V_{in})&=\rho_{(sym)}(\vp_{in},U_{in};\phi_{in},V_{in}) \,,  \\
\rho_{d,in}(\vp_{in},U_{in};\phi_{in},V_{in})&=\dfrac{N+1}{2N}\rho_{(1)}(\vp_{in},U_{in};\phi_{in},V_{in}) \nonumber \\
& \qquad +\dfrac{N-1}{2N}\rho_{(adj)}(\vp_{in},U_{in};\phi_{in},V_{in}) \,.
\end{align}
where $\rho_{(sym)}$ satisfies the relation \eqref{subeqalgrho2a}, while $\rho_{(1)}$ and $\rho_{(adj)}$ satisfy the relation 
\eqref{subeq:algrho}.  
Note that $\rho_c$ does not have $\rho_{(asym)}$ since identical scalars cannot form anti-symmetric combinations. 
Similar calculations as before  give the entanglement entropy as
\bea
&&S_{c,EE}=\log\dfrac{N(N+1)}{2} \,, \\
&&S_{d,EE}= -  \left\{ \frac{N+1}{2N}\log\left(\frac{N+1}{2N}\right)
+ \frac{N-1}{2N}\log\left(\frac{N-1}{2N}\right) \right\} \nn\\
&& \qquad\qquad + \, \frac{N-1}{2N}\log (N^2-1) 
\,. 
\eea
For the case (c) in the same direction,
the entanglement entropy is given solely by the color entanglement of the symmetric representation without Shanon entropy for
the superselection sector distribution,
while for the case (d) in the opposite direction, both Shannon part and the color entanglement part appear.
Again there is no Bell pair in both  cases.

\subsubsection{Case (e) and (f): only 2 (anti)quarks at the same position} 
To make the classification complete, 
we consider the following wave functions,  
\begin{align}
\Psi_e(\vp_i,U_{ij})&=\dfrac{1}{\mathcal{N}_e}\left(\vp_{2\rightarrow3}^\dagger U_{34} \vp_{4\rightarrow5}\right)\left(\vp_{4\rightarrow5}^\dagger U^\dagger_{34} \vp_{3}\right)\prod^{7}_{n=1}e^{-\frac{\gamma}{2}\vp^\dagger_n\vp_n} \,, \\
\Psi_f(\vp_i,U_{ij})&=\dfrac{1}{\mathcal{N}_f}\left(\vp_{2\rightarrow3}^\dagger U_{34} \vp_{4\rightarrow5}\right)\left(\vp_{3}^\dagger U_{34} \vp_{4\rightarrow5}\right)\prod^{7}_{n=1}e^{-\frac{\gamma}{2}\vp^\dagger_n\vp_n} \,, \\
{|\mathcal{N}_e|^{-2}}&= {|\mathcal{N}_f|^{-2}} = \dfrac{\gamma^4}{N(N+1)}\left(\dfrac{\pi}{\gamma}\right)^{-7N} \,,
\end{align} 
where mesons are in the opposite and the same directions with 2 (anti)quarks are at the same position, 
corresponding to Fig. \ref{fig:2meson} (e) and (f), respectively. 

The reduced density matrices become
\bea
&& \rho_{e,in}(\vp_{in},U_{in};\phi_{in},V_{in})=\dfrac{1}{N}\rho_{(1)}(\vp_{in},U_{in};\phi_{in},V_{in}) \nn \\
&& \qquad \qquad \qquad \qquad \qquad  \quad+\left(1-\dfrac{1}{N}\right)\rho_{(adj)}(\vp_{in},U_{in};\phi_{in},V_{in}) \,, \quad\\
&& \rho_{f,in}(\vp_{in},U_{in};\phi_{in},V_{in})=\rho_{(sym)}(\vp_{in},U_{in};\phi_{in},V_{in}) \,, 
\eea
where $\rho_{(1)}$, $\rho_{(adj)}$ and $\rho_{(sym)}$ satisfy the relation \eqref{subeq:algrho} and \eqref{subeqalgrho2a}. 
The resultant entanglement entropy becomes
\bea
&&S_{e,EE} = - \left\{ \frac{1}{N}\log\frac{1}{N} 
+ \left(1-\frac{1}{N}\right)\log\left(1-\frac{1}{N}\right) \right\} \nn \\
&& \qquad \qquad +\left(1-\frac{1}{N}\right)\log(N^2-1) \,, \\
&&S_{f,EE}=\log\dfrac{N(N+1)}{2} \,. 
\eea
Again, for the case (f) in the same direction, 
 is given solely by the color entanglement of the symmetric representation without Shanon entropy part,
 while for the case (e) in the opposite direction, both the Shannon part and the color entanglement part appear. 
Both states have no genuine entanglement.

\subsection{Four mesons at the same position}\label{subsec:4m}
\begin{figure}[tbp]
\begin{center}
\resizebox{44mm}{!}{
\includegraphics{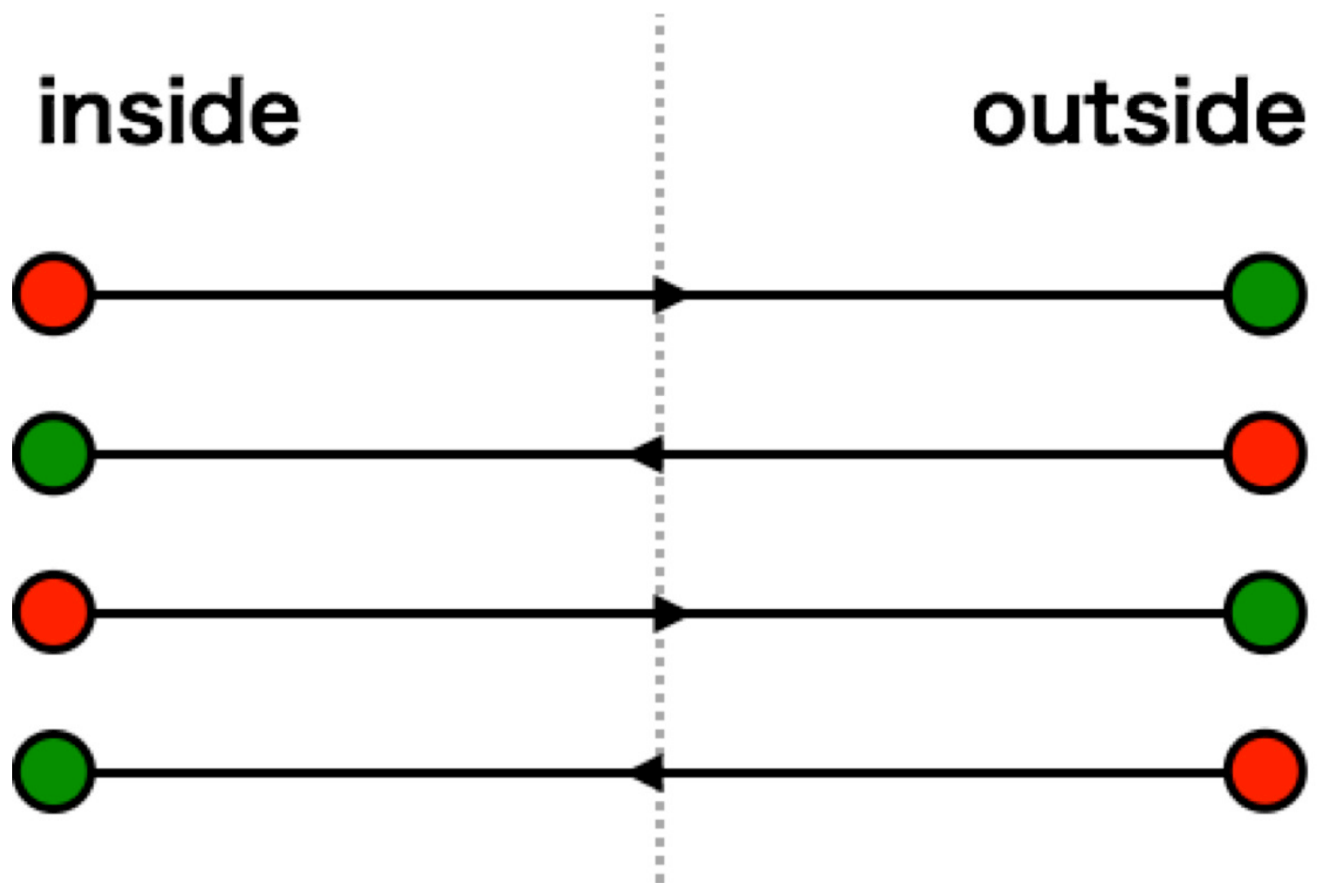}}
\caption{Four mesons we consider in this subsection. For simplicity, we consider the case that 
a distance between quark and antiquark is one lattice spacing for all four mesons. 
}\label{fig:4q}
\end{center}
\end{figure}

Let us consider a more complicated example,  four mesons at same position as is given in Fig.  \ref{fig:4q}. 
Our wave function is
\begin{align}
\Psi(\vp,U)&=\dfrac{1}{\mathcal{N}}\left[\vp_3^\dagger U_{34}\vp_{4}\right]^2\left[\vp_{4}^\dagger U^\dagger_{34}\vp_{3}\right]^2\prod^{7}_{i=1}e^{-\frac{\gamma}{2}\vp^\dagger_i\vp_i} \nn\\
&=\dfrac{1}{\mathcal{N}}(\vp^\dagger\phi)^2(\phi^\dagger\vp)^2 \prod^{7}_{i=1}e^{-\frac{\gamma}{2}\vp^\dagger_i\vp_i}
 \,, 
\label{4mesonswavefn}\\
 |\mathcal{N}|^{-2} &=\frac{\gamma^8}{4!N(N+1)(N+2)(N+3)} \left(\frac{\pi}{\gamma} \right)^{-7N} \nn \,,
\end{align}
where we define $\vp\equiv\vp_3$ and $\phi\equiv U_{34}\vp_{4}$. 
We will omit the damping factor such as $\prod^{7}_{i=1}e^{-\frac{\gamma}{2}\vp^\dagger_i\vp_i}$ from now on just for simplicity.  

One can decompose our wave function as follows.
\be
\Psi(\vp,U) = \dfrac{1}{\mathcal{N}} \left( \Psi_\bold{1}(\vp,U)+\Psi_\bold{N^2-1}(\vp,U)+\Psi_{\bold{\frac{1}{4}N^2(N-1)(N+3)}}(\vp,U)  \right)\,, 
\label{eq:4qdec}
\ee
where
\bea
&& \quad \, \,\, \Psi_\bold{1}(\vp,U)  =\dfrac{2}{N(N+1)}(\vp^\dagger_3\vp_3)^2(\vp^\dagger_{4}\vp_{4})^2 \,, \quad 
\label{wavefnforsinglet4}\\
&&\Psi_\bold{N^2-1}(\vp,U)=\dfrac{4}{N+2}(\vp^\dagger_3\vp_3)(\vp^\dagger_{4}\vp_{4})\left[(\vp^\dagger_3U_{34}\vp_{4})(\vp^\dagger_{4}U^\dagger_{34}\vp_3)\right. 
\quad
\nn\\
&&\hspace{7cm} \,\, \left.-\dfrac{1}{N}(\vp^\dagger_3\vp_3)(\vp^\dagger_{4}\vp_{4})\right] \,, \quad \\
&&\Psi_\bold{\bold{\frac{1}{4}N^2(N-1)(N+3)}}(\vp,U)=
(\vp^\dagger_3U_{34}\vp_{4})^2(\vp^\dagger_4 U^\dagger_{34}\vp_3)^2 
-\dfrac{2}{N+2} (\vp^\dagger_3\vp_3)(\vp^\dagger_{4}\vp_{4}) \nn\\
&& \qquad \qquad \qquad  \times \left[2(\vp^\dagger_3U_{34}\vp_{4})(\vp^\dagger_{4}U^\dagger_{34}\vp_3) 
-\dfrac{1}{N+1}(\vp^\dagger_3\vp_3)(\vp^\dagger_{4}\vp_{4})\right] \,. \qquad  
\label{wavefnforN2134}
\eea
Here $\bold{R}$ of $\Psi_\bold{R}(\vp,U)$ denotes the irreducible representation of $SU(N)$. Note that 
above $\Psi_\bold{1}$, $\Psi_\bold{N^2-1}$ and $\Psi_\bold{\bold{\frac{1}{4}N^2(N-1)(N+3)}}$ are not normalized at this moment.  
Like the case (c) before for two mesons at the same position,  
anti-symmetric combinations disappear. See appendix \ref{app:4qdec} for the derivation of \eqref{eq:4qdec}.  
Since these wave functions are mutually orthogonal, our reduced density matrix also becomes the sum of each sector as 
\be
\label{singeleadjointotherdecomposision}
\rho(\vp,\phi)=\sum_{\bold{R}}p_{\bold{R}}\rho_\bold{R}(\vp,\phi) \,,
\ee
where $\bold{R} = \bold{1},\bold{N^2-1}$ and $\bold{\frac{1}{4}N^2(N-1)(N+3)}$.

From \eqref{wavefnforsinglet4} - \eqref{wavefnforN2134}, together with the normalization that $\Tr \,\rho_{\bold{R}}=1$ for each $\rho_{\bold{R}}$, we obtain\footnote{Here we omit the damping factor $\prod^{3}_{i=1}e^{-\frac{\gamma}{2}(\vp^\dagger_i\vp_i+\phi^\dagger_i\phi_i)}$.} 
\bea
&&\hspace{-1.2cm} \rho_{\bold{1}}(\vp,\phi)=\dfrac{\gamma^4}{N(N+1)(N+2)(N+3)}(\vp^\dagger_3\vp_3)^2(\phi^\dagger_3\phi_3)^2\left(\dfrac{\pi}{\gamma}\right)^{-3N} \,, \\
&&\hspace{-1.2cm} \rho_{\bold{N^2-1}}(\vp,\phi)=\dfrac{\gamma^4}{(N^2-1)(N+2)(N+3)}\left(\dfrac{\pi}{\gamma}\right)^{-3N}
(\vp^\dagger_3\vp_3)(\phi^\dagger_3\phi_3) \nn\\
&& \qquad \qquad \qquad\qquad \qquad \times \left[(\vp^\dagger_3\phi_3)(\phi^\dagger_3\vp_3)-\dfrac{1}{N}(\vp^\dagger_3\vp_3)(\phi^\dagger_3\phi_3)\right],\\
&&\hspace{-1.2cm} \rho_{\bold{\frac{1}{4}N^2(N-1)(N+3)}}(\vp,\phi) = \dfrac{\gamma^4}{N^2(N-1)(N+3)} 
\left(\dfrac{\pi}{\gamma}\right)^{-3N} 
\Biggl[(\vp^\dagger_3\phi_3)^2(\phi^\dagger_3\vp_3)^2  
\nn\\
&&\hspace{-0.9cm}  -\dfrac{4}{N+2}(\vp^\dagger_3\vp_3)(\phi^\dagger_3\phi_3)\left((\vp^\dagger_3\phi_3)(\phi^\dagger_3\vp_3)-\dfrac{1}{2(N+1)}(\vp^\dagger_3\vp_3)(\phi^\dagger_{3}\phi_{3})\right)\Biggr] \,, \,\, \hspace{0.2cm}
\eea
while \eqref{4mesonswavefn} directly gives the reduced density matrix as 
\begin{align}
\rho(\vp,\phi)&=\dfrac{\gamma^4}{6N(N+1)(N+2)(N+3)}\left(\dfrac{\pi}{\gamma}\right)^{-3N} 
 \Biggl[ (\vp^\dagger_3\phi_3)^2(\phi^\dagger_3\vp_3)^2  \nn \\
 & \qquad +4(\vp^\dagger_3\phi_3)(\phi^\dagger_3\vp_3)(\vp^\dagger_3\vp_3)(\phi^\dagger_3\phi_3)+(\vp^\dagger_3\vp_3)^2(\phi^\dagger_3\phi_3)^2 \Biggr] \,.
\end{align}
A comparison of these with the formula \eqref{singeleadjointotherdecomposision} yields 
\bea
p_{\bold{1}}=\dfrac{1}{6}\dfrac{(N+2)(N+3)}{N(N+1)} \,, \qquad 
p_{\bold{N^2-1}}=\dfrac{2}{3}\dfrac{(N-1)(N+3)}{N(N+2)} \,,\\
p_{\bold{\frac{1}{4}N^2(N-1)(N+3)}}=\dfrac{1}{6}\dfrac{N(N-1)}{(N+1)(N+2)} \,. \qquad  \qquad  \,\,\,
\eea
Since these density matrices satisfy the relation
\be
\rho_{\bold{R}}\,\rho_{\bold{R}^\prime}=\dfrac{1}{d_{\bold{R}}}\delta_{\bold{R}\bold{R}^\prime}\rho_{\bold{R}} \,,
\ee
the resulting entanglement entropy for this state is given by
\begin{align}
S_{EE} &= - \lim_{n\to1} \frac{\partial }{\partial n} \Tr \rho^n \nn = - \lim_{n \to 1}\sum_{\bold{R}}  \frac{\partial }{\partial n} \left( 
\frac{p_{\bold{R}}^n}{d_{\bold{R}}^{n-1}} 
\right) \\
&= -\sum_{\bold{R}}p_{\bold{R}}\log p_{\bold{R}}+\sum_{\bold{R}}p_{\bold{R}}\log d_{\bold{R}} \,,
\label{eq:4mesonEE}
\end{align}
where $d_{\bold{R}}$ is the dimension of the irreducible representation $\bold{R}$. Eq.~\eqref{eq:4mesonEE} corresponds to 
eq.~\eqref{colorEE} with the vanishing Bell pair term.

\section{Comments on three contributions to the entanglement entropy in the extended Hilbert space}
\label{commentsection}

We have discussed the entanglement entropy for the 1+1 dimensional non-Abelian gauge theory on the lattice. In the extended Hilbert space definition, we have three contributions to the entanglement entropy as \eqref{colorEE}. In this section, we illustrate these three contributions, by considering the following three examples: 
1) two spins, 2) the $Z_2$ gauge theory on 1d spatial lattice and 3) the $SU(2)$ gauge theory with the fundamental scalar field on 1d spatial lattice. 
All of these examples give the same mathematical structure in the extended Hilbert space definition and result in the same values of entanglement entropy; however, the interpretation differs for each cases. These viewpoints is probably not new for experts, but we think it is still useful to present it here. 

\subsection{Two spins}
Let us consider two spins, whose Hilbert space is a tensor product of left and right spins and both of which takes two values ($\pm$), which is
\be
{\cal{H}}  = \ket{\pm}_{\textrm{left}} \otimes \ket{\pm}_{\textrm{right}} =  \{\ket{++} \,, \ket{+-} \,, \ket{-+} \,, \ket{--} \,\}. 
\label{eq:2spinH}
\ee 
If we consider following specific state, 
\be
\ket{\psi} = \frac{1}{\sqrt{2}} \ket{++} +  \frac{1}{\sqrt{2}} \ket{--}  \,,\label{eq:sstate1}
\ee
clearly this gives entanglement entropy $S_{EE} = \log 2$. This represents the genuine entanglement since one can extract this by entanglement distillation.  

\subsection{$Z_2$ pure gauge theory on 1d spatial lattice}
\label{Z2gaugetheory}
\begin{figure}[tbp]
\begin{center}
\includegraphics[height=1.2cm, width=12.5cm]{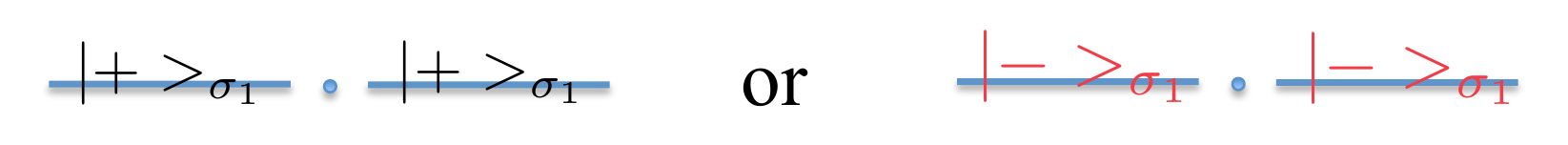}
\caption{Due to the Gauss's law constraint, the physical Hilbert space is 2-dimensional for the $Z_2$ pure gauge theory on the 1d spatial lattice. Here $\pm$ represent ``electric fluxes'', which label superselection sectors. \label{Z2lattice}}  
\end{center}
\end{figure}

Instead of above two spins, let us consider a $Z_2$ pure gauge theory on the 1d spatial lattice. To simplify the argument, we take an extreme situation that the space is composed of only two links, (12) and (23), with the periodic boundary condition (vertices 1 and 3 are identical). See Fig. \ref{Z2lattice}. 
Each link variable $U_{ij}$ takes $\pm$ values and the corresponding basis is denoted by $\ket{\pm}_{\sigma_3}$, which satisfy $\hat{\sigma}_3 \ket{\pm}_{\sigma_3} = \pm \ket{\pm}_{\sigma_3}$, where Pauli $\hat{\sigma}_3$ is a link operator. The non-trivial gauge transformation is given by acting $\hat{\sigma}_1$ on both links (12) and (23). Here $\hat{\sigma}_1$ is the electric flux operator. See \S2 of \cite{Ghosh:2015iwa} for more detail. We have eigenfunctions of $\hat{\sigma}_1$,  
\bea
\ket{\pm}_{\sigma_1} \equiv \frac{1}{\sqrt{2}} \left( \ket{+}_{\sigma_3} \pm \ket{-}_{\sigma_3} \right) \,,
\eea
where the eigenvalue $\pm$ represents the electric flux on the corresponding link. Therefore, there exist only two independent gauge invariant states in this setup, which are given by $\ket{++}_{\sigma_1}$ or $\ket{--}_{\sigma_1}$. 
Note that states such as $\ket{+-}_{\sigma_1}$ or $\ket{-+}_{\sigma_1}$ are not allowed. It is not gauge invariant due to Gauss's law; the electric flux cannot take different values between (12) and (23).

However, in the extended Hilbert space, we allow non-gauge invariant states, then the Hilbert space becomes 
\be
{\cal{H}}=\{ \ket{++}_{\sigma_1} \,, \ket{+-}_{\sigma_1} \,, \ket{-+}_{\sigma_1} \,, \ket{--}_{\sigma_1} \,\},\label{eq:Z2H}
\ee
which gives the same structure as \eqref{eq:2spinH}. 
Under this setup, let us consider the following specific state 
\bea
\label{Z2entangledstate}
\ket{\psi} = \frac{1}{\sqrt{2}}  \ket{++}_{\sigma_1} + \frac{1}{\sqrt{2}}   \ket{--}_{\sigma_1}  \,. \label{eq:sstate2}
\eea
Clearly the state \eqref{Z2entangledstate} shows the entanglement entropy $S_{EE} = \log 2$ in the extended Hilbert space definition due to mathematically the same structure as two spins case. However physical interpretation is different. 

In {\it physical} Hilbert space, the two physical states $\ket{++}_{\sigma_1}$ and $\ket{--}_{\sigma_1}$ can not be mixed with each other by any ``local'' gauge invariant operation\footnote{For example, if one want to convert $\ket{++}_{\sigma_1}$ into $\ket{--}_{\sigma_1}$ by using only {\it local} operations, one must have unphysical $\ket{+-}_{\sigma_1}$ or $\ket{-+}_{\sigma_1}$ as a intermediate state.}.  
This means that $\ket{++}_{\sigma_1}$ and $\ket{--}_{\sigma_1}$ belong to different superselection sectors. 
In addition, in physical Hilbert space, allowed states are $\ket{++}_{\sigma_1}$ and $\ket{--}_{\sigma_1}$ only. Therefore 
once we fix the superselection sector (either $+$ or $-$), then physical Hilbert space shows manifestly a tensor product structure between inside and outside.  
Therefore $S_{EE} = \log 2$ is not the ``genuine'' entanglement entropy, but rather 
should be interpreted as the distribution entropy associated with superselection sectors, which is given by
\bea
S = - \sum_i p_i \log p_i \quad \mbox{with} \quad p_\pm = \frac{1}{2}
\eea
where $\pm$ represents the electric flux at the boundary. 
This is a typical example of the first contribution in \eqref{colorEE}.

\subsection{$SU(2)$ gauge theory with fundamental matter} 
Let us consider again 1d spatial lattice in Fig. \ref{sevenvertexmodel}, where 
vertices 1, 2, 3 and link (71), (12), (23) are ``inside'' and the rest is ``outside''.
We consider the wave function for the excited meson which is given by  
\begin{align}
\Psi(\vp_i,U_{ij})& = \dfrac{1}{\mathcal{N}}\;\left[\vp_3^\dagger U_{34} \vp_{4}\right]  =  \dfrac{1}{\mathcal{N}}\;\left[\vp_{inside\, a}^{\dagger} \vp^a_{outside}\right]  \,,
\label{mesonexcitedsinglet}
\end{align}
where we omit the Gaussian factor for the normalization, but keep explicitly the color index  $a = \pm$ in the fundamental representation. We denote $\vp_3 = \vp_{inside}$ 
and $\vp_{outside} = U_{34} \vp_{4}$, where link (34) and vertex 4 are both outside. 
This is the one we studied in \S \ref{singlemesonEE}. 
Note that this wave function is gauge-singlet. 
We focus on the color degrees of freedom for $\vp_{inside}^{\dagger}$ and $\vp_{outside}$.
Taking a map as
\begin{subequations}
\bea
\vp_{inside\,\pm}^{\dagger}\mapsto \ket{\pm}_{inside}, \\
\vp^{\pm}_{outside}\mapsto \ket{\pm}_{outside}. 
\eea
\end{subequations}
then the meson wave function \eqref{mesonexcitedsinglet} becomes
\bea
\ket{\psi} =\frac{1}{\sqrt{2}} \ket{++} +  \frac{1}{\sqrt{2}} \ket{--}  \,. \label{eq:sstate3}
\eea
\begin{figure}[tbp]
\begin{center}
\includegraphics[height=1.1cm, width=9cm]{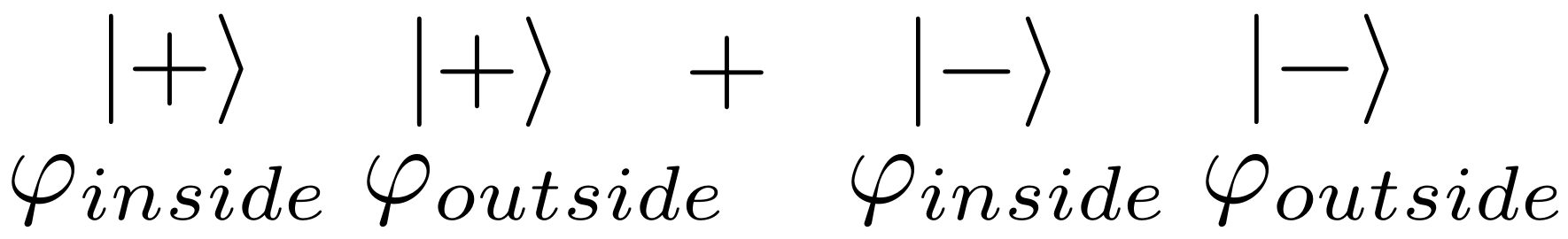}
\caption{Due to gauge singlet condition for mesons, Hilbert space for quark-anti-quark color configuration is 1-dimensional. $\pm$ represent color charge.\label{Z2color}}  
\end{center}
\end{figure} 
Mathematical structure is the same as \S \ref{Z2gaugetheory}. 
In extended Hilbert space, we include gauge-non-singlet $\ket{+-}$ and $\ket{-+}$ states and the Hilbert space becomes 
\be
{\cal{H}}=\{ \ket{++} \,, \ket{+-} \,, \ket{-+} \,, \ket{--} \,\},
\label{eq:SU2H}
\ee
then we have $S_{EE} = \log 2$ for the entanglement entropy. 
However physics is different again;   
This $\log 2$ is due to the ``color'' entanglement, which is associated with the color singlet meson 
between inside and outside (anti)quarks. 
Note that here we have only one superselection sector, the fundamental representation at the boundary. 
Thus the first contribution of \eqref{colorEE} vanishes. 
In this way, one obtain color entanglement associated with all the 
boundary with the dimension of color representation as 
\be
S = \sum_{i} \log d_i  
\ee
in each superselection sector, where $i$ represents all boundary vertex and $d_i$ is the dimension of the color representation. 

Note that since gauge singlet condition prohibits the color configuration $\ket{+-}$ and $\ket{-+}$, one cannot destroy the color entanglement by LOCC. This implies that one cannot extract the entanglement by the distillation, just as the same as superselection sector prohibits $\ket{+-}_{\sigma_1}$ and $\ket{-+}_{\sigma_1}$ in \S \ref{Z2gaugetheory} and one cannot extract the entanglement by the distillation in that case. 

\paragraph{- Summary -}
We calculate the entanglement entropy for a specific state in three cases. For all cases, the extended Hilbert space is constructed as a tensor product of 2-valued ($\pm$) degrees of freedom at inside (left side) and outside (right side), giving the same structure \eqref{eq:2spinH}, \eqref{eq:Z2H} and \eqref{eq:SU2H}. Thus the states \eqref{eq:sstate1}, \eqref{eq:sstate2}, and \eqref{eq:sstate3} automatically give the same entanglement entropy $\log 2$. However the interpretations for the results are different. 

In the two spin model, there is no constraint in the system, {\it i.e.} the extended Hilbert space is just the physical Hilbert space itself. In other words, there is no extension of the Hilbert space. Then we can interpret the entanglement entropy as just the number of Bell pairs, the 3rd contribution in \eqref{colorEE}.  

In the $Z_2$ pure gauge theory case, the states $\ket{++}_{\sigma_1}$ and $\ket{--}_{\sigma_1}$ are separated by the gauge constraint, {\it i.e.} these two belong to different superselection sector. Then the entanglement entropy just originates from the probability distribution for the each sector, becoming the Shannon entropy, the 1st contribution in \eqref{colorEE}. 

In the $SU(2)$ gauge theory case, although $\ket{++}_{\sigma_1}$ and $\ket{--}_{\sigma_1}$ belong to the same superselection sector (fundamental representation at the boundary), the color degrees of freedom $\pm$ is {\it not} 
observable. Since the entanglement entropy here is associated with color, it should be non-extractable, and it corresponds to  the 2nd contribution in \eqref{colorEE}. 

Lesson from the second and third examples is that 
there appears the entanglement which can {\it not} be extracted by local operations 
when we consider entanglement in gauge theories. 
This is because gauge theories prohibit the local operations which break the gauge invariance.

\section{Transfer matrix and hopping parameter expansion}
\label{TmatrixandHPE}
In the previous sections, we consider the entanglement entropy for various states, which are chosen by hand, in order to demonstrate how the first and the second contributions in  \eqref{colorEE} appear in the 1+1 lattice gauge theories with scalar fields.
Our next task is to calculate the entanglement entropy for the grand state of the 1+1 dimensional $SU(N)$ gauge theories with the fundamental scalar field on the lattice.
We are particularly interested in how the genuine entanglement, {\it i.e.,} the third contribution in \eqref{colorEE} shows up in this theory.
In this section, we give several definitions and formula useful for this purpose.
The calculation of the entanglement entropy will be given in the next section.

\subsection{Lattice action and Transfer matrix}

Actions for gauge field and fundamental scalar field on the 2-d lattice are denoted as
\beqa
S &=& S_G + S_M.
\label{eq:action_GM}
\eeqa
Explicitly the pure gauge action  $S_G$ is given by
\beqa
S_G &=& \beta \sum_{\vec{n}} \tr  \left( \plaq{\vec{n}}(F)+\plaq{\vec{n}}^\dagger(F)-{\bf 2}\right) \,, \quad   \beta \equiv \frac{1}{g_{YM}^2 a^2}  \,,
\label{eq:plaquette} 
\eeqa
where the plaquette $U_{P,\vec{n}}(F)$ is defined as a minimal closed loop in the 2-dimensional (Euclidean) space-time as
\beqa
\plaq{\vec{n}} &=& U_{\vec n,0} U_{\vec n+\hat{0},1} U_{\vec n +\hat{1},0}^\dagger U_{\vec n,\hat{1}}^\dagger ,
\eeqa
$F$ stands for the fundamental representation, {\it i.e.}, $U(F)$ is an $N\times N$ unitary matrix for $G =$ SU(N), $\hat{\mu}$ is the unit vector in the $\mu$ direction ($\mu=0,1$ represent Euclidean time direction and space direction, respectively),  $g_{YM}$ is the bare gauge coupling constant  and $a$ is the lattice spacing.
The gauge invariant action for the fundamental scalar field is given by
\beqa
S_M &=& a^2\sum_{\vec n} \varphi_{\vec{n}}^\dagger\left( \nabla^2 -m^2\right) \varphi_{\vec n} \,,\\ 
a^2\nabla^2\varphi_{\vec n} &=&  \sum_{\mu=0,1}\left\{ U_{\vec{n},\mu} (F) \varphi_{\vec n + \hat\mu} +
U_{\vec{n}-\hat{\mu},\mu}^\dagger (F) \varphi_{\vec n-\hat\mu} - 2\varphi_{\vec{n}}\right\} \,,
\eeqa
where $m$ is the mass of the scalar field. 

The entanglement entropy for the grand state of the theory is often calculated in the path integral formalism using the replica method.
In this paper, however, in order to distinguish all three contributions in \eqref{colorEE}, we employ the operator formalism, as in the previous case for the pure gauge theories \cite{Aoki:2016lma}, where the transfer matrix and its eigenstates (instead of the Hamiltonian) were used to calculate the entanglement entropy. 
The transfer matrix $\hat T$ is defined to generate the time translation by one (temporal) lattice unit 
 \cite{Creutz:1976ch,Luscher:1976ms} and thus is symbolically denoted as
\be
\hat{T}(a_t,a) \equiv e^{- a_t H_L(a_t,a)} \,.
\ee
where $a_t$  ($a$) is the lattice spacing in the temporal (spatial) direction and $H_L(a_t,a)$ is the lattice ``Hamiltonian" for the discrete time.
In the $a_t\rightarrow 0$ limit while keeping the spatial lattice spacing $a$ non-zero, we recover the lattice Hamiltonian \eqref{eq:H_KS}
for the continuous time as
\beqa
H &=& \lim_{a_t\rightarrow 0} H_L(a_t,a) =  - \lim_{a_t\rightarrow 0} \frac{1}{a_t} \log \hat T(a_t,a) \,.
\eeqa
Although eigenvalues and eigenstates are different between $H$ and $\hat T$ at non-zero $a_t$, they agree in the continuum limit that $(a_t, a) \rightarrow (0, 0)$.  In particular, the eigenstate for the largest eigenvalue of $\hat T$ corresponds to the ground state of the theory at $a_t=a\not=0$ in one to one, and it approaches to the ground state of the continuum theory as $a\rightarrow 0$.  
 Hereafter we simply write $\hat T = \hat T (a,a)$.
 
To derive the transfer matrix from the path integral with the given action \eqref{eq:action_GM},
we first take the temporal gauge $U_{\vec{n},0}={\bf 1}$ for ${}^\forall\vec{n}$, and then define $\hat T$ as
\beqa
\langle \Psi_{\rm out} \vert (\hat T)^{N_t} \vert \Psi_{\rm in} \rangle &=&
\int_{\Psi_0=\Psi_{\rm in}}^{\Psi_{N_t}=\Psi_{\rm out}} \prod_{n_0=1}^{N_t-1} {\cal D}\Psi_{n_0} \ e^{S_G + S_M},
\label{eq:TM}
\eeqa
where $\Psi_{n_0}  =\{ U_{n_0}, \varphi_{n_0}\}$ represents the gauge field $U_{n_0} = \{ U_{\vec n,1}\}$
and the scalar fields $\varphi_{n_0}=\{ \varphi_{\vec{n}} \}$ at a give time slice $n_0$, and
we fix them to $\Psi_{\rm in}$ at $n_0=0$ and $\Psi_{\rm out}$ at $n_0=N_t$. 

We next rewrite the left-hand side of \eqref{eq:TM} as
\beqa
\int_{\Psi_0=\Psi_{\rm in}}^{\Psi_{N_t}=\Psi_{\rm out}} \prod_{n_0=1}^{N_t-1} {\cal D}\Psi_{n_0}
\prod_{n_0=0}^{N_t-1} \langle \Psi_{n_0+1} \vert \hat T \vert \Psi_{n_0} \rangle, 
\label{eq:TM2}
\eeqa
which must be equal to the right-hand side. 

We thus obtain
\beqa
\langle \Psi^A \vert \hat T \vert \Psi^B \rangle &\equiv& T( \Psi^A,  \Psi^B)
= T_0(\Psi^A) c_G T_G(U^A,U^B) T_M(\varphi^A,\varphi^B) T_0(\Psi^B), \nn \\
\label{eq:TM3}
\eeqa
where
\beqa
 T_0(\Psi) &=&\prod_{n=0}^{N_l-1} \exp\left[\frac{1}{2} \left\{\varphi_{n}^\dagger U_{n} \varphi_{n+1} + \varphi_{n+1}^\dagger U_{n}^\dagger \varphi_{n} - (m^2a^2+2) \varphi_{n}^\dagger \varphi_{\vec n} \right\}\right], \nn \\
 \\
c_G T_G(U,V)  &=&\prod_{n=0}^{N_l-1} \exp\left\{\beta\, \tr \left( U_{n} V_{n}^\dagger + V_{n} U_{n}^\dagger - \bf{2}\right)\right\}, \\
T_M(\varphi, \phi) &=& \prod_{n=0}^{N_l-1} \exp\left[ \phi_{n}^\dagger \varphi_{n}  + \varphi_{n}^\dagger\phi_{n} \right],
\eeqa
with the periodic BC in space that $U_{N_l}=U_{0}$, $V_{N_l} =V_{0}$ and $\varphi_{N_l}=\varphi_{0}$, $\phi_{N_l}=\phi_0$,
where $n$ represents the 1-dimensional spatial lattice point, we suppress an index for the direction $\mu=1$ of $U_{n,1}$ and $V_{n,1}$ is omitted for simplicity, and $c_G$ is a normalization factor such that the largest eigenvalue of $T_G$ is one (see \eqref{cGdeflambda}).

Note that the expression of $ \hat T$ which satisfies \eqref{eq:TM}  is not unique. Instead of \eqref{eq:TM3}, 
the asymmetric choice,
\beqa
\langle \Psi^A \vert \hat T \vert \Psi^B \rangle 
&=&  c_G T_G(U^A,U^B) T_M(\varphi^A,\varphi^B) T_0^2(\Psi^B) \,,
\label{eq:TM4}
\eeqa
also satisfies it, and thus can be used equally well.
We use \eqref{eq:TM4} rather than \eqref{eq:TM3} for our convenience. 

\subsection{Character expansion}
In Ref.~\cite{Aoki:2016lma}, the character expansion is applied to
the pure gauge part of the transfer matrix $T_G$ 
as
\beqa
c_G T_G(U,V)  &=& \prod_{n=0}^{N_l-1} \sum_{\bf R} d_{\bf R} \lambda_{\bf R}(\beta) \chi_{\bf R}(U_n V_n^\dagger)\\
&=& \lambda_{\bf 1}^{N_l}(\beta)  \prod_{n=0}^{N_l-1} \left\{ 1 + \sum_{{\bf R}\not= {\bf 1}} d_{\bf R} \frac{\lambda_{\bf R}(\beta)}{\lambda_{\bf 1}(\beta)}
\chi_{\bf R}(U_n V_n^\dagger) \right\}
\label{cGdeflambda}
\eeqa 
where $\chi_{\bf R}(U) = \tr \, U({\bf R}) $ is a character for the  irreducible representation ${\bf R}$ with its dimension $d_{\bf R}=\chi_{\bf R}({\bf 1})$,  and ${\bf R}={\bf 1}$ denotes the trivial representation, and  
$c_G =   \lambda_{\bf 1}^{N_l}(\beta) $. 
The expansion coefficient is given by 
\beqa
\lambda_{\bf R}(\beta) &\equiv& \frac{1}{d_{\bf R}} \int dU \chi_{{{\bf R}}}(U) \exp\left[\beta \chi_{\bf F}\left( U + U^\dagger -\bf{2}\right)\right] ,
\eeqa
which satisfies  
\beqa
0\le \frac{\lambda_{\bf R}(\beta)}{\lambda_{\bf 1}(\beta)}  \le 1 \,, 
\qquad  \lim_{\beta \to \infty}\frac{\lambda_{\bf R}(\beta)}{\lambda_{\bf 1}(\beta)} =1 \,.  
\eeqa
Note that $ \chi_{\bf R}(U^\dagger) = \chi_{\bar {\bf R}}(U)$ and $\lambda_{\bf R}(\beta) =  \lambda_{\bar{\bf R}}(\beta)$.
We take $c_G =  \lambda_{\bf 1}^{N_l}(\beta)$ for the normalization.

There are several useful formula for the group integral as follows.
\beqa
\int [dU]\, \chi_{\bf R} (AU)\chi_{{\bf R'}} (U^{\dagger}B)&=&\frac{1}{d_{\bf R}}\delta_{{\bf RR'}}  \chi (AB) \,,\label{formula1}\\
\int [dU]\, \chi_{\bf R} (AUBU^{\dagger})&=&\frac{1}{d_{\bf R}}  \chi_{\bf R} (A)  \chi_{\bf R} (B) \,,\label{formula2}
\eeqa
\beqa
&& \int [dU]\, \chi_{\bf R} (AUBU^\dagger)\chi_{{\bf R}} (CUDU^{\dagger}) 
\nn\\
&& \, = \frac{1}{d_{\bf R}^2-1} \biggl[ \chi_{\bf R}(A)\chi_{\bf R}(C)\chi_{\bf R}(B)\chi_{\bf R}(D) 
  +\chi_{\bf R}(AC) \chi_{\bf R}(BD) \nn \\
&& \qquad -\frac{1}{d_{\bf R}}\Bigl\{ \chi_{\bf R}(A)\chi_{\bf R}(C)\chi_{\bf R}(BD)
+\chi_{\bf R}(AC)\chi_{\bf R}(B)\chi_{\bf R}(D)\Bigr\}\biggr]  \,, \qquad 
\label{formula3} \\
&&\int [dU]\, \chi_{\bf R} (AUBU^\dagger CUDU^{\dagger}) 
\nn \\
&& 
= \, \frac{1}{d_{\bf R}^2-1}  \biggl[ \chi_{\bf R}(AC)\chi_{\bf R}(B)\chi_{\bf R}(D) +\chi_{\bf R}(A) \chi_{\bf R}(C)\chi_{\bf R}(BD) \nn \\
&& \qquad -\frac{1}{d_{\bf R}}\Bigl\{ \chi_{\bf R}(AC)\chi_{\bf R}(BD)
+\chi_{\bf R}(A)\chi_{\bf R}(C)\chi_{\bf R}(B)\chi_{\bf R}(D)\Bigr\}\biggr]  \,. \qquad 
\label{formula4} 
\eeqa

\subsection{Hopping parameter expansion (HPE)} 
\label{HPEandT}
We rescale $\hat{T} \to c_G \hat{T}$ so that $c_G$ does not appear any more. 
We also rescale scalar fields as $\varphi_n \rightarrow \sqrt{K}\varphi_n$ and $\phi_n \rightarrow \sqrt{K}\phi_n$ with
the hopping parameter $K=1/(m^2 a^2 +2)$, so that
$T_0^2$ and $T_M$ becomes
\beqa
T_0^2(\Psi) &=&\prod_{n=0}^{N_l-1} \exp\left[ - \varphi_{n}^\dagger \varphi_{n} + K
 \left\{\varphi_{n}^\dagger U_{n} \varphi_{n+1} + \varphi_{n+1}^\dagger U_{n}^\dagger \varphi_{n} \right\}\right], \\
 T_M(\varphi, \phi) &=& \prod_{n=0}^{N_l-1} \exp\left[ K\left(\varphi_{n} \phi_{n}^\dagger +  \varphi_{n}^\dagger\phi_{n}\right)\right].
\eeqa

Assuming that $K$ is small, we can expand the transfer matrix around $K=0$, which is called the hopping parameter expansion (HPE) \cite{Kawamoto:1980fd,Blairon:1980pk}.
 In this case, the Feynman rule for the scalar field is given by
 \begin{eqnarray}
& \langle (\varphi_n^\dagger)_a \varphi_m^b \rangle = \delta_{nm} \delta^{b}{}_a, \qquad \langle \phi_n^a\phi_m^b \rangle = \langle (\phi_n^\dagger)_a(\phi_m^\dagger)_b \rangle = 0 \,, \qquad \\
& \langle  (\varphi_{n_a}^\dagger)_a \varphi_{n_b}^b  (\varphi_{n_c}^\dagger)_c \varphi_{n_d}^d \rangle  = \delta^{b}{}_a\delta^{d}{}_c\delta_{n_a,n_b}\delta_{n_c,n_d} +  \delta^{d}{}_a\delta^{b}{}_c\delta_{n_a,n_d}\delta_{n_c,n_b} .
 \end{eqnarray}
 
We define states as 
 \beqa
\langle \Phi^B\vert n, m \rangle &=&  \phi_n^\dagger V_{n\to m} \phi_m, \qquad V_{n\to m} \equiv V_n V_{n+1} \cdots V_{m-1}  \\
\langle \Phi^B\vert 0 \rangle &=& 1 .
\eeqa

We then calculate $\hat{T} \vert 0 \rangle$ up to the order $K^4$ and  $\hat{T} \vert n,m \rangle$ up to the order $K^3$,
which are given below.
\bea
T\vert 0\rangle &=&\left(1+K^2NN_l+\dfrac{3}{2}K^4NN_l+\dfrac{1}{2}K^4N^2N_l^2\right)\vert 0\rangle \nn \\
&&\, +\sum_n (K^2+2K^4+K^4NN_l)\vert n,n\rangle \nn\\
&&\,\, +\dfrac{1}{2}K^4 \sum_n\ket{n,n}\ket{n,n}+ K^4\sum_{n\not=m}\ket{n,n}\ket{m,m}\nn\\
&&\,\,\, +\sum_n K^3\left(\frac{\lambda_{\bf F}}{\lambda_{\bf 1}}\right)\left\{\vert n, n+1\rangle +\vert n, n-1\rangle\right\} \nn \\
&& \,\,\,\, +\sum_nK^4\left(\dfrac{\lambda_{\bf F}}{\lambda_{\bf 1}}\right)^2\left\{\ket{n,n+2}+\ket{n,n-2}\right\} \,,\label{Transferonzero} 
\eea
\bea
T\vert n, n\rangle &=&N\{1+2K^2(N+1) + K^2 N (N_l -2)\}\vert 0\rangle +K^2 \ket{n,n}  \nn\\
&& \,+ K^2 N \sum_{m} \ket{m,m} \nn\\
&& \, \, \,\, +K^3\left( \frac{\lambda_{\bf F}}{\lambda_{\bf 1}} \right) \Big( \ket{n,n+1}+\ket{n+1,n} \nn \\
&& \,\,\,\,\, +\ket{n,n-1}+\ket{n-1,n} \Big) \nn\\
&& 
\,\,\,\,\,\,
+K^3N\left( \frac{\lambda_{\bf F}}{\lambda_{\bf 1}}\right)\sum_m \Big(\ket{m,m+1} 
+\ket{m+1,m} \Big) \,, 
\label{Transferonnn}
\eea
\bea
T\vert n, n+1\rangle &=&NK\left\{1+4(N+1)K^2+N(N_{l}-3)K^2\right\}\vert 0\rangle \nn \\
&& \, +K^2\left( \frac{\lambda_{\bf F}}{\lambda_{\bf 1}}\right)\vert n,n+1\rangle\nn\\
&&\,\, +K^3\left( \frac{\lambda_{\bf F}}{\lambda_{\bf 1}}\right)^2\left\{\vert n,n+2\rangle +\vert n-1,n+1\rangle\right\}\nn \\
&&\,\,\, + {K^3\left\{\vert n,n\rangle +\vert n+1,n+1\rangle\right\}} 
\nn \\
&&\,\,\, 
+ K^3N\sum_{m} \vert m,m\rangle \,, 
\eea
\bea
T\vert n, n-1\rangle &=&NK\left\{1+4(N+1)K^2+N(N_{l}-3)K^2\right\}\vert 0\rangle \nn \\
&&\,+K^2\left( \frac{\lambda_{\bf F}}{\lambda_{\bf 1}}\right)\vert n,n-1\rangle\nn\\
&&\,\,+K^3\left( \frac{\lambda_{\bf F}}{\lambda_{\bf 1}}\right)^2\left\{\vert n,n-2\rangle +\vert n+1,n-1\rangle\right\}\nn \\
&&\,\,\, + {K^3\left\{\vert n,n\rangle +\vert n-1,n-1\rangle\right\}} \nn \\
&&\,\,\,\, + K^3N\sum_{m} \vert m,m\rangle  \,, 
\eea
\bea
T\vert n, n+2\rangle &=&NK^2\vert 0\rangle +K^2\left( \frac{\lambda_{\bf F}}{\lambda_{\bf 1}}\right)^2\vert n,n+2\rangle \nn\\
&&\, +K^3\left( \frac{\lambda_{\bf F}}{\lambda_{\bf 1}}\right)^3\left\{\vert n,n+3\rangle +\vert n-1,n+2\rangle\right\} \nn\\
&& \,\, +K^3\left( \frac{\lambda_{\bf F}}{\lambda_{\bf 1}}\right)\left\{\vert n,n+1\rangle +\vert n+1,n+2\rangle\right\} \,, 
\eea
\bea
T\vert n, n-2\rangle &=&NK^2\vert 0\rangle +K^2\left( \frac{\lambda_{\bf F}}{\lambda_{\bf 1}}\right)^2\vert n,n-2\rangle \nn\\
&&\, +K^3\left( \frac{\lambda_{\bf F}}{\lambda_{\bf 1}}\right)^3\left\{\vert n,n-3\rangle +\vert n+1,n-2\rangle\right\} \nn\\
&&\,\, +K^3\left( \frac{\lambda_{\bf F}}{\lambda_{\bf 1}}\right)\left\{\vert n,n-1\rangle +\vert n-1,n-2\rangle\right\} \,,
\eea
\bea
T\vert n, n+3\rangle &=&NK^3\vert 0\rangle +K^2\left( \frac{\lambda_{\bf F}}{\lambda_{\bf 1}}\right)^3\vert n,n+3\rangle \nn\\
&&\, +K^3\left( \frac{\lambda_{\bf F}}{\lambda_{\bf 1}}\right)^4\left\{\vert n,n+4\rangle +\vert n-1,n+3\rangle\right\} \nn\\
&& \,\, +K^3\left( \frac{\lambda_{\bf F}}{\lambda_{\bf 1}}\right)^2\left\{\vert n,n+2\rangle +\vert n+1,n+3\rangle\right\} \,,
\eea
\bea
T\vert n, n-3\rangle &=&NK^3\vert 0\rangle +K^2\left( \frac{\lambda_{\bf F}}{\lambda_{\bf 1}}\right)^3\vert n,n-3\rangle \nn\\
&&\, +K^3\left( \frac{\lambda_{\bf F}}{\lambda_{\bf 1}}\right)^4\left\{\vert n,n-4\rangle +\vert n+1,n-3\rangle\right\} \nn\\
&& \,\, +K^3\left( \frac{\lambda_{\bf F}}{\lambda_{\bf 1}}\right)^2\left\{\vert n,n-2\rangle +\vert n-1,n-3\rangle\right\} \,, 
\eea
\bea
T\vert n, n+l\rangle &=&K^2\left( \frac{\lambda_{\bf F}}{\lambda_{\bf 1}}\right)^l\vert n,n+l\rangle \nn \\
&&\, +K^3\left( \frac{\lambda_{\bf F}}{\lambda_{\bf 1}}\right)^{l+1}\left\{\vert n,n+l+1\rangle +\vert n-1,n+l\rangle\right\} \nn\\
&&\,\, +K^3\left( \frac{\lambda_{\bf F}}{\lambda_{\bf 1}}\right)^{l-1}\left\{\vert n,n+l-1\rangle +\vert n+1,n+l\rangle\right\}, \nn \\
&&\quad \qquad\qquad\qquad\qquad\qquad\qquad\qquad({\rm for} \ l >3)~~~~~~~ 
\eea
\bea
T\vert n, n-l\rangle &=&K^2\left( \frac{\lambda_{\bf F}}{\lambda_{\bf 1}}\right)^l\vert n,n-l\rangle \nn \\
&&\,+K^3\left( \frac{\lambda_{\bf F}}{\lambda_{\bf 1}}\right)^{l+1}\left\{\vert n,n-l-1\rangle +\vert n+1,n-l\rangle\right\} \nn\\
&&\,\, +K^3\left( \frac{\lambda_{\bf F}}{\lambda_{\bf 1}}\right)^{l-1}\left\{\vert n,n-l+1\rangle +\vert n-1,n-l\rangle\right\}, \nn\\
&&\quad \qquad\qquad\qquad\qquad\qquad\qquad\qquad({\rm for} \ l >3)~~~~~~~ \,.
\label{Transferonnnl}
\eea
There are mixings among states, therefore we have to diagonalize them. 
Up to the $K^2$ order, the states $\vert n, n+l\rangle$ and $\vert n,n-l\rangle$ for $l\geq 3$ 
are the eigenstates for the transfer matrix, since
\be
T\vert n, n\pm l\rangle =K^2\left( \frac{\lambda_{\bf F}}{\lambda_{\bf 1}}\right)^l\vert n,n\pm l\rangle ,\quad({\rm for} \ l \geq 3) .
\ee
Thus at this order,
all we have to do is to diagonalize the mixing among $\vert 0\rangle , \vert n, n\rangle,\vert n,n\pm 1\rangle$, and $\vert n, n\pm 2\rangle$ states. 


\section{Entanglement entropy for the ground state by the HPE}  
\label{Bellsection}
\subsection{Taking into higher order corrections in $K$}
In \S \ref{puregaugeEE} and \ref{singlemesonEE}, we have seen that a single Wilson loop or a single meson state holds nonzero entanglement entropy due to the second term of (\ref{colorEE}), which is associated with the color entanglement. 
In \S \ref{multimesonEE}, we discussed multiple meson states, whose fluxes connect quarks-antiquarks  through the boundary. In this case, by decomposing the wave function into irreducible representations, we obtain multiple superselection sectors, and as a result, nonzero entanglement entropy associated with the first term (the classical Shannon entropy for the probability distribution of each irreducible representation) as well as the second term (the color entanglement part) of \eqref{colorEE}
appear. 
We have shown these explicit examples, in order to illustrate how  we obtain these non-Bell terms in the entanglement entropy in the extended Hilbert space definition. 

One might wonder 
whether the Bell pair part  of the entanglement,  third term of \eqref{colorEE}, never appears 
in 2-dimensional gauge theory.
In the pure gauge theory, we cannot have any Bell pairs due to the absence of local degrees of freedom \cite{Aoki:2016lma}.
In gauge theories with matter fields, of course, we can always prepare an appropriate linear combination of meson states {by hand}, which produces the  Bell pair part in  \eqref{colorEE}.  
Our main interest/concern here, however, is how the ground state of the gauge theory (the strong coupling ground state) 
acquires  entanglements including Bell pairs from matter fields, and how entanglements for the ground state of the continuum gauge theory can be understood in terms of the lattice ground state.

In the 2-dimensional gauge theory without matter fields, which corresponds  to the leading order of the HPE  ($K=0$), the ground state can be calculated exactly at an arbitrary coupling without strong coupling expansion,\footnote{In 2-dimensions, there is no plaquette term ({\it i.e.,} magnetic field), therefore its Hamiltonian has a similar structure to the strong coupling limit of higher dimensional ones.} and it is written by the tensor product of a trivial state on each link
satisfying $\hat{J}^2_{ij} \ket{0}_{ij} = 0$ as
\bea
\ket{0}_{\rm strong} = \bigotimes_{ij} \ket{0}_{ij} \,.
\eea
Thus the entanglement entropy of the strong coupling ground state $\ket{0}_{\rm strong}$ vanishes at $K=0$.\footnote{This state corresponds to the wave function  $\chi_{\bf 1}(U)$, while the wave function $\chi_{\bf R}(U)$ with  ${\bold{R}} \neq {\bf 1}$ describes an excited state, which yields nonzero entanglement entropy as \eqref{Aokietalresult}.} 

Therefore, in this section, we study how the higher order in $K$ of the HPE makes the strong coupling ground state entangled, and which part of \eqref{colorEE} appears. We will show the following properties. 
\begin{itemize}
\item The strong coupling ground state has no entanglement up to order $K^2$ in HPE (\S \ref{Teigeneigen}). 
\item The first term (the Shannon part for the superselection sector distribution) and the second term (the color entanglement part) first appear  at the order $K^3$ for the ground state (\S \ref{subsec:K3}).
\item The third term (the Bell pair part) first appears at the order $K^6$ for the ground state (\S \ref{subsec:K6}).
\end{itemize}
Since all these contributions are positive definite order by order in the HPE, they never cancel each other. Therefore, the above observations imply that the 2-dimensional Yang-Mills theory with matter fields keeps all three types of entanglements in \eqref{colorEE}  in the continuum limit.

From now on, we simply denote the strong coupling ground state $\ket{0}_{\rm strong} $ as $\ket{0}$.

\subsection{Eigenstates and eigenvalues of $\hat{T}$ up to $\mathcal{O}(K^2)$}\label{Teigeneigen}
We first  consider contributions at $\mathcal{O}(K^2)$, and  diagonalize the transfer matrix $\hat{T}$. 
At this order, 
the generic state $\ket{\Psi}_K$ which mixes with the strong coupling ground state $\ket{0}$ can be expressed as 
\bea
\label{genericstateomega}
\ket{\Psi}_K &\equiv& f_0\ket{0}+\sum_{n}a_n\ket{n,n}+\sum_{n}b_n\ket{n,n+1}+\sum_{n}c_n\ket{n,n-1} \nn \\
&&\,\,\, \qquad +\sum_{n}d_n\ket{n,n+2}+\sum_{n}e_n\ket{n,n-2}.
\eea
We thus determine the $K$ dependent coefficients $a_n, b_n, c_n, d_n, e_n$, and $f_0$ in such a way that 
\bea
\label{eigeneqs}
\hat{T} \ket{\Psi}_K  \propto \ket{\Psi}_K 
\eea
is satisfied. 
As long as the HPE converges, the ground state in the HPE must contain $\ket{0}$, so that we will  consider the state with $f_0 \neq 0$. 
We can set $f_0\equiv1$ without loss of generality,
and we denote it as 
\begin{align}
\ket{G^+}_K\equiv &\ket{0}+\sum_{n}a_n\ket{n,n}+\sum_{n}b_n\ket{n,n+1}+\sum_{n}c_n\ket{n,n-1} \nn \\
&\qquad +\sum_{n}d_n\ket{n,n+2}+\sum_{n}e_n\ket{n,n-2} \,. 
\end{align}

At the $O(K^2)$, using the transfer matrix $\hat{T}$ given in \S \ref{HPEandT}, the ground state is given by
\begin{align}
\label{Gpmketdef}
\ket{G^+}_K&=\ket{0}+\sum_na_n^+\ket{n,n} ,\quad \mbox{where}  \quad
a^+_n =\dfrac{K^2}{G^+_K-(1+NN_\ell)K^2} \,, \\
\label{Gpmdef}
G^+_K&=\dfrac{1}{2}\{1+K^2(1+2NN_\ell)\}\nn\\
&\quad +\dfrac{1}{2}\sqrt{1-2(1-2NN_{\ell})K^2+\{1+4N(NN_\ell+2)N_{\ell}\}K^4}.
\end{align}
The complete list of all other eigenstates and eigenvalues at this order are given in the appendix \ref{eigenstatesfortransfer}. 

In the $K\rightarrow 0$ limit, this state $\ket{G^+}_K$ 
has a maximum eigenvalue of the transfer matrix, $G_K^+=1$,  which corresponds to
 ``zero energy'',  since  the transfer matrix is related to the  ``Hamiltonian" as $T\approx e^{-aH}$. 
We therefore  identify this state as the ground state at $O(K^2)$,  which is composed of the {\it strong coupling} ground state $\ket{0}$ and {\it lattice} point-like exited meson states $\ket{ n,n}$.
It is thus clear that this state does not have any entanglement.
More precisely, we can write this ground state as a product state as 
\be
\ket{G^+}_K=\left(\ket{0}_{\rm in}+K^2\sum_{\rm in}\ket{n,n}\right)\left(\ket{0}_{\rm out}+K^2\sum_{\rm out}\ket{n,n}\right)+\mathcal{O}(K^3). \label{eq:gsK2} \, 
\ee
This means that there is no correlation between inside and outside and thus  no entanglement at this order.

On the other hand, the vacuum state $\ket{0}_{\rm cont.}$ in the continuum gauge theory  is expected to have non-zero entanglement. 
So there still remains a qualitative difference (whether it is entangled or not) between the ground state $\ket{G^+}_K$ at $O(K^2)$ and the continuum ground state $\ket{0}_{\rm cont.}$.  
This indicates we need higher order of the HPE than $K^2$. Indeed,  since the vacuum state in the continuum theory is realized in the continuum limit as    
\bea
\lim_{\substack{ K \to 1/2, \\ \beta \to \infty }} \ket{G^+}_K \to \ket{0}_{\rm cont.} \,, 
\eea
where $ K = \left({2+(ma)^2}\right)^{-1}\to 1/2 $  and $\beta =\left(g_{\rm YM}^2 a^2\right)^{-1} \to \infty$ as $a\to 0$ for finite mass $m$ and coupling $g_{\rm YM}$, 
the higher order terms in the HPE become more and more important as we approach the continuum limit.
Note that our calculations include all order of the gauge coupling constant at each order of the HPE.
What we will see next is that once we take into account higher order corrections, $\ket{G^+}_K$ contains various contributions of the entanglement in \eqref{colorEE}. 

\subsection{Entanglement appear at $\mathcal{O}(K^3)$ corrections}\label{subsec:K3}
As a next step, we check how $K^3$ order effects modify the properties of $\ket{G^+}_K$. 
At the order $K^3$, $\ket{G^+}_K$  becomes
\be
\ket{G^+}_K=\ket{0}+K^2\sum_n\ket{n,n}+K^3\dfrac{\lambda_{\bf F}}{\lambda_{\bf 1}}\sum_n(\ket{n,n+1}+\ket{n,n-1})+\mathcal{O}(K^4). \label{eq:orderK3} 
\ee
We therefore see that the $\mathcal{O}(K^3)$ contributions (quark-antiquark pairs separated with unit length) give the entanglement, once we divide the system into inside and outside. 
 
Before  we will see that the first and the second terms of (\ref{colorEE}) for the entanglement entropy becomes nonzero at this order, let us first  explain how we obtain the above result.
The eigenvalue equation is given by
\be
\hat{T}\ket{G^+}_K=G^{+}_K\ket{G^+}_K, \label{eq:eveq}
\ee
which must be solved order by order.
 Expanding $\hat{T}, \ket{G^+}_K$, and $G^+_K$ in power series of $K$, and using the resuts at $O(K^2)$ in \eqref{Gpmketdef} and \eqref{Gpmdef}, we have  
\begin{align}
\hat{T}&=\hat{T}_0+K^1\hat{T}_1+K^2\hat{T}_2+K^3\hat{T}_3+\mathcal{O}(K^4) \,, \label{Kexp1}\\
\ket{G^+}_K&=\ket{G^+_0}+K^1\ket{G^+_1}+K^2\ket{G^+_2}+K^3\ket{G^+_3}+\mathcal{O}(K^4)\nn\\
&=\ket{0}+0+K^2\sum_n\ket{n,n}+K^3\ket{G^+_3}+\mathcal{O}(K^4) \,, \label{Kexp2}\\
G^+_K&=G^+_0+K^1G^+_1+K^2 G^+_2+K^3 G^+_3+\mathcal{O}(K^4)\nn\\
&=1+0+K^2 2NN_\ell + K^3 G^+_3+\mathcal{O}(K^4)\label{Kexp3} \,, 
\end{align}
and solve the equations at each order in $K$. 

Since \eqref{Gpmketdef} and \eqref{Gpmdef} satisfy eigenvalue equation \eqref{eq:eveq} up to  ${\mathcal{O}}(K^2)$, it is enough to consider only $\mathcal{O}(K^3)$ terms. 
Left hand side of \eqref{eq:eveq} becomes
\begin{align}
 K^3(\hat{T}_3\ket{G^+_0}+\hat{T}_2\ket{G^+_1}+\hat{T}_1\ket{G^+_2}+\hat{T}_0\ket{G^+_3}) \,.
\end{align}
while
the right hand side of \eqref{eq:eveq} is
\begin{align}
K^3(G_3\ket{G^+_0}+G_2\ket{G^+_1}+G_1\ket{G^+_2}+G_0\ket{G^+_3}) \,.
\end{align}
We therefore obtain 
\begin{align}
\hat{T}_3\ket{0}+\hat{T}_0\ket{G^+_3}=G^+_3\ket{0}+\ket{G^+_3}, 
\end{align}
where we used $\ket{G^+_1}=0$ and  $G^+_1=0$, which are seen from \eqref{Gpmketdef} and \eqref{Gpmdef},
and $\hat{T}_1=0$ for $\ket{n,n}$ from \eqref{Transferonnn}. 
Since $\hat{T}_3\ket{0}=\displaystyle\frac{\lambda_{\bf F}}{\lambda_{\bf 1}}\sum_n(\ket{n,n+1}+\ket{n,n-1})$ from \eqref{Transferonzero}, 
the above equation is equivalent to
\begin{align}
\hat{T}_0\ket{G^+_3}+\dfrac{\lambda_{\bf F}}{\lambda_{\bf 1}}\sum_n(\ket{n,n+1}+\ket{n,n-1})=G^+_3\ket{0}+\ket{G^+_3}. \label{eq:evorderK3}
\end{align}
By substituting the ansatz that 
\be
\ket{G^+_3}=\omega\ket{0}+\sum_n\alpha_n\ket{n,n}+\sum_n\beta_n\ket{n,n+1}+\sum_n\gamma_n\ket{n,n-1},
\ee
into \eqref{eq:evorderK3},  together with the relation 
\bea
\hat{T}_0 \ket{0} = \ket{0} \,, \quad \hat{T}_0 \ket{n,n} = N \ket{n,n} \,, \quad  \mbox{(and the rest is zero)}
\eea
from \eqref{Transferonzero} - \eqref{Transferonnnl}, 
we have 
\begin{align}
&\hspace{-5mm}\omega\ket{0}+N\sum_n\alpha_n\ket{0}+\dfrac{\lambda_{\bf F}}{\lambda_{\bf 1}}\sum_n(\ket{n,n+1}+\ket{n,n-1})\nn\\
&=(G_3^++\omega)\ket{0}+\sum_n\alpha_n\ket{n,n}+\sum_n\beta_n\ket{n,n+1}+\sum_n\gamma_n\ket{n,n-1}. 
\end{align}
Comparing  l.h.s. and r.h.s., we finally obtain, 
\begin{align}
\beta_n = \gamma_n=\dfrac{\lambda_{\bf F}}{\lambda_{\bf 1}} \,, \quad 
G_3^+&= \alpha_n =0 \,.
\end{align}
while $\omega$ is an arbitrary constant. 

In conclusion,  
we have obtained the eigenstate at the order of $K^3$ as 
\begin{align}
\ket{G^+}_K&=(1+\omega K^3)\ket{0}+K^2\sum_n\ket{n,n}+K^3\dfrac{\lambda_{\bf F}}{\lambda_{\bf 1}}\sum_n(\ket{n,n+1}+\ket{n,n-1})+\mathcal{O}(K^4)\nn\\
&=(1+\omega K^3)\Big[\ket{0}+\frac{1}{(1+\omega K^3)} \Big\{K^2\sum_n\ket{n,n}  \nn \\
&\qquad \qquad \qquad \qquad + K^3\dfrac{\lambda_{\bf F}}{\lambda_{\bf 1}}\sum_n(\ket{n,n+1}+\ket{n,n-1})+\mathcal{O}(K^4)\Big\}\Big]\nn\\
&\propto\ket{0}+K^2\sum_n\ket{n,n}+K^3\dfrac{\lambda_{\bf F}}{\lambda_{\bf 1}}\sum_n(\ket{n,n+1}+\ket{n,n-1})+\mathcal{O}(K^4), \\
G^+&=1+2NN_\ell K^2+\mathcal{O}(K^4) \,.
\end{align}
This  exactly gives eq.~\eqref{eq:orderK3}. 

At this order, the ground state includes terms such as $\ket{i,i+1}$ and $\ket{i+1,i}$, where $i$-th vertex is located in the inside and $(i+1)$-th vertex is located in the outside. 
Thus there appears the non-trivial electric flux penetrating the boundary, so that we have a nontrivial superselection sector distribution. Namely, the term $\ket{i,i+1}(\ket{i+1,i})$ belongs to a (anti-)fundamental sector, wheres the other terms  to a singlet sector. Then the state makes the non-zero entanglement entropy corresponding to the first and second terms in (\ref{colorEE}).

We can confirm that there is {\it no} Bell pairs at this order by investigating each superselection sector. For simplicity, we here assume that there is only one boundary between $i$-th inner vertex and ($i+1$)-th outer vertex with the {\it outer} link variable $U_{i,i+1}$.  

The singlet sector for the ground state still shows the tensor product structure, 
\begin{align}
\hspace{-0.1cm}\left.\ket{G^+}_K\right|_{\textrm{singlet}}&=\left(\ket{0}_{\rm in}+K^2\sum_{\rm in}\ket{n,n}+K^3\dfrac{\lambda_{\bf F}}{\lambda_{\bf 1}}\sum_{\rm in}(\ket{n,n+1}+\ket{n,n-1})\right)\nn\\
&\quad \otimes\left(\ket{0}_{\rm out}+K^2\sum_{\rm out}\ket{n,n}+K^3\dfrac{\lambda_{\bf F}}{\lambda_{\bf 1}}\sum_{\rm out}(\ket{n,n+1}+\ket{n,n-1})\right)\nn
\\
&\qquad +\mathcal{O}(K^4). 
\label{eq:gsK3}
\end{align}
Thus the singlet sector is {\it not} entangled at all.

Next let us focus on the fundamental sector (the discussion for the anti-fundamental sector is almost same). In this sector the state is simply $\ket{i,i+1}$ up to its normalization. If we explicitly denote the color degree of freedom $a (=1,2,\dots N)$, the state can be represented as
\begin{align}
\left.\ket{G^+}_K\right|_{\textrm{fundamental}}\propto & \, K^3\dfrac{\lambda_{\bf F}}{\lambda_{\bf 1}}\ket{i,i+1}+\mathcal{O}(K^4)\nn\\
=&\, K^3\dfrac{\lambda_{\bf F}}{\lambda_{\bf 1}}\sum_a\left(\ket{i,\textrm{bdy}}_{a\;{\rm in}}\otimes\ket{\textrm{bdy},i+1}^a_{\rm out}\right) +\mathcal{O}(K^4),
\label{eq:fundamental_K3}
\end{align}
where $\ket{i,\textrm{bdy}}_a$ corresponds to a quark at $i$-th vertex with flux going to outside area, and $\ket{\textrm{bdy},i+1}^a$  to the similar object. (As the wave function, these objects are represented as $(\vp^{\dagger}_i)_a$ and $(U_{i,i+1}\vp_{i+1})^a$, respectively.) Clearly the state gives the entanglement entropy $\log N$  originating entirely from the color degree of freedom. For each color, the state shows the tensor product structure, indicating the absence of Bell pairs.

Before closing this subsection, we calculate the entanglement entropy for this ground state, which is given by
\begin{eqnarray}
\ket{G^+}_K = \left.\ket{G^+}_K\right|_{\textrm{singlet}} + \left.\ket{G^+}_K\right|_{\textrm{fundamental}} 
+ \left.\ket{G^+}_K\right|_{\textrm{anti-fundamental}}  \,, \quad  
\end{eqnarray}
up to $\mathcal{O}(K^4)$, where the state in the singlet sector $\left.\ket{G^+}_K\right|_{\textrm{singlet}} $ is given by eq.~\eqref{eq:gsK3} while the one in the fundamental sector $\left.\ket{G^+}_K\right|_{\textrm{fundamental}}$ by eq.~\eqref{eq:fundamental_K3}.
The corresponding reduced density matrix $\rho_{\rm red.}$ becomes
\be
\rho_{\rm red.} = p_{\bf 1} \rho_{\bf 1} + p_{\bf F} \rho _{\bf F} + p_{\bf \bar F} \rho _{\bf \bar F} \,,
\ee 
where
\bea
&& p_{\bf 1} = \frac{|{\cal N}_{\rm in}|^2 |{\cal N}_{\rm out}|^2 }{|{\cal N}|^2} \,, \qquad
  p_{\bf F} = p_{\bf \bar F} =\frac{   c_F^2 N}{|{\cal N}|^2}  \,, \\
&&  \rho_{\bf 1} = \frac{1}{|{\cal N}_{\rm in}|^2 } {}_{\rm in} \hspace{-1mm}\ket{\bf 1}  \bra{\bf 1}_{\rm in} , \quad
 \rho_{\bf F} = \frac{1}{N}{}_{\rm in} \hspace{-1mm}\ket{\bf F} \bra{\bf F}_{\rm in} , 
 \quad  \rho_{\bf \bar F} =\frac{1}{N}{}_{\rm in} \hspace{-1mm} \ket{\bf \bar F} \bra{\bf \bar F}_{\rm in} \,,\qquad
\eea
with
\beqa
&&\hspace{-8mm} |{\cal N}_{\rm in/out}|^2 =( 1 + K^2 N N_{\rm in/out})^2 + K^4 N N_{\rm in/out}  + 2 c_F^2N( N_{\rm in/out} -1) \,, \qquad \\
&& |{\cal N}|^2 = |{\cal N}_{\rm in}|^2 |{\cal N}_{\rm out}|^2 + 2 c_F^2 N \,, \qquad
c_F \equiv K^3\frac{\lambda_{\bf F}}{\lambda_{\bf 1}} \,, \quad \\
&& \ket{\bf 1}_{\rm in} = \ket{0}_{\rm in} +K^2 \sum_{\rm in} \ket{n,n} + {c_F}\sum_{\rm in} \left(\ket{n,n+1}+\ket{n+1,n}\right),\\
&& \ket{\bf F}_{\rm in} = \sum_a\ket{i, {\rm bdy}}_{a\, {\rm in}}, \quad
\ket{\bf \bar F}_{\rm in} = \sum_{\bar a}\ket{{\rm bdy} , i}_{\rm in}^{\bar a}.
\eeqa
Here $N_{\rm in (out)}$ is a number of sites in the inside (outside) region, thus $N_l = N_{\rm in}+N_{\rm out}$, and  $|{\cal N}|^2$ and 
$|{\cal N}_{\rm in/out}|^2$  are defined as $|{\cal N}|^2 = {}_K \hspace{-1mm} \braket{G^+|G^+}_K$, $|{\cal N}_{\rm in/out}|^2 = {}_{\rm in/out} \hspace{-1mm} \braket{\bf 1|1}_{\rm in/out}$. 
It is easy to see 
\beqa
&& \rho_{\bf 1}^2 = \rho_{\bf 1} \,,\quad
\rho_{\bf F}^2 = \frac{1}{N}\rho_{\bf F} \,, \quad
\rho_{\bf \bar F}^2 = \frac{1}{N}\rho_{\bf \bar F} \,.
\eeqa
The total entanglement entropy $S_{EE}$ for this state is given by
\beqa
S_{EE} &=&\sum_{\bf R=1,F , \bar F} \left\{ - p_{\bf R} \log p_{\bf R} +  p_{\bf R}\log d_{\bf R} \right\} ,
\eeqa
where $d_{\bf 1}=1, d_{\bf F}=d_{\bf \bar F} = N$. 


\subsection{$\mathcal{O}(K^4)$ and $\mathcal{O}(K^5)$ corrections}\label{subsec:K4}
By almost the same way as the previous subsection, we obtain $\mathcal{O}(K^4)$ correction to the state $\ket{G^+}_K$ and eigenvalue $G^+$ as 
\begin{align}
\ket{G^+}_K=&\, \ket{0}+K^2\sum_n\ket{n,n}+K^3\dfrac{\lambda_{\bf F}}{\lambda_{\bf 1}}\sum_n(\ket{n,n+1}+\ket{n,n-1}) \nn\\
&\,\,+K^4\left(3\sum_n\ket{n,n}+2\sum_n\ket{n,n}\ket{n,n}+\sum_{n\not=m}\ket{n,n}\ket{m,m}\right.\nn\\
&\,\,\,\,+\left.\left(\dfrac{\lambda_{\bf F}}{\lambda_{\bf 1}}\right)^2\sum_n(\ket{n,n+2}+\ket{n,n-2})\right)+\mathcal{O}(K^5) \,, \label{eq:orderK4}\\
G^+=&\,1+2NN_\ell K^2 \nn\\
&\,\,+\left(7+2\dfrac{\lambda_{\bf F}}{\lambda_{\bf 1}}+2NN_l\right)NN_l K^4+\mathcal{O}(K^5)\,\blue{,}\label{eq:eigenvalueK4}
\end{align}
again having three sectors (singlet, fundamental, and anti-fundamental).

This is obtained from the equation \eqref{eq:eveq} at order $K^4$ as follows. Using expansions \eqref{Kexp1},  \eqref{Kexp2} and  \eqref{Kexp3} at order $K^4$, we obtain
\begin{align}
&\hat{T}_4\ket{0}+\hat{T}_2\sum_n\ket{n,n}+\hat{T}_1\dfrac{\lambda_{\bf F}}{\lambda_{\bf 1}}\sum_n(\ket{n,n+1}+\ket{n,n-1})+\hat{T}_0\ket{G_4^+}\nn\\
&\, \quad= \, G^+_4\ket{0}+2NN_l\sum_n\ket{n,n}+\ket{G_4^+}
\label{eq:eigenK4}.
\end{align}
A comparison between  the l.h.s and r.h.s. in (\ref{eq:eigenK4}), 
together with the formula \eqref{Transferonzero}
and the ansatz
\begin{align}
\ket{G^+_4}=&\,\, \omega\ket{0}+\sum_n\alpha_n\ket{n,n}+\sum_n\alpha_{n,n}\ket{n,n}\ket{n,n}+\sum_{n\not= m}\alpha_{n,m}\ket{n,n}\ket{m,m}\nn\\
&\,\, +\sum_n\beta_n\ket{n,n+1}+\sum_n\gamma_n\ket{n,n-1}\nn\\
&\,\,\,\, \,\,+\sum_n\delta_n\ket{n,n+2}+\sum_n\varepsilon_n\ket{n,n-2} \,, 
\end{align}
gives 
\begin{align}
& \delta_n=\varepsilon_n=\left(\dfrac{\lambda_{\bf F}}{\lambda_{\bf 1}}\right)^2 \,,  \quad 
\beta_n=\gamma_n=0 \,, \\
& \alpha_{n,m}=1\;\;(\mbox{for}\,\,\, n \neq m) \,, \quad 
\alpha_{n,n}=\dfrac{1}{2} \,, \quad
\alpha_n=3 \,, \label{c4unitresult}\\
& G_4^+=\left(7+2\dfrac{\lambda_{\bf F}}{\lambda_{\bf 1}}+2NN_l\right)NN_l \,.
\end{align}
These lead to results (\ref{eq:orderK4}) and (\ref{eq:eigenvalueK4}). 


Let us consider whether the ground state wave function \eqref{eq:orderK4} at  $\mathcal{O}(K^4)$ in the HPE  contains the Bell pair part of the entanglement entropy \eqref{colorEE}. To see this, we examine singlet sector and (anti-)fundamental sector separately. Again we assume a single boundary between the $i$-th inner vertex and the ($i+1$)-th outer vertex.

We first analyze the singlet sector in the following way. 
If we assume that the Bell pair part is absent,  we immediately notice 
that the term  $\ket{n,n}_{\rm in}\ket{m,m}_{\rm out}$, where
the $n$-th vertex is in the  inside and the $m$-th vertex is in the outside, must appear in the ground state as 
\be
\left.\ket{G^+}_K\right|_{\textrm{singlet}} \supset c_4 \, K^4 \ket{n,n}_{\rm in}\ket{m,m}_{\rm out}
\ee
with the coefficient $c_4 = 1$, which is determined from the result at the lower order given in  \eqref{eq:gsK3}, 
since such a term must be a part of the tensor product of inside-only excited states and outside-only excited states.
Inversely, if $c_4 \neq 1$, such a state can {\it not} be written as a tensor product state given in  \eqref{eq:gsK3}.
The result \eqref{c4unitresult} indeed shows $c_4 = 1$  
for our wave function \eqref{eq:orderK4} at  $\mathcal{O}(K^4)$.
Therefore {\it no} Bell pair part appears in this sector.

In the higher orders, we can employ the similar analysis.
With the assumption on the tensor product structure,
we can predict coefficients of  new terms at the higher order from results at lower orders.
At $\mathcal{O}(K^5)$, for instance, the term $\ket{n,n}\ket{m,m}$ cannot exist since there is no corresponding inside-only or outside-only excited terms at lower orders. 
Indeed we cannot construct $\ket{n,n}\ket{m,m}$ states from $\ket{0}$ by the $\mathcal{O}(K^5)$ part of $\hat{T}$,
since we need at least $\mathcal{O}(K^6)$ terms, which consist of  two ``U''-shaped contributions.\footnote{Each ``U''-shape is $\mathcal{O}(K^3)$, see appendix  \ref{yokoyafeynman} for details.}

The (anti-)fundamental sector at $K^4$ order has almost the same structure as the $K^3$ order case, where only difference is the distance of (anti-)quark from the boundary. As is the case of $\mathcal{O}(K^3)$, we can explicitly represent the state as
\begin{align}
\left.\ket{G^+}_K\right|_{\textrm{fundamental}}\propto & \, K^3\dfrac{\lambda_{\bf F}}{\lambda_{\bf 1}}\ket{i,i+1}+K^4\left(\dfrac{\lambda_{\bf F}}{\lambda_{\bf 1}}\right)^2\left(\ket{i,i+2}+\ket{i-1,i+1}\right) \nonumber \\
& \qquad +\mathcal{O}(K^5)\nn\\
=& \, K^3\dfrac{\lambda_{\bf F}}{\lambda_{\bf 1}}\sum_a\left(\ket{i,\textrm{bdy}}_{a\;{\rm in}}\otimes\ket{\textrm{bdy},i+1}^a_{\rm out}\right)\nn\\
&+K^4\left(\dfrac{\lambda_{\bf F}}{\lambda_{\bf 1}}\right)^2\sum_{a'}\left(\ket{i,\textrm{bdy}}_{a'\;{\rm in}}\otimes\ket{\textrm{bdy},i+2}^{a'}_{\rm out}\right)\nn\\
& \,\, +K^4\left(\dfrac{\lambda_{\bf F}}{\lambda_{\bf 1}}\right)^2\sum_{a''}\left(\ket{i-1,\textrm{bdy}}_{a''\;{\rm in}}\otimes\ket{\textrm{bdy},i}^{a''}_{\rm out}\right) +\mathcal{O}(K^5)\nn\\
=&\, K^3\dfrac{\lambda_{\bf F}}{\lambda_{\bf 1}}\sum_a\left(\ket{i,\textrm{bdy}}_a+K\dfrac{\lambda_{\bf F}}{\lambda_{\bf 1}}\ket{i-1,\textrm{bdy}}_{a}\right)_{\rm in}\nn\\
&\otimes \left(\ket{\textrm{bdy},i+1}^{a}+K\dfrac{\lambda_{\bf F}}{\lambda_{\bf 1}}\ket{\textrm{bdy},i+2}^{a}\right)_{\rm out}+\mathcal{O}(K^5),
\end{align}
again without producing any Bell pairs.

We can apply the similar analysis to the $\mathcal{O}(K^5)$ case, and get the tensor product structure. With the fact that there appears {\it no} new superselection sector at $\mathcal{O}(K^5)$,\footnote{At the $\mathcal{O}(K^6)$, a new adjoint sector appears.} we thus conclude that there is no Bell pair at this order.

In the next subsection 
we will see that once we take into account $\mathcal{O}(K^6)$ corrections,  the ground state can {\it not} be written as a tensor product state predicted from lower order results. As a consequence, we obtain the Bell pair part at $O(K^6)$.

\subsection{Bell pair appears at $\mathcal{O}(K^6)$ corrections}\label{subsec:K6}

To show that the Bell pair part appears in the ground state at $\mathcal{O}(K^6)$, 
we perform the same analysis. 

Suppose again that 
the $i$-th vertex is located in the inside while the  $(i+1)$-th vertex is in the outside. 
We focus on the singlet sector of the ground state, and 
we thus look 
at the coefficient $c_6$, which is associated with the  term at $\mathcal{O}(K^6)$ as
\be
\label{K6orderc6}
\left.\ket{G^+}_K\right|_{\textrm{singlet}} \supset c_6 \, K^6 \ket{i,i}_{\rm in}\ket{i+1,i+1}_{\rm out}. 
\ee
As was discussed in the previous subsection, if there is no Bell pair, $\left.\ket{G^+}_K\right|_{\textrm{singlet}}$ must be the tensor product of the inside-only excited state and the outside-only excited state, and vice versa. 
Then, the term $\ket{i,i}_{\rm in}\ket{i+1,i+1}_{\rm out}$ must come from the product of $\ket{i,i}_{\rm in}$ and $\ket{i+1,i+1}_{\rm out}$ at lower order in the HPE. 
Eq.~\eqref{eq:orderK4} and the absence of terms such as  $\ket{i,i}_{\rm in}$ or  $\ket{i+1,i+1}_{\rm out}$ at $\mathcal{O}(K^5)$ imply that the $c_6$ term at $\mathcal{O}(K^6)$ in \eqref{K6orderc6} must be obtained 
from lower orders as  
\begin{align}
\label{K6inotimesout}
&\quad \Bigl[ \ket{0}_{\rm in} + K^2 \ket{i,i}_{\rm in}  +3K^4 \ket{i,i}_{\rm in} +\mathcal{O}(K^6)  \Bigr]_{\rm in}\nn\\
& \qquad \quad \otimes \Bigl[ \ket{0}_{\rm out} + K^2 \ket{i+1,i+1}_{\rm out}  +3K^4 \ket{i+1,i+1}_{\rm out} +\mathcal{O}(K^6)  \Bigr]_{\rm out}  \nn \\
& \supset \hspace{-0.1cm} \, K^2\ket{i,i}_{\rm in} \otimes 3K^4\ket{i+1,i+1}_{\rm out} + 3K^4\ket{i,i}_{\rm in} \otimes K^2\ket{i+1,i+1}_{\rm out} \nn \\ 
& \, \hspace{-0.1cm} = 6  K^6 \ket{i,i}_{\rm in}\ket{i+1,i+1}_{\rm out} \,, 
\end{align}
which gives $c_6=6$. Inversely if $c_6 \neq 6$, which is the case we will see, there are Bell pairs in this ground state. 

To calculate $c_6$, we consider the corresponding terms in the eigenstate equation,
\be
\label{K6ordereigenstateisthis}
\hat{T}\ket{G^+}_K=G^+_K\ket{G^+}_K \,.
\ee
Since at least the forth order part of the transfer matrix in the HPE is needed to generate  the $\ket{i,i}\ket{i+1,i+1}$ state in the future time, together with  $\ket{G^+_1}=0$, the relevant part of the left hand side can be calculated as
\begin{align}
\left.\left(\hat{T}_6\ket{G^+_0}+\hat{T}_4\ket{G^+_2}\right)\right|_{K^6,\ket{i,i}\ket{i+1,i+1}}=&\left.\left(\hat{T}_6\ket{0}+\hat{T}_4\sum_n\ket{n,n}\right)\right|_{K^6,\ket{i,i}\ket{i+1,i+1}}\nn\\
=&\, 6+2NN_l+\frac{1}{N} \,.
\label{eigenlhs}
\end{align}
See \S \ref{ii+1} for the explicit calculation to derive this result.

On the other hand,
since  $\ket{i,i}\ket{i+1,i+1}$ term appears only at $K^n\;(n\geq 4)$ order and $G^+_1=0$, 
the right hand side is evaluated as
\begin{align}
\left.\left(G^+_0\ket{G^+_6}+G^+_2\ket{G^+_4}\right)\right|_{K^6,\ket{i,i}\ket{i+1,i+1}}=1\times c_6+2NN_l\times 1=c_6+2NN_l.
\end{align}
Thus eq.~\eqref{K6ordereigenstateisthis} leads to 
\be
c_6=6+\frac{1}{N} \Rightarrow c_6 \neq 6 \,.  
\ee
We therefore conclude that there is the Bell pair part of the entanglement entropy in the singlet sector for the ground state.  

Finally, we estimate the Bell pair part of the entanglement on the singlet sector at $K^6$ order.
Since the ground state $\ket{G^+}_K$ in eq.~\eqref{eq:orderK4}  has the following structure
\begin{align}
&\left.\ket{G^+}_K\right|_{\textrm{Non-singlet}} =  O(K^3) \,,
\end{align}
the probability distribution $p_{\bf 1}$  for the singlet sector ($\bf k = {\bf 1}$) and $p_{{\bf k} \ne {\bf 1}}$ for the non-singlet sector  ($\bf k \neq {\bf 1} $) 
are given by
\be
p_{\bf 1} = 1 + \mathcal{O}(K^{6})  \,, \quad p_{{\bf k}  \neq {\bf 1} }= \mathcal{O}(K^6)  \,.
\ee
Therefore, the Bell pair part, the third term of \eqref{colorEE}, is estimated in the HPE as
\be
S^{\rm Bell}_{EE} \equiv  -\sum_{\bf{k}} p_{\bf{k}}\Tr_{{\hat {\cal H}}^{\bf{k}}_{\rm in}}{ \rho}_{\rm in}^{\bf{k}} \log {\rho}_{\rm in}^{\bf{k}} 
= - \Tr_{{\hat {\cal H}}^{\bf{1}}_{\rm in}}{ \rho}_{\rm in}^{\bf{1}} \log {\rho}_{\rm in}^{\bf{1}}  + \mathcal{O}(K^6) \,,
\ee
In fact one can explicitly show that for the ground state wave function up to $\mathcal{O}(K^6)$, the Bell pair part of the entanglement appears only from the singlet sector. 
Therefore we here focus on the singlet sector of the ground state $\ket{G^+}_K$ and evaluate 
the leading contribution of the Bell pair part in the HPE.

As discussed, the singlet sector of the ground state has the following structure. 
\be
\label{approxforGroundK6}
\left.\ket{G^+}_K\right|_{\textrm{singlet}} = \ket{\Psi}_{\rm in} \otimes \ket{\Psi}_{\rm out}  + \frac{K^6}{N} \ket{i,i}_{\rm in} \otimes \ket{i+1,i+1}_{\rm out} + \mathcal{O}(K^7) \,,
\ee
Here $\ket{\Psi}_{\rm in} \otimes \ket{\Psi}_{\rm out}$ corresponds to the l.h.s. of \eqref{K6inotimesout} if we focus only on the 
$i$-th and $i+1$-th vertices.  
In addition, $\ket{\Psi}_{\rm in}$ and $\ket{\Psi}_{\rm out}$ of course contain also purely inside only and outside only excitations, respectively. 
In particular, $\ket{\Psi}_{\rm in/out}$ becomes $\ket{0}_{\rm in/out}$ at $K=0$ as we have seen in previous section. Since the first term of \eqref{approxforGroundK6} has a tensor product structure, the second term is crucial to generate the Bell pair part of the entanglement.

From \eqref{approxforGroundK6}, we can obtain 
the reduced density matrix $\rho_{\rm red.}$ neglecting $\mathcal{O}(K^7)$ for the singlet state as 
\begin{eqnarray}
&\hspace{-5mm}{|{\cal N_{\textrm{singlet}}}|^2} \rho_{\rm red.} = \ket{\Psi}_{\rm in} {}_{\rm out}\hspace{-1mm}\braket{\Psi\vert\Psi}_{\rm out} {}_{\rm in} \hspace{-1mm}\bra{\Psi} 
+ \frac{K^6}{N}  \ket{\Psi}_{\rm in} {}_{\rm out}\hspace{-1mm}\braket{i+1,i+1\vert\Psi}_{\rm out} {}_{\rm in}\hspace{-1mm}\bra{i,i} 
\nonumber\\
& \hspace{-15mm}+ \frac{K^6}{N}  \ket{i,i}_{\rm in} {}_{\rm out}\hspace{-1mm}\braket{\Psi\vert i+1,i+1}_{\rm out} {}_{\rm in}\hspace{-1mm}\bra{\Psi}\nonumber \\
& \qquad \qquad \qquad  + \left(\frac{K^6}{N} \right)^2 \ket{i,i}_{\rm in} {}_{\rm out}\hspace{-1mm}\braket{i+1,i+1\vert i+1,i+1}_{\rm out} {}_{\rm in} \hspace{-1mm} \bra{i,i} \,. \qquad
\label{singletsectorreducedd}
\end{eqnarray}
Here the norm ${|{\cal N_{\textrm{singlet}}}|^2}$ is 
\begin{eqnarray}
& \hspace{-85mm} {|{\cal N_{\textrm{singlet}}}|^2} \equiv \left. {}_K\hspace{-1mm}\braket{G^+ \vert G^+}_K \right\vert_{\rm singlet} \nonumber \\
&= {}_{\rm in} \hspace{-1mm}\braket{\Psi \vert \Psi}_{\rm in} {}_{\rm out}\hspace{-1mm}\braket{\Psi\vert\Psi}_{\rm out} 
+ \frac{K^6}{N}   {}_{\rm in}\hspace{-1mm}\braket{i,i \vert \Psi}_{\rm in} {}_{\rm out}\hspace{-1mm}\braket{i+1,i+1\vert\Psi}_{\rm out}
\nonumber\\
& \hspace{-15mm}+ \frac{K^6}{N} {}_{\rm in}\hspace{-1mm}\braket{\Psi \vert i,i}_{\rm in} {}_{\rm out}\hspace{-1mm}\braket{\Psi\vert i+1,i+1}_{\rm out}  \nonumber \\
& \qquad \qquad \qquad  + \left(\frac{K^6}{N} \right)^2 {}_{\rm in} \hspace{-1mm} \braket{i,i \vert i,i}_{\rm in} {}_{\rm out}\hspace{-1mm}\braket{i+1,i+1\vert i+1,i+1}_{\rm out}  \,. \quad 
\end{eqnarray}
To diagonalize the reduced density matrix \eqref{singletsectorreducedd}, 
we would like to solve the following eigenvalue problem
\beqa
\rho_{\rm red.} \ket{P} &=&  p \ket{P} \,,\quad
\ket{P} = \alpha \ket{\Psi}_{\rm in} + \beta \ket{i,i}_{\rm in} \,,
\eeqa
which leads to
\beqa
\left(
\begin{array}{cc}
\rho_{11} - p & \rho_{12} \\
\rho_{21} & \rho_{22} - p \\
\end{array}
\right)
\left(\begin{array}{c}
\alpha \\
\beta \\
\end{array}
\right) &=& 0\,, 
\eeqa
where
\begin{eqnarray}
&&\hspace{-10mm}{|{\cal N_{\textrm{singlet}}}|^2} \rho_{11} = {}_{\rm in}\hspace{-1mm} \braket{\Psi\vert\Psi}_{\rm in} {}_{\rm out} \hspace{-1mm} \braket{\Psi\vert\Psi}_{\rm out}
+  \frac{K^6}{N} \, {}_{\rm in}\hspace{-1mm} \braket{i,i\vert\Psi}_{\rm in} {}_{\rm out}\hspace{-1mm} \braket{i+1,i+1\vert\Psi}_{\rm out},~~~~\\
&&\hspace{-10mm}{|{\cal N_{\textrm{singlet}}}|^2} \rho_{12}= {}_{\rm in}\hspace{-1mm} \braket{\Psi\vert i,i}_{\rm in} {}_{\rm out} \hspace{-1mm}\braket{\Psi\vert\Psi}_{\rm out}
+  \frac{K^6}{N} \, {}_{\rm in}\hspace{-1mm} \braket{i,i\vert i,i}_{\rm in} {}_{\rm out} \hspace{-1mm}\braket{i+1,i+1\vert\Psi}_{\rm out},~~~~~~~~\\
&&\hspace{-10mm}{|{\cal N_{\textrm{singlet}}}|^2} \rho_{21}=  \frac{K^6}{N}\,  {}_{\rm in} \hspace{-1mm}\braket{\Psi\vert\Psi}_{\rm in} {}_{\rm out}\hspace{-1mm} \braket{\Psi \vert i+1,i+1}_{\rm out}\nonumber \\
&& \qquad \qquad + \, \left( \frac{K^6}{N}\right)^2\,  {}_{\rm in} \hspace{-1mm}\braket{ i,i \vert \Psi}_{\rm in} {}_{\rm out} \hspace{-1mm}\braket{i+1,i+1\vert i+1,i+1}_{\rm out},\\
&&\hspace{-10mm}{|{\cal N_{\textrm{singlet}}}|^2} \rho_{22}=  \frac{K^6}{N}\, {}_{\rm in} \hspace{-1mm}\braket{\Psi \vert i,i}_{\rm in} {}_{\rm out} \hspace{-1mm}\braket{\Psi \vert i+1,i+1}_{\rm out}\nonumber \\
&& \qquad \qquad + \, \left( \frac{K^6}{N}\right)^2\,  {}_{\rm in} \hspace{-1mm}\braket{i,i\vert i,i}_{\rm in} {}_{\rm out} \hspace{-1mm}\braket{i+1,i+1\vert i+1,i+1}_{\rm out} \,. 
\end{eqnarray}
Thus, the eigenvalue is given by
\beqa
\label{redpeigen}
p &=& \frac{\rho_{11}+\rho_{22} \pm \sqrt{(\rho_{11}-\rho_{22})^2+4\rho_{12}\rho_{21}}}{2} \,.
\eeqa
To evaluate this, we use 
\beqa
\label{normconditions7begin}
{}_{\rm in/out}\hspace{-1mm}\braket{\Psi\vert\Psi}_{\rm in/out} &=& 1 +\mathcal{O}(K^2)\,, \\
{}_{\rm in/out}\hspace{-1mm}\braket{n,n \vert \Psi}_{\rm in/out} &=& N +\mathcal{O}(K^2)\,,\\
{}_{\rm in/out}\hspace{-1mm}\braket{n,n\vert n,n}_{\rm in/out} &=& N(N+1) \,,
\label{normconditions7end}
\eeqa
which can be obtained by 
recalling ${}_{\rm in/out}\hspace{-1mm}\braket{n,n|0}_{\rm in/out} = N$ and ${}_{\rm in/out}\hspace{-1mm}\braket{0|0}_{\rm in} = 1$,
together with the fact that in the leading order in HPE, we have 
$\ket{\Psi}_{\rm in} = \ket{0}_{\rm in}   +  \mathcal{O}(K^2)$. 
%
%
Then the leading contribution of \eqref{redpeigen} yields
\beqa
p &\simeq& 1 - K^{12} \,, \, \, K^{12} \,.
\eeqa
We therefore obtain the entanglement entropy $S_{EE}^{\rm Bell}$ for the singlet state as
\beqa
S_{EE}^{\rm Bell} &=& - (1-K^{12}) \log(1-K^{12}) - K^{12}\log K^{12} +\mathcal{O}(K^{14})\nn \\
&=& \left( 1 - \log K^{12} \right) K^{12} +\mathcal{O}(K^{14}) \,. 
\eeqa
Note that we obtain entangled $\mathcal{O}(K^{12} N^0)$ Bell pairs 
in the HPE from the $\mathcal{O}(K^6 N^{-1})$ term in the wave function \eqref{approxforGroundK6}.

\section{Summary and discussions}
\label{discussionsession}
In this paper, we studied 1+1 dimensional $SU(N)$ gauge theories with matter fields, mainly in the  fundamental representation of the gauge group. 
In the first part of this paper,
the entanglement entropy for various meson states is evaluated using the extended Hilbert space formalism \cite{Ghosh:2015iwa,Aoki:2015bsa,Soni:2015yga}.
We show that the entanglement entropy has 
two different contributions.
One is the 
classical Shannon entropy for various different superselection sector distribution, which is the first term in \eqref{colorEE}, and 
the other is  the sum over the logarithm of the dimensions for the irreducible representation at all boundaries, which is the second term in \eqref{colorEE}.  
In the second part, 
we consider the ground state in the HPE and show that the first term and 
the second term in \eqref{colorEE} appear from the ground state at the $\mathcal{O}(K^3)$, while the third term, which corresponds to the number of Bell pairs obtained by the entanglement distillation, appears at $\mathcal{O}(K^6)$. 
Since all terms in \eqref{colorEE} are positive definite,  
they also remain positive even in the continuum limit ($\beta=1/(g_{\rm YM}^2 a^2)\to \infty$ and $K=1/(m^2a^2+2) \to 1/2$). 
This means that the continuum vacuum of  gauge theories with the fundamental  matter fields in 1+1 dimensions contains all terms in \eqref{colorEE}.  
Unfortunately, it is very hard to calculate these three contributions precisely in the continuum limit, since higher and higher order terms in the HPE are needed toward the continuum limit.

Even though precise values are unknown, it is certain that
the true vacuum state contains not only the strong coupling ground state $\ket{0}_{\rm strong}$, which is the ground state of the pure gauge theories, but also gauge invariant meson states, which consist of multiple pairs of scalar and anti-scalar fields.
Since there are no contributions to the entanglement entropy from the strong coupling ground state $\ket{0}_{\rm strong}$, all of the positive values of three terms in \eqref{colorEE} are caused by  multiple meson states.
Therefore the entanglement entropy for the true ground state comes mainly from the meson pair 
with small separation (a few lattice spacings) at small $K$.
In the continuum limit ($\beta \to\infty$ and $K\to 1/2$), however,  
the separation $n$ between the entangled meson pair (the Bell pair) can become infinitely large
due to the higher order of the HPE, so that   $r = n a$ becomes non-zero in the continuum limit. 
These suggest that the continuum vacuum entanglement is due to the ``condensation'' of multiple meson states. More precisely, 
the continuum vacuum is fully filled with lattice meson states.  
This is the key picture we obtain through the analysis in this paper. We end this paper with several comments.  

Our results also imply an interesting property. If we take the continuum limit as $\beta\to\infty$ but $K < 1/2$,  the matter field becomes infinitely heavy, and thus decouples from the low energy physics in the continuum limit, so that the continuum theory is the pure gauge theory.
This infinitely heavy matter, however, produces non-zero (genuine) entanglement of the pure gauge vacuum. This means that the entanglement might be very sensitive to degrees of freedom at high energy, which can not be detected at low energy. 
This left-over entanglement might be much smaller than the entanglement of the continuum gauge theory with matters, which could be divergent.
Even though the entanglement is not observable in the strict sense,
it is interesting if this left-over entanglement can be detected by some mathematical means.

To make the above picture for the EE in the continuum limit more quantitative,  
 we have to perform some kind of resummation for the HPE. 
 At this moment, unfortunately, we do not have an explicit idea how to do this generically 
and we are not sure if this is possible. However in 1+1 dimensions, the gauge theory with matter fields is 
in principle solvable at least in the large $N$ limit \cite{tHooft:1974pnl}. 
We therefore have a good chance to obtain the EE for this model in the large $N$ limit.
This direction is worth investigating furthermore in future. 
Note that in this paper we focus especially on the ground state but it is also interesting to study excited state entanglement entropy and their time evolution.  It is also interesting to generalize our analysis to higher dimension.

Finally all of above results suggest interesting points in holography. 
In the gauge theory side, the natural extended Hilbert space definition gives three different terms for the entanglement entropy. In the gravity side, however, we have only a minimal area term (RT formula \cite{Ryu:2006bv}), at least in the large $N$ limit.  Therefore, 
which term dominates in the large $N$ limit among three terms in \eqref{colorEE}
is an important question, when we compare the results with those in the gravity side. It is interesting that 
the ``genuine'' entanglement part (=the Bell pair part) may not be dominant one in the large $N$ limit. 
In order to deepen our understanding of the holographic meaning of the entanglement entropy, 
it is important to find the corresponding gravity dual to all of these three terms in extended Hilbert space entanglement entropy. Last but not least, it is interesting to ask what corresponds to the extended Hilbert space in the dual gravity side. We hope to come back to these questions in near future.


\bigskip
\goodbreak
\centerline{\bf Acknowledgments}
\noindent
NI would like to thank YITP for nice hospitality during the long term workshop, ``Quantum Information in String Theory and Many-body Systems''. NI would also want to thank Sandip Trivedi for very stimulating conversation during the workshop, which motivate him to think of entanglement in non-Abelian gauge theories. 
The work of NI was supported in part by JSPS KAKENHI Grant Number 25800143. 
SA is supported in part by the Japanese Grant-in-Aid for Scientific
Research (No.  JP16H03978 ), by MEXT as ``Priority
Issue on Post-K computer'' (Elucidation of the Fundamental Laws and Evolution of
the Universe) and by Joint Institute for Computational Fundamental Science
(JICFuS).  

\appendix

\section{Useful formulas}
We use $a$, $b$, $c$, $d$, $\cdots$, and $i$, $j$, $\cdots$ as color indices in fundamental representation (which run $1, \cdots, N$) of the $SU(N)$ gauge group.

\subsection{Matter fields}
\paragraph{\underline{Scalar field $\vp$ in Fundamental representation}}

For the scalar field $\vp^c$ in the fundamental representation with $\vp^\dagger\vp \equiv\vp^\dagger_{c}\vp^c$,
we have following useful Gaussian integral formulas:
\begin{align}
\int [d\vp]\;e^{-a\vp^\dagger\vp}
&=\left(\sqrt{ \dfrac{\pi}{a} } \right)^{2N} \,, \label{eq:gauss0}\\
\int [d\vp]\;\vp^\dagger_c\vp^de^{-a\vp^\dagger\vp}
&=\delta^d{}_c\,\dfrac{1}{a}\left(\sqrt{ \dfrac{\pi}{a} } \right)^{2N}\label{eq:gauss2} \,, \\
\int [d\vp]\;  \vp^\dagger_a\vp^b \vp^\dagger_c\vp^d e^{-a\vp^\dagger\vp}
& =\left({ \delta^b{}_a\delta^d{}_c +\delta^d{}_a\delta^b{}_c}\right) \frac{1}{a^2}   \left(\sqrt{ \dfrac{\pi}{a} } \right)^{2N}\label{eq:gauss3a}\,.
\end{align}
The last formula gives
\bea
\int [d\vp]\; (\sum_{b=1}^N \vp^\dagger_b\vp^b) (\sum_{d=1}^N \vp^\dagger_d\vp^d)  e^{- \sum_{c=1}^N a \vp^\dagger_{c}\vp^{c}} 
&= \dfrac{N (N + 1)}{a^2}   \left(\sqrt{ \dfrac{\pi}{a} } \right)^{2N}\label{eq:gauss3}\,.
\eea

\paragraph{\underline{Hermitian $N \times N$ matrix scalar $X^c{}_d$ field}}  
Next we consider the Gaussian integral for
the Hermitian $N\times N$ matrix field. 
This is an adjoint representation matter field for gauge group $U(N)$, whose  
Gaussian integral becomes  
\bea
\int [dX] \exp \left( - a \Tr X^2 \right) 
&=& \left(\sqrt{\frac{\pi}{a}} \right)^{N^2} \,, \\
\int [d X] 
X^a{}_b X^c{}_d\exp \left( -a \Tr X^2 \right) &=& \delta^a{}_d\delta^c{}_b\frac{1}{2 a}\left(\sqrt{\frac{\pi}{a}}\right)^{N^2}\label{XX1} \,,
\eea
while the Gaussian integral for the field in the adjoint representation of the gauge group $SU(N)$
leads to  
\begin{align}
\int [d X] 
&X^a{}_bX^c{}_d\exp \left( -a \Tr X^2 \right) \nn \\=
& \left( \delta^a{}_d\delta^c{}_b - \frac{1}{N} \delta^a{}_b \delta^c{}_d \right) \frac{1}{2 a}\left(\sqrt{\frac{\pi}{a}}\right)^{N^2-1}  \,,
\label{eq:sunadjg}
\end{align}
where the traceless condition is used.
The above formulae are obtained by expanding 
\beqa
X &=& \sum_{A=0}^{N^2-1} t^A X_A\,, \qquad  \sum_{A=0}^{N^2-1} (t^A)^a{}_b(t^A)^c{}_d =\delta^a{}_d\delta^c{}_b
\eeqa
for $U(N)$ and
\beqa
X &=& \sum_{A=1}^{N^2-1} t^A X_A\,, \qquad  \sum_{A=1}^{N^2-1} (t^A)^a{}_b(t^A)^c{}_d =\delta^a{}_d\delta^c{}_b-\frac{1}{N}\delta^a{}_b\delta^c{}_d
\eeqa
for $SU(N)$, where $X_A$ is real and $\tr\, (t^A t^B) =\delta^{AB}$. 

\subsection{Link variables (= exponential of gauge fields)}
For link variables $U^a{}_b, U^c{}_d, \cdots$ in the fundamental representation ($a,b,c,d = 1 , \cdots , N$), 
the integration over the group with the invariant Haar measure $[dU]$ gives 
\be
\int [dU]\;U^a{}_bU^{\dagger c}{}_d=\dfrac{1}{N}\delta^a{}_d\delta^c{}_b\label{eq:UUd} \,,
\ee
which can be derived  from the symmetry under group transformation $U \to L U R$ \cite{Creutz:1983}. 
Similarly, one can show \cite{Creutz:1983} 
\bea
&&\int [dU]\;U^a{}_bU^c{}_dU^{\dagger i}{}_jU^{\dagger k}{}_\ell 
\nn \\
&&=\dfrac{1}{N^2-1}\Big[\delta^a{}_j\delta^i{}_b\delta^c{}_\ell\delta^k{}_d+\delta^a{}_\ell\delta^k{}_b\delta^c{}_j\delta^i{}_d-\dfrac{1}{N}\left(\delta^a{}_j\delta^k{}_b\delta^c{}_\ell\delta^i{}_d+\delta^a{}_\ell\delta^i{}_b\delta^c{}_j\delta^k{}_d\right)\Big] \,. \nn \\
\label{eq:U4d}
\eea
where not only $a, b, c, d$ but also  $i, j, k, l$ are indices of the fundamental/anti-fundamental representation and thus run from $1$ to $N$.

For generic representations ${\bf R}$ and ${\bf R'}$, 
eq.~(\ref{eq:UUd}) is replaced with
\be
\int [dU]\;U^a{}_b({\bf R}) U^{\dagger c}{}_d({\bf R'})=\dfrac{1}{d_{\bf R}}\delta_{{\bf RR'}} \delta^a{}_d\delta^c{}_b \,, \label{eq:UUdgene}
\ee
where $d_{\bf R}$ is the dimension of the representation ${\bf R}$ ($d_{\bf R}= N$
for the fundamental and $d_{\bf R} = N^2 -1$ for the adjoint) and  $a,b,c,d = 1 , \cdots , d_{\bf R}$ in this case. 
Furthermore eq.~(\ref{eq:U4d}) becomes
 \bea
&&\int [dU]\;U^a{}_b({\bf R})U^c{}_d({\bf R})U^{\dagger i}{}_j({\bf R})U^{\dagger k}{}_\ell({\bf R}) 
\nn \\
&&=\dfrac{1}{d_{\bf R}^2-1}\Big[\delta^a{}_j\delta^i{}_b\delta^c{}_\ell\delta^k{}_d+\delta^a{}_\ell\delta^k{}_b\delta^c{}_j\delta^i{}_d-\dfrac{1}{d_{\bf R}}\left(\delta^a{}_j\delta^k{}_b\delta^c{}_\ell\delta^i{}_d+\delta^a{}_\ell\delta^i{}_b\delta^c{}_j\delta^k{}_d\right)\Big] \,. \nn \\
\eea

\section{Characters for link variables}
\label{characternature}
Characters are very useful in order to handle link variables for gauge theories, and 
we review briefly in this appendix. 

For a gauge group element $g \in G$ and its representation {$g({\bf R})$}, the character is defined as 
\be
\chi_{\bf R}(g)\equiv\underset{{\bf R}}{\Tr}[g({\bf R})] \,.
\ee
The character satisfies several important properties. One of them is that the product of characters can be expressed as the sum of characters. The other important property of character is that 
different representation characters are orthogonal under the group integral.

To illustrate these, 
let us consider the group $SU(2)$ as an example. One can label representations by their spin $j$. Their dimensions are given by $d_j = 2 j + 1$.  Since characters are invariant under the group transformation, one can always choose a basis such that $R(g)$ becomes a rotation along the ``$z$''-axis. Then it is clear that a number of parameters for each character is given by the dimension of its Cartan sub-algebra. 
More explicitly, characters for spin-$j$ representations of the $SU(2)$ are given by 
\begin{align}
\chi_{\frac{1}{2}}(\theta)&=2\cos\frac{\theta}{2}\hspace{1cm}\left(R^{(\frac{1}{2})}_z(\theta)=\bbm e^{i\frac{\theta}{2}} & 0 \\ 0 & e^{-i\frac{\theta}{2}}\ebm\right) \,, \\
\chi_{1}(\theta)&=1+2\cos\theta\hspace{1cm}\left(R^{(1)}_z(\theta)=\bbm \cos\theta & -\sin\theta & 0 \\ \sin\theta & \cos\theta & 0 \\ 0 & 0 & 1\ebm\right)  \,, \\
\chi_j(\theta)&=\underset{j}{\Tr}[e^{iJ_z\theta}]=\sum^j_{m=-j}e^{im\theta}=\dfrac{\sin\left(j+\frac{1}{2}\right)\theta}{\sin\frac{\theta}{2}} \,.
\end{align}

For the $SU(2)$, the fact that the product of characters can be expressed as the sum of characters is equivalent to a familiar Clebsch-Gordan expansion in quantum mechanics:   
\be
\chi_{j_1}(\theta)\chi_{j_2}(\theta)=\sum_{j=|j_1-j_2|}^{j_1+j_2}\chi_j(\theta) \,. \label{eq:characterCG}
\ee
For example, 
\begin{align}
\chi_{\frac{1}{2}}(\theta)\chi_{\frac{1}{2}}(\theta)&=4\cos^2\frac{\theta}{2} 
=4\times\dfrac{1+\cos\theta}{2} =\chi_{0}(\theta)+\chi_{1}(\theta) \,.
\end{align}

The property that different representations are orthogonal is expressed as 
\be
\int [dg]\; \chi_{j}(g)\chi^\ast_{j^\prime}(g)=\delta_{jj^\prime}
\label{orthogonalSU2characters}
\ee
This can be seen as follows. For the $SU(2)$, due to its pseudo-reality, $\chi_{j}(g)=\chi^\ast_{j}(g)$. 
Using \eqref{eq:characterCG}, above integrand can be expressed as a sum over different representations of characters. From the invariance of the measure, $\forall h\in G, [dg]=d[(hg)]=d[(gh)]$, it is clear that only the singlet representation gives nonzero value after the integral. We take $\int[dg]=1$ as the normalization condition.

\section{Tensor product decomposition of the wave function}\label{app:4qdec}
In this appendix, we discuss the decomposition of the wave function given in section \ref{subsec:4m} as 
\begin{align}
\Psi(\vp,U)&=\left[\vp_b^\dagger U_{b,b+1}\vp_{b+1}\right]^2\left[\vp_{b+1}^\dagger U^\dagger_{b,b+1}\vp_{b}\right]^2 \nn\\
&\equiv\left[(\vp^\dagger\phi)(\phi^\dagger\vp)\right]^2 \,, 
\end{align}
where $\vp\equiv\vp_b$ and $\phi\equiv U_{b,b+1}\vp_{b+1}$. We regard $\vp$ and $\phi$ as objects in inside and outside regions, respectively. 
Since there are 2 sets of ``fundamental $\otimes$ anti-fundamental'' matters in the inside, one can decompose it into 2 sets of ``adjoint $\oplus$ singlet''  as
\be
(\vp^\dagger\phi)(\phi^\dagger\vp)=\Tr (XY)+\dfrac{1}{N}(\vp^\dagger\vp)(\phi^\dagger\phi) \,, \label{dec1}
\ee
where we define ``adjoint matters'' by
\begin{align}
X^a{}_b&\equiv\vp^\dagger_{b}\vp^{a}-\dfrac{1}{N}\delta^a{}_b(\vp^\dagger\vp) \,, \nn\\
Y^a{}_b&\equiv\phi^\dagger_{b}\phi^{a}-\dfrac{1}{N}\delta^a{}_b(\phi^\dagger\phi) \,. 
\end{align}
In this notation, we can rewrite our wave function as follows.
\be
\Psi(\vp,U)=\left[\dfrac{1}{N^2}(\vp^\dagger\vp)^2(\phi^\dagger\phi)^2+\dfrac{2}{N}\Tr (XY)(\vp^\dagger\vp)(\phi^\dagger\phi)+\left[\Tr (XY)\right]^2\right]. 
\ee
The first term belongs to the singlet sector, which comes from ``singlet $\otimes$ singlet'', while the second term to the adjoint sector from ``singlet $\otimes$ adjoint''.    The third term corresponds to ``adjoint $\otimes$ adjoint'', and  we therefore  need to further decompose this term into irreducible representations. 
For $SU(N)$, the tensor product decomposition of adjoint $\otimes$ adjoint is
\begin{align}
\bold{N^2-1}\,\otimes\,\bold{N^2-1}&=\bold{1}\oplus(\bold{N^2-1})_s\oplus(\bold{N^2-1})_a \nn \\
& \quad  \oplus\bold{\frac{1}{4}N^2(N-1)(N+3)}  \oplus\bold{\frac{1}{4}N^2(N+1)(N-3)} \nn \\
&\quad \oplus\bold{\frac{1}{4}(N^2-1)(N^2-4)}\oplus\overline{\bold{\frac{1}{4}(N^2-1)(N^2-4)}}\, , \label{eq:sun_adjadj}
\end{align}
where $\bold{\frac{1}{4}N^2(N-1)(N+3)}$ and $\bold{\frac{1}{4}N^2(N+1)(N-3)}$ are totally symmetric and anti-symmetric traceless combination for original indices, respectively, while  
two adjoint representations, $(\bold{N^2-1})_s$ and $(\bold{N^2-1})_a$, comes from these ``$s$''\,ymmetric and ``$a$''\,nti-symmetric representations by the partial trace.
Finally 
$\bold{\frac{1}{4}(N^2-1)(N^2-4)}$ and its conjugates are mixed symmetric and traceless. 
In $SU(2)$ and $SU(3)$, some of these representations are absent but our final result is true for these special cases. 
\begin{figure}[tbp]
\begin{center}
\label{adjadjpic}
\includegraphics[height=7.5cm, width=10cm]{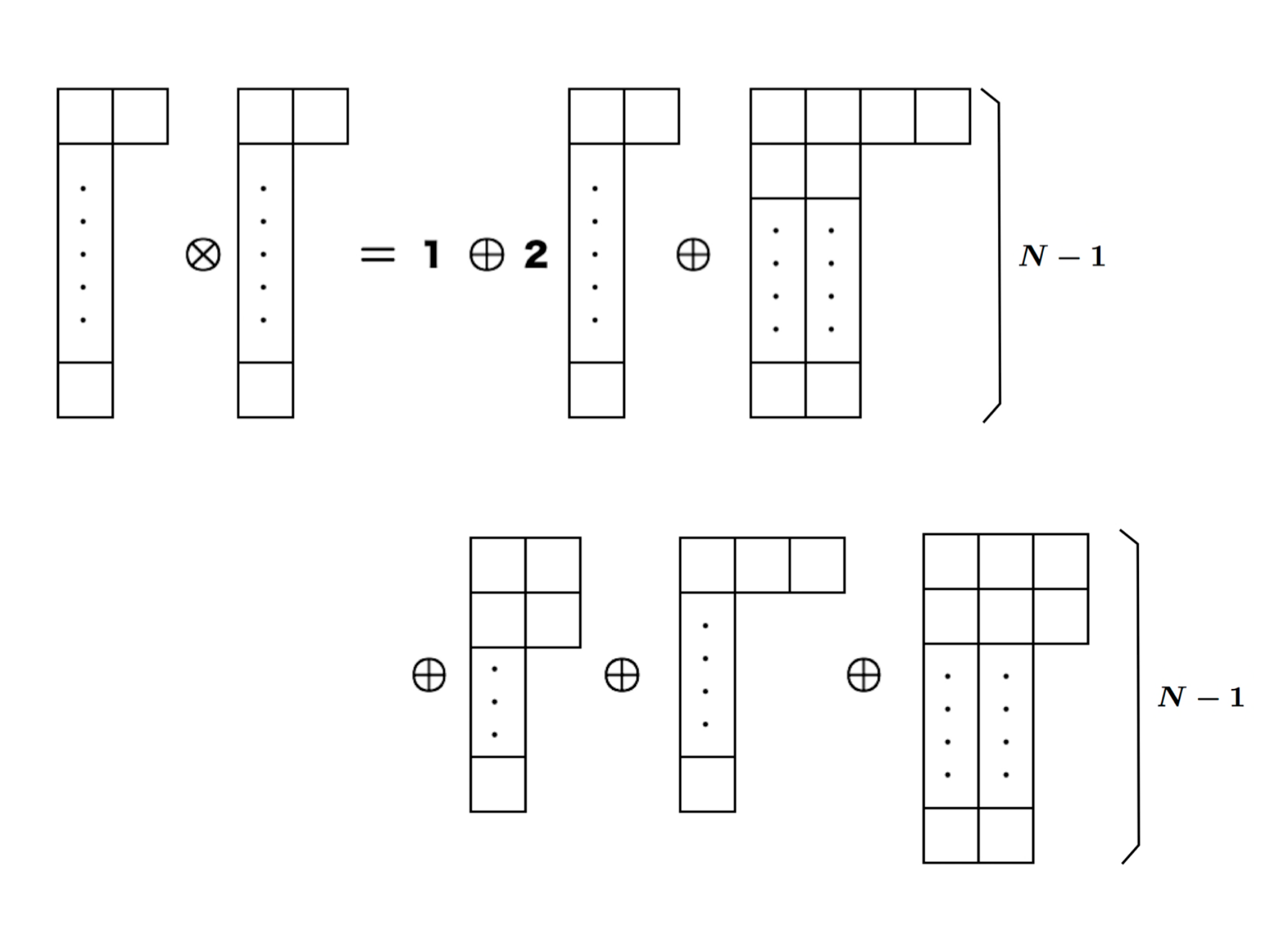}
\caption{Young diagrams for the tensor decomposition \eqref{eq:sun_adjadj}.}
\end{center}
\end{figure} 
\subsection{Tensor product decomposition for components}\label{subsec:XoX}
As a first step, we decompose $X^a{}_bX^i{}_j$ into irreducible combinations. 
We follow the standard procedure in the representation theory of $SU(N)$. Namely, we first symmetrize and anti-symmetrize $X^a{}_bX^i{}_j$, and then remove the trace of these combinations. We continue this manipulation until we obtain the trivial representation. From now on we assume $N\geq4$.\footnote{Although intermediate steps cannot be applied directly to $N=2, 3$, our final result is valid even in these cases.} The final result becomes
\begin{align}
X^a{}_bX^i{}_j&=(\bold{1})^{ai}{}_{bj}+(\bold{N^2-1}_s)^{ai}{}_{bj}+(\bold{N^2-1}_a)^{ai}{}_{bj}+\left(\bold{\frac{1}{4}N^2(N-1)(N+3)}\right)^{ai}{}_{bj}\nn\\
&\qquad +\left(\bold{\frac{1}{4}N^2(N+1)(N-3)}\right)^{ai}{}_{bj} \,, 
\end{align}
where
\begin{align}
(\bold{1})^{ai}{}_{bj}&=\dfrac{1}{(N^2-1)}\Tr (X^2)\left[\delta^a{}_j\delta^i{}_b-\dfrac{1}{N}\delta^a{}_b\delta^i{}_j\right] \,, 
\end{align}
\begin{align}
(\bold{N^2-1}_s)^{ai}{}_{bj}&=\dfrac{1}{2(N+2)}\left[\delta^i{}_b(X^2)^a{}_j+\delta^a{}_j(X^2)^i{}_b+(X^2)^a{}_b\delta^i{}_j+(X^2)^i{}_j\delta^a{}_b\right.\nn\\
&\left.-\dfrac{2}{N}\Tr (X^2)(\delta^a{}_j\delta^i{}_b+\delta^a{}_b\delta^i{}_j)\right] \,, \\
(\bold{N^2-1}_a)^{ai}{}_{bj}&=\dfrac{1}{2(N-2)}\left[\delta^i{}_b(X^2)^a{}_j+\delta^a{}_j(X^2)^i{}_b-(X^2)^a{}_b\delta^i{}_j-(X^2)^i{}_j\delta^a{}_b\right.\nn\\
&\left.-\dfrac{2}{N}\Tr (X^2)(\delta^a{}_j\delta^i{}_b-\delta^a{}_b\delta^i{}_j)\right] \,, 
\end{align}
\begin{align}
&\hspace{-1cm}\left(\bold{\frac{1}{4}N^2(N-1)(N+3)}\right)^{ai}{}_{bj}\nn\\
&=\dfrac{1}{2}\left[X^a{}_bX^i{}_j+X^a{}_jX^i{}_b-\dfrac{1}{N+2}\left(\delta^i{}_b(X^2)^a{}_j+\delta^a{}_j(X^2)^i{}_b+(X^2)^a{}_b\delta^i{}_j+(X^2)^i{}_j\delta^a{}_b\right)\right.\nn\\
&\left.+\dfrac{1}{(N+1)(N+2)}\Tr (X^2)(\delta^a{}_b\delta^i{}_j+\delta^a{}_j\delta^i{}_b)\right] \,, \\
&\hspace{-1cm}\left(\bold{\frac{1}{4}N^2(N+1)(N-3)}\right)^{ai}{}_{bj}\nn\\
&=\dfrac{1}{2}\left[X^a{}_bX^i{}_j-X^a{}_jX^i{}_b-\dfrac{1}{N-2}\left(\delta^i{}_b(X^2)^a{}_j-\delta^i{}_j(X^2)^a{}_b+(X^2)^i{}_b\delta^a{}_j-(X^2)^i{}_j\delta^a{}_b\right)\right.\nn\\
&\left.+\dfrac{1}{(N-1)(N-2)}\Tr (X^2)(\delta^a{}_j\delta^i{}_b-\delta^a{}_b\delta^i{}_j)\right] \,.
\end{align}
One can obtain the same decomposition for $Y^a{}_bY^i{}_j$ just replacing $X$ to $Y$. 
There are no contributions from the last 2 terms of \eqref{eq:sun_adjadj}. 
This is simply because our wave function is ``real''.  

\subsection{Decomposition for $[\Tr(XY)]^2$}
By using previous results, we can decompose ``adjoint $\otimes$ adjoint'' into irreducible representations. Since the contraction with different representations vanish, one can decompose $[\Tr(XY)]^2$ as follows.
\begin{align}
\hspace{-0.1cm}[\Tr(XY)]^2&=\left.[\Tr(XY)]^2\right|_{\bold{1}}+\left[\Tr(XY)]^2\right.|_{\bold{N^2-1}_s}+\left[\Tr(XY)]^2\right|_{\bold{N^2-1}_a}\nn\\
&\,\,\, \left.+[\Tr(XY)]^2\right|_{\bold{\frac{1}{4}N^2(N-1)(N+3)}}+\left.[\Tr(XY)]^2\right|_{\bold{\frac{1}{4}N^2(N+1)(N-3)}}, 
\end{align}
where
\begin{align}
\left[\Tr(XY)]^2\right|_{\bold{1}}&=\dfrac{1}{N^2-1}\Tr X^2\Tr Y^2 \,,\\
\left[\Tr(XY)]^2\right|_{\bold{N^2-1}_s}&=\dfrac{1}{N+2}\left[\Tr(X^2Y^2)-\dfrac{1}{N}(\Tr X^2)(\Tr Y^2)\right] \, ,\\
\left[\Tr(XY)]^2\right|_{\bold{N^2-1}_a}&=\dfrac{1}{N-2}\left[\Tr(X^2Y^2)-\dfrac{1}{N}(\Tr X^2)(\Tr Y^2)\right] \,,
\end{align}
\begin{align}
&\hspace{-1cm}\left[\Tr(XY)]^2\right|_{\bold{\frac{1}{4}N^2(N-1)(N+3)}}\nn\\
&=\dfrac{1}{2}\left[\Tr(XY)\Tr(XY)+\Tr(XYXY)-\dfrac{2}{N+2}\Tr(X^2Y^2)+\dfrac{1}{(N+1)(N+2)}\Tr X^2\Tr Y^2\right],\\
&\hspace{-1cm}\left[\Tr(XY)]^2\right|_{\bold{\frac{1}{4}N^2(N+1)(N-3)}}\nn\\
&=\dfrac{1}{2}\left[\Tr(XY)\Tr(XY)-\Tr(XYXY)-\dfrac{2}{N-2}\Tr(X^2Y^2)+\dfrac{1}{(N-1)(N-2)}\Tr X^2\Tr Y^2\right]\nn\\
&=0. 
\end{align}
Each symbol $|_{\bold{R}}$ denotes the projection into each irreducible representation $\bold{R}$. 
A reason why the last vanishes is the same as the case of
two mesons at the same position with the same direction, which do not have the totally anti-symmetric combination.

Explicitly we have 
\begin{align}
(X^2)^a{}_b&=(\vp^\dagger\vp)\left[\left(1-\dfrac{2}{N}\right)\vp^a\vp^\dagger_b+\dfrac{1}{N^2}\delta^a{}_b(\vp^\dagger\vp)\right] \,,\\
(XY)^a{}_b&=(\vp^\dagger\phi)\phi^\dagger_b\vp^a-\dfrac{1}{N}\left\{(\vp^\dagger\vp)\phi^\dagger_b\phi^a+\vp^\dagger_b\vp^a(\phi^\dagger\phi)\right\}+\dfrac{1}{N^2}\delta^a{}_b(\vp^\dagger\vp)(\phi^\dagger\phi) \,, 
\end{align}
\begin{align}
\Tr X^2&=\left(1-\dfrac{1}{N}\right)(\vp^\dagger\vp)^2 \,, \\
\Tr(XY)&=(\vp^\dagger\phi)(\phi^\dagger\vp)-\dfrac{1}{N}(\vp^\dagger\vp)(\phi^\dagger\phi) \,, \\
\Tr(X^2Y^2)&=(\vp^\dagger\vp)(\phi^\dagger\phi)\left[\left(\dfrac{N-2}{N}\right)^2(\phi^\dagger\vp)(\vp^\dagger\phi)+\dfrac{1}{N^3}(2N-3)(\vp^\dagger\vp)(\phi^\dagger\phi)\right] \,,\\
\Tr(XYXY)&=(\vp^\dagger\phi)^2(\phi^\dagger\vp)^2-\dfrac{4(N-1)}{N^2}(\vp^\dagger\phi)(\phi^\dagger\vp)(\vp^\dagger\vp)(\phi^\dagger\phi)\nn\\
& \qquad +\dfrac{1}{N^3}(2N-3)(\vp^\dagger\vp)^2(\phi^\dagger\phi)^2 \,, \\
&\hspace{-3cm}\left[\Tr(X^2Y^2)-\dfrac{1}{N}(\Tr X^2)(\Tr Y^2)\right]=\left(\dfrac{N-2}{N}\right)^2(\vp^\dagger\vp)(\phi^\dagger\phi)\Tr(XY) \,.
\end{align}

\subsection{Summary}
To summarize, the final results are explicitly given as  
\be
\Psi(\vp,U)=\Psi_\bold{1}(\vp,U)+\Psi_\bold{N^2-1}(\vp,U)+\Psi_{\bold{\frac{1}{4}N^2(N-1)(N+3)}}(\vp,U) \,, 
\ee
where
\begin{align}
\Psi_\bold{1}(\vp,U)&=\dfrac{1}{N^2}(\vp^\dagger\vp)^2(\phi^\dagger\phi)^2+\left[\Tr(XY)]^2\right|_{\bold{1}}\nn\\
&=\dfrac{2}{N(N+1)}(\vp^\dagger_b\vp_b)^2(\vp^\dagger_{b+1}\vp_{b+1})^2 \,, \\
\Psi_\bold{N^2-1}(\vp,U)&=\dfrac{2}{N}(\vp^\dagger\vp)(\phi^\dagger\phi)\Tr(XY)+\left[\Tr(XY)]^2\right|_{\bold{N^2-1}_s}+\left[\Tr(XY)]^2\right|_{\bold{N^2-1}_a}\nn\\
&=\dfrac{4}{N+2}(\vp^\dagger_b\vp_b)(\vp^\dagger_{b+1}\vp_{b+1})\left[(\vp^\dagger_bU_{b,b+1}\vp_{b+1})(\vp^\dagger_{b+1}U^\dagger_{b,b+1}\vp_b)\right.\nn\\
&\hspace{6cm}\left.-\dfrac{1}{N}(\vp^\dagger_b\vp_b)(\vp^\dagger_{b+1}\vp_{b+1})\right] \,,\\
\Psi_\bold{\bold{\frac{1}{4}N^2(N-1)(N+3)}}(\vp,U)&=\left.[\Tr(XY)]^2\right|_{\bold{\frac{1}{4}N^2(N-1)(N+3)}}\nn\\
&=(\vp^\dagger_bU_{b,b+1}\vp_{b+1})^2(\vp^\dagger_{b+1}U^\dagger_{b,b+1}\vp_b)^2\nn\\
&-\dfrac{2}{N+2}(\vp^\dagger_b\vp_b)(\vp^\dagger_{b+1}\vp_{b+1})\left[2(\vp^\dagger_bU_{b,b+1}\vp_{b+1})(\vp^\dagger_{b+1}U^\dagger_{b,b+1}\vp_b)\right.\nn\\
&\hspace{5.5cm}\left.-\dfrac{1}{N+1}(\vp^\dagger_b\vp_b)(\vp^\dagger_{b+1}\vp_{b+1})\right] \,.
\end{align}
Note that this result also holds for the  $N=2$ case. 



\section{Feynman diagrams for transfer matrix in the HPE}
\label{yokoyafeynman}
The hopping parameter expansions (HPE) for the transfer matrix can be evaluated efficiently using Feynman diagrams.
We consider the $SU(N)$ gauge theory with fundamental scalar fields in 2-dimensional lattice space-time, where the horizontal direction corresponds to the spatial direction while the vertical direction corresponds to the Euclidean time direction, respectively. 

The transfer matrix is defined in~\S \ref{TmatrixandHPE}. As is clear from the expression, it represents a transition from a ``current state'' (which we denote as  $\Psi^B=\{\phi ,V$\}) to a ``future state'' (which we denote as $\Psi^A=\{\varphi ,U$\}) by unit time shift.   
As mentioned, we take the temporal gauge, therefore all gauge link variables along the time direction are set to unity.


\subsection{Diagrams}
\subsubsection{States}
The gauge invariant ``quark-antiquark''  states $\ket{n,m}$ labeled by site positions $(n,m)$ are defined as
\bea
\braket{\Psi^A\vert n,n} &=&\varphi^{\dagger}_n\varphi_n  \,,\nn\\
\braket{\Psi^A\vert n,m}&=&\varphi^{\dagger}_nU_{n\rightarrow m}\varphi_m\;\;(n<m),\nn\\
\braket{\Psi^A\vert n,m}&=&\varphi^{\dagger}_{n}U^{\dagger}_{m\rightarrow n}\varphi_{m}\;\;(n> m),
\eea
where
\be
U_{n\rightarrow m}=U_{n,n+1}U_{n+1,n+2}\cdots U_{m-1,m} \,.
\ee
%
These states can be represented graphically as 
\bea
\braket{\Psi^B\vert n,n}= \,\,
\underset{n}{
\begin{tikzpicture}
[q/.style={circle,draw,scale=0.5},
aq/.style={circle,draw,scale=0.5,fill}]
\node at ( 0,0) [q] {};
\node at ( 0.2,0) [aq] {};
\end{tikzpicture}
} \quad \,, \qquad  
\braket{\Psi^B\vert n,n+1}=
\underset{\;\;n\;\;\;\;\;\;\;\;n+1}{
\begin{tikzpicture}
[q/.style={circle,draw,scale=0.5},
aq/.style={circle,draw,scale=0.5,fill}]
\node at ( 0,0) [circle, draw,scale=0]{};
\node at ( 0,0) [q] {};
\draw[->] (0.1,0) -- (0.5,0);
\draw (0.5,0) -- (0.9,0);
\node at ( 1,0) [aq] {};
\end{tikzpicture}
}
\eea
for the ``current'' states, and 
\bea
\braket{\Psi^A\vert n,n}= \,\,
\underset{n}{
\begin{tikzpicture}
[q/.style={circle,draw,scale=0.5},
aq/.style={circle,draw,scale=0.5,fill}]
\node at ( 0,0) [circle, draw,scale=0]{};
\node at ( 0,1) [q] {};
\node at ( 0.2,1) [aq] {};
\end{tikzpicture}
} \quad \,, \qquad  
\braket{\Psi^A\vert n+2,n}=
\underset{\;\;\;n\;\;\;\;\;\;\;\;\;\;\;\;\;\;\;\;\;\;\;\;n+2}{
\begin{tikzpicture}
[q/.style={circle,draw,scale=0.5},
aq/.style={circle,draw,scale=0.5,fill}]
\node at ( 0,0) [circle, draw,scale=0]{};
\node at ( 0,1) [aq] {};
\draw (0.5,1) -- (0.1,1);
\draw[<-] (0.5,1) -- (1,1);
\draw (1,1) -- (1.5,1);
\draw[<-] (1.3,1) -- (1.9,1);
\node at ( 2,1) [q] {};
\end{tikzpicture}
}
\eea
for the ``future'' states. Here a matter field is represented as a white or black circle for $\vp^\dagger$ or $\vp$ respectively, while a (spatial) gauge field is a line with direction. 
``Current'' fields are on the bottom and ``future'' fields are on the top such that the (Euclidean) time  goes upward. 


The ground state $\ket{0}$ is represented as an empty diagram. 

\subsubsection{Transfer matrix}
The transfer matrix $\hat T$ is given by\footnote{In this appendix, we use rescaled $\hat{T}$ which is used after \S \ref{HPEandT}. Therefore $c_G$ does not appear here.}
\beqa
\langle \Psi^A \vert \hat T \vert \Psi^B \rangle &=& T_G(U,V) T_M(\varphi,\phi){T_0^2(\Psi^B)}.\label{transfermatrix}
\eeqa
Using hopping parameter $K$, we can represent $T_0(\Psi^B), T_M(U,V)$ as 
\begin{align}
{T_0^2(\Psi^B)} &=\prod_{n=0}^{N_l-1} \left( \exp\left[ - \phi_{n}^\dagger \phi_{n} + K
 \left\{\phi_{n}^\dagger V_{n} \phi_{n+1} + \phi_{n+1}^\dagger V_{n}^\dagger \phi_{n} \right\}\right] \right) \nn\\
 &=\prod_{n=0}^{N_l-1} A_n\exp\left[K
 \left\{
\underset{\;\;n\;\;\;n+1}{
\begin{tikzpicture}
[q/.style={circle,draw,scale=0.5},
aq/.style={circle,draw,scale=0.5,fill}]
\node at ( 0,0) [circle, draw,scale=0]{};
\node at ( 0,0) [q] {};
\draw[->] (0.1,0) -- (0.25,0);
\draw (0.25,0) -- (0.4,0);
\node at ( 0.5,0) [aq] {};
\end{tikzpicture}
}+
\underset{\;\;n\;\;\;n+1}{
\begin{tikzpicture}
[q/.style={circle,draw,scale=0.5},
aq/.style={circle,draw,scale=0.5,fill}]
\node at ( 0,0) [circle, draw,scale=0]{};
\node at ( 0,0) [aq] {};
\draw (0.1,0) -- (0.25,0);
\draw [<-](0.25,0) -- (0.4,0);
\node at ( 0.5,0) [q] {};
\end{tikzpicture}
}
 \right\}\right],\nn\\
&=A\prod_{n=0}^{N_l-1} \sum_{h_n=0}^{\infty}\frac{K^{h_n}}{h_n!}
 \left(
\underset{\;\;n\;\;\;n+1}{
\begin{tikzpicture}
[q/.style={circle,draw,scale=0.5},
aq/.style={circle,draw,scale=0.5,fill}]
\node at ( 0,0) [circle, draw,scale=0]{};
\node at ( 0,0) [q] {};
\draw[->] (0.1,0) -- (0.25,0);
\draw (0.25,0) -- (0.4,0);
\node at ( 0.5,0) [aq] {};
\end{tikzpicture}
}+
\underset{\;\;n\;\;\;n+1}{
\begin{tikzpicture}
[q/.style={circle,draw,scale=0.5},
aq/.style={circle,draw,scale=0.5,fill}]
\node at ( 0,0) [circle, draw,scale=0]{};
\node at ( 0,0) [aq] {};
\draw (0.1,0) -- (0.25,0);
\draw [<-](0.25,0) -- (0.4,0);
\node at ( 0.5,0) [q] {};
\end{tikzpicture}
}
 \right)^{h_n},\\
 T_M(\varphi, \phi) &= \prod_{n=0}^{N_l-1} \exp\left[ K\left(\varphi_{n}^\dagger \phi_{n} + \phi_{n}^\dagger \varphi_{n}\right)\right]\nn\\
&= \prod_{n=0}^{N_l-1} \exp\left[ K\left(
\underset{n}{
\begin{tikzpicture}
[q/.style={circle,draw,scale=0.5},
aq/.style={circle,draw,scale=0.5,fill}]
\node at ( 0,0) [circle, draw,scale=0]{};
\node at ( 0,0) [q] {};
\draw (0,0.1) -- (0,0.4);
\node at ( 0,0.5) [aq] {};
\end{tikzpicture}
}+
\underset{n}{
\begin{tikzpicture}
[q/.style={circle,draw,scale=0.5},
aq/.style={circle,draw,scale=0.5,fill}]
\node at ( 0,0) [circle, draw,scale=0]{};
\node at ( 0,0) [aq] {};
\draw (0,0.1) -- (0,0.4);
\node at ( 0,0.5) [q] {};
\end{tikzpicture}
}
\right)\right]\nn\\
&= \prod_{n=0}^{N_l-1} \sum_{v_n=0}^{\infty}\frac{K^{v_n}}{v_n!}
 \left(
\underset{n}{
\begin{tikzpicture}
[q/.style={circle,draw,scale=0.5},
aq/.style={circle,draw,scale=0.5,fill}]
\node at ( 0,0) [circle, draw,scale=0]{};
\node at ( 0,0) [q] {};
\draw (0,0.1) -- (0,0.4);
\node at ( 0,0.5) [aq] {};
\end{tikzpicture}
}+
\underset{n}{
\begin{tikzpicture}
[q/.style={circle,draw,scale=0.5},
aq/.style={circle,draw,scale=0.5,fill}]
\node at ( 0,0) [circle, draw,scale=0]{};
\node at ( 0,0) [aq] {};
\draw (0,0.1) -- (0,0.4);
\node at ( 0,0.5) [q] {};
\end{tikzpicture}
}
 \right)^{v_n},
\end{align}
In the last line of both equations, we expand them in the power series of $K$ (HPE). Here we define
{$A_n =e^{- \phi_{n}^\dagger \phi_{n}}$ and $A=\prod_n A_n$, which give damping factors 
under the $\phi$ integral for normalization\footnote{{We here ignore irrelevant constants such as powers of $\pi$'s.}}.}

Ignoring the difference {between a meson and its Hermitian conjugation, }
we have two types of diagrams,  
horizontal pairs and vertical pairs. Notice that vertical lines have no direction, due to the temporal gauge we take.  Vertical lines are simply connecting color degrees of freedom on both ends {in the (anti)fundamental representation. }  

%
%
%
\subsection{Evaluating the transfer matrix in HPE}
{In this subsection, we explicitly evaluate the action of the transfer matrix to some states.} 
At the $\mathcal{O}(K^3)$ in the HPE, 
generic matrix elements are given by $\braket{\Psi^A\vert \hat{T}\vert \alpha}$ where $\ket{\alpha}=\{\ket{0},\ket{n,m}\}$.  In other words, the ground state $\ket{0}$ mix with at most  a single meson state, and one can neglect multi-meson states at this order\footnote{The multi-meson states are important once we take into account higher order corrections in the HPE.}. 

By inserting the completeness relation, we get
\be
\braket{\Psi^A\vert \hat{T}\vert \alpha}=\int d\Psi^B \braket{\Psi^A\vert \hat{T}\vert \Psi^B}\braket{\Psi^B\vert\alpha}.
\ee
We thus get $\braket{\Psi^A\vert \hat{T}\vert \alpha}$ at $\mathcal{O}(K^s)$ order from the following rules, 
\begin{enumerate}
 \item Start from the diagram representing $\braket{\Psi^B\vert\alpha}$. 
 \item Expand $T_0$ and $T_M$ in terms of $K$ and pick up all allowed terms, {\it i.e.,} terms which satisfy
$ \sum_n(h_n+v_n)\leq s$, 
 where $h_n$ and $v_n$ are numbers of horizontal and vertical pairs, respectively. 
 Then act these terms on the above $\braket{\Psi^B\vert\alpha}$ (graphically {putting corresponding} diagrams), and integrate {$\phi$ (=current matter fields) in the total diagrams. }
 \item Finally act $T_G$ on the diagrams, and {integrate $V$ (=current link variables).} 
\end{enumerate}

We have several comments for integrals of matters and link variables.
\begin{itemize}
\item The integration of $\phi$ can be done by using correlation functions for scalar fields such as
\begin{eqnarray}
& \langle (\phi_n^\dagger)_a \phi_m^b \rangle = \delta_{nm} \delta^{b}{}_a \,, \quad \langle \phi_n^a\phi_m^b \rangle = \langle (\phi_n^\dagger)_a(\phi_m^\dagger)_b \rangle = 0 \,, \;\nn\\
& \langle  (\phi_{n_a}^\dagger)_a \phi_{n_b}^b  (\phi_{n_c}^\dagger)_c \phi_{n_d}^d \rangle  = \delta^{b}{}_a \delta^{d}{}_c \delta_{n_a,n_b}\delta_{n_c,n_d} +  \delta^{d}{}_a \delta^{b}{}_c \delta_{n_a,n_d}\delta_{n_c,n_b} \,, \qquad 
\label{cof}
\end{eqnarray}
where $a,b,c,d=1,2,\dots N$ are color index\footnote{{We take the irrelevant multiplicative constant of $T_0$ to normalize the first equation} }. 
{Non-zero contributions can be obtained if and only if}  the integrand {contains} same number of \tikz{\node at ( 0,0) [circle, draw,scale=0.5]{};} and \tikz{\node at ( 0,0) [circle, draw,scale=0.5, fill]{};} at each site {at the bottom (``current")}. In addition we see that the number of 
\begin{tikzpicture}
[q/.style={circle,draw,scale=0.5},
aq/.style={circle,draw,scale=0.5,fill}]
\node at ( 0,0) [circle, draw,scale=0]{};
\node at ( 0,0) [q] {};
\draw (0,0.1) -- (0,0.4);
\node at ( 0,0.5) [aq] {};
\end{tikzpicture} and 
\begin{tikzpicture}
[q/.style={circle,draw,scale=0.5},
aq/.style={circle,draw,scale=0.5,fill}]
\node at ( 0,0) [circle, draw,scale=0]{};
\node at ( 0,0) [aq] {};
\draw (0,0.1) -- (0,0.4);
\node at ( 0,0.5) [q] {};
\end{tikzpicture}
 must be globally {equal} and the total number of vertical pairs must be even. 
\item In the diagrammatic representation, the integration by $\phi$ {at the bottom (``current")
connects a line attaching to a white circle with a line attaching to a black circle at the same site, and then remove these circles. 
For example, }
\begin{align}
\tikz{\node at ( 0,0) [circle, draw,scale=0.5,fill]{};\node at ( 0,0.5) [circle, draw,scale=0.5]{};\node at ( 0.5,0.5) [circle, draw,scale=0.5]{};\node at ( 0.5,0) [circle, draw,scale=0.5,fill]{};\draw (-0.5,0) -- (-0.1,0);\draw (-0.5,0.5) -- (-0.1,0.5);\draw (0.6,0) -- (1,0);\draw (0.6,0.5) -- (1,0.5);} \qquad \longrightarrow \qquad 
\tikz{\node at ( 0,0) [circle, draw,scale=0.7,white]{};\draw (-0.5,0) -- (0,0);\draw (-0.5,0.5) -- (0,0.5);\draw (0.5,0) -- (1,0);\draw (0.5,0.5) -- (1,0.5);\draw (0,0) -- (0,0.5);\draw (0.5,0) -- (0.5,0.5);} \qquad + \qquad \tikz{\node at ( 0,0) [circle, draw,scale=0.7,white]{};\draw (-0.5,0) -- (0,0);\draw (-0.5,0.5) -- (0,0.5);\draw (0.5,0) -- (1,0);\draw (0.5,0.5) -- (1,0.5);\draw (0,0) -- (0.5,0.5);\draw (0.5,0) -- (0,0.5);} \,\,.\label{rule1}
\end{align}
{If a closed loop or a shrunk point without links appear after the integral, a factor $N$ must be attached as}
\begin{align}
\tikz{\node at ( 0,0) [circle, draw,scale=0.5]{};\node at ( 0,0.5) [circle, draw,scale=0.5,fill]{};\node at ( 1,0.5) [circle, draw,scale=0.5]{};\node at ( 1,0) [circle, draw,scale=0.5,fill]{};\draw [->](0.1,0) -- (0.5,0);\draw (0.5,0) -- (0.9,0);\draw [->](0.9,0.5) -- (0.5,0.5);\draw (0.5,0.5) -- (0.1,0.5);} \quad \longrightarrow \quad 
\tikz{\draw [->](0,0) -- (0.5,0);\draw (0.5,0) -- (1,0);\draw [->](1,0.5) -- (0.5,0.5);\draw (0.5,0.5) -- (0,0.5);\draw (0,0.5) -- (0,0);\draw (1,0.5) -- (1,0);}=N \,,\;\; \qquad 
\tikz{\node at ( 0,0) [circle, draw,scale=0.5]{};\node at ( 0,0.2) [circle, draw,scale=0.5,fill]{};} \quad \longrightarrow N \,.\label{rule2}
\end{align}
{We can explicitly check the above rules using (\ref{cof}).} 
\item As explained in {Section~\S \ref{TmatrixandHPE}}, $T_G$ {can be expanded in terms of characters as} 
\be
T_G(U,V) = \prod_{n=0}^{N_l-1} \sum_{\bf R} d_{\bf R} {\frac{\lambda_{\bf R}(\beta)}{\lambda_{\bf 1}(\beta)}} \chi_{\bf R}( U_{n,1} V_{n,1}^\dagger) \,.
\ee
With the orthogonality condition \eqref{formula3}, one can 
{easily perform the gauge field integration on each link.}
For example, {if $T_{G}(U,V)$ acts on gauge fields $(V_{n,n+3})^a{}_b$ and $V$'s are integrated, 
we can represent this procedure graphically as }
\begin{align}
&\int \left( \Pi_{n=1, \cdots , N_l} dV_{n,n+1} \right) T_G(U,V) \left(V_{n,n+3}\right)^a{}_b\nn\\
&=\Pi_{s=0,1,2} \left( \int dV_{n+s,n+1+s}\left\{ \left(\frac{d_F\lambda_{\bf F}}{\lambda_{\bf 1}}\right)\Tr\left(
\underset{\;\;n+s\;\;\;n+1+s}{
\begin{tikzpicture}
\draw (0,0) -- (0.5,0);
\draw[<-] (0.5,0) -- (1,0);
\draw (1,0) -- (1,1);
\draw (1,1) -- (0.5,1);
\draw[<-] (0.5,1) -- (0,1);
\draw (0,1) -- (0,0);
\end{tikzpicture}
}\right)
\underset{\;\;n+s\;\;\;n+1+s}{
\begin{tikzpicture}
\node at (-0.1,0)[scale=0.7]{$a$};
\node at (1.1,0)[scale=0.7]{$b$};
\draw[->] (0,0) -- (0.5,0);
\draw (0.5,0) -- (1,0);
\end{tikzpicture}
}\right\} \right) \nn\\
&=\left(\frac{d_F\lambda_{\bf F}}{\lambda_{\bf 1}}\right)^3 \Pi_{s=0,1,2}\int dV_{n+s,n+1+s}
\underset{\;\;n+s\;\;\;n+1+s}{
\begin{tikzpicture}
\draw (0,0.2) -- (0.5,0.2);
\draw[<-] (0.5,0.2) -- (1,0.2);
\draw (1,0.2) -- (1,1);
\draw (1,1) -- (0.5,1);
\draw[<-] (0.5,1) -- (0,1);
\draw (0,1) -- (0,0.2);
\node at (-0.1,1)[scale=0.7]{$c$};
\node at (1.1,1)[scale=0.7]{$d$};
\node at (-0.1,0.2)[scale=0.7]{$c$};
\node at (1.1,0.2)[scale=0.7]{$d$};
\node at (-0.1,0)[scale=0.7]{$a$};
\node at (1.1,0)[scale=0.7]{$b$};
\draw[->] (0,0) -- (0.5,0);
\draw (0.5,0) -- (1,0);
\end{tikzpicture}
}\nn\\\nn\\\nn\\
&=\left(\frac{\lambda_{\bf F}}{\lambda_{\bf 1}}\right)^3
\underset{\;\;n\;\;\;n+3}{
\begin{tikzpicture}
\node at ( 0,0) [circle, draw,scale=0]{};
\node at (-0.1,1)[scale=0.7]{$a$};
\node at (1.1,1)[scale=0.7]{$b$};
\draw (1,1) -- (0.5,1);
\draw[<-] (0.5,1) -- (0,1);
\end{tikzpicture}
}\;.
\end{align}
We see in this case that $T_G$ plays a role of uplifting gauge fields with the factor $\lambda_{\bf F}{/\lambda_{\bf 1}}$ for each link.
\item More generally, acting on links which belong to the irreducible representation $\bold{R}$, $T_G$ uplifts gauge fields with the factor $\lambda_{R}{/\lambda_{\bf 1}}$. 
\item For more complicated links which do not belong to one irreducible representation {such as a product of links in some representations}, we should decompose them into irreducible representations before the integration. For example, $V^a{}_bV^{\dagger d}{}_c$, which belongs to fundamental $\times$ anti-fundamental representations, can be decomposed into singlet and adjoint part as
\be
V^a{}_bV^{\dagger d}{}_c=\left(\dfrac{1}{N}\delta^a{}_c\delta^d{}_b\right)+\left(V^a{}_bV^{\dagger d}{}_c-\dfrac{1}{N}\delta^a{}_c\delta^d{}_b\right). 
\ee 
We can visualize this as
\be
\tikz{\node at (-0.1,0)[scale=0.7]{$a$};\node at (-0.1,0.2)[scale=0.7]{$c$};\node at (1.1,0)[scale=0.7]{$b$};\node at (1.1,0.2)[scale=0.7]{$d$};\draw [->](0.1,0) -- (0.5,0);\draw (0.5,0) -- (0.9,0);\draw [->](0.9,0.2) -- (0.5,0.2);\draw (0.5,0.2) -- (0.1,0.2);} 
\;\;=\;\; \dfrac{1}{N}\delta^a{}_c\;\;\;\;\delta^d{}_b+\tikz{\node at (-0.3,0)[scale=0.7]{$(a,c)$};\node at (1.3,0)[scale=0.7]{$(b,d)$};\draw [double](0.1,0) -- (0.9,0);} ,\label{gaugedec}
\ee
where the doubled line without direction represents the gauge field in the adjoint representation. {Each pair $(a,c)$ or $(b,d)$ correspond to an index of the adjoint representation of the gauge field, whose dimension is $N^2-1$. }
\item With matter fields, we can represent the decomposition(\ref{dec1}) as:
\be
\tikz{\node at ( 0,0) [circle, draw,scale=0.5]{};\node at ( 0,0.5) [circle, draw,scale=0.5,fill]{};\node at ( 1,0.5) [circle, draw,scale=0.5]{};\node at ( 1,0) [circle, draw,scale=0.5,fill]{};\draw [->](0.1,0) -- (0.5,0);\draw (0.5,0) -- (0.9,0);\draw [->](0.9,0.5) -- (0.5,0.5);\draw (0.5,0.5) -- (0.1,0.5);} \;\; =\;\;\dfrac{1}{N}\tikz{\node at ( 0,0) [circle, draw,scale=0.5]{};\node at ( 0,0.2) [circle, draw,scale=0.5,fill]{};\node at ( 1,0.2) [circle, draw,scale=0.5]{};\node at ( 1,0) [circle, draw,scale=0.5,fill]{};}\;\;+\;\;
\tikz{\node at ( 0,0) [rectangle, draw,scale=0.7]{};\node at ( 1,0) [rectangle, draw,scale=0.7]{};\draw [double](0.1,0) -- (0.9,0);}\;\;,
\ee
where squares corresponds to the adjoint parts of the matter field. {This leads to} the following relation we will use later.  
\begin{align}
\int dVT_G(U,V)
\begin{tikzpicture}
[q/.style={circle,draw,scale=0.5},
aq/.style={circle,draw,scale=0.5,fill}]
\node at ( 0,1) [q] {};
\draw (0,0.9) -- (0,0);
\draw [->](0,0) -- (0.5,0);
\draw (0.5,0) -- (1,0);
\draw (1,0) -- (1,0.9);
\node at ( 1,1) [aq] {};
\node at ( 0.2,1) [aq] {};
\draw (0.2,0.9) -- (0.2,0.1);
\draw (0.2,0.1) -- (0.5,0.1);
\draw [<-](0.5,0.1) -- (0.8,0.1);
\draw (0.8,0.1) -- (0.8,0.9);
\node at ( 0.8,1) [q] {};
\node at ( 1,1) [aq] {};
\end{tikzpicture}
=&\int dVT_G(U,V)\left[\dfrac{1}{N}
\begin{tikzpicture}
[q/.style={circle,draw,scale=0.5},
aq/.style={circle,draw,scale=0.5,fill}]
\node at ( 0,0) [circle, draw,scale=0]{};
\node at ( 0,1) [q] {};
\node at ( 1,1) [aq] {};
\node at ( 0.2,1) [aq] {};
\node at ( 0.8,1) [q] {};
\node at ( 1,1) [aq] {};
\end{tikzpicture}
+
\begin{tikzpicture}
[a/.style={rectangle,draw,scale=0.7}]
\node at ( 0,1) [a] {};
\draw [double](0,0.9) -- (0,0);
\draw [double](0,0) -- (1,0);
\draw [double](1,0) -- (1,0.9);
\node at ( 1,1) [a] {};
\end{tikzpicture}
\right]\nn\\\nn\\\nn\\
=&\dfrac{1}{N}
\begin{tikzpicture}
[q/.style={circle,draw,scale=0.5},
aq/.style={circle,draw,scale=0.5,fill}]
\node at ( 0,0) [circle, draw,scale=0]{};
\node at ( 0,1) [q] {};
\node at ( 1,1) [aq] {};
\node at ( 0.2,1) [aq] {};
\node at ( 0.8,1) [q] {};
\node at ( 1,1) [aq] {};
\end{tikzpicture}
+\dfrac{\lambda_{\bf Adj}}{\lambda_{\bf 1}}
\begin{tikzpicture}
[a/.style={rectangle,draw,scale=0.7}]
\node at ( 0,0) [circle, draw,scale=0]{};
\node at ( 0,1) [a] {};
\draw [double](0.1,1) -- (0.9,1);
\node at ( 1,1) [a] {};
\end{tikzpicture}
.\label{formuladec}
\end{align}
\end{itemize}
\subsection{Some examples}
\subsubsection{$\braket{\Psi^A\vert \hat{T}\vert n,n}$ at $\mathcal{O}(K^3)$}

We derive the explicit form of $\braket{\Psi^A\vert \hat{T}\vert n,n}$ at $\mathcal{O}(K^3)$. 
We start from the diagram \tikz{\node at ( 0,0) [circle, draw,scale=0.5]{};\node at ( 0.2,0) [circle, draw,scale=0.5, fill]{};}. 

At $K^0$ order we only have
\be
\braket{\Psi^A\vert \hat{T}\vert n,n}\vert_{K^0}=\int d\Psi^BA
\overset{aa}{\underset{n}{
\begin{tikzpicture}
[q/.style={circle,draw,scale=0.5},
aq/.style={circle,draw,scale=0.5,fill}]
\node at ( 0,0) [q] {};
\node at ( 0.2,0) [aq] {};
\end{tikzpicture}
}}
=\delta^{a}{}_{a}=N, 
\ee
where we denote color indices explicitly.  We thus obtain
\be
\hat{T}\ket{n,n}\vert_{K^0}=N\ket{0}.
\ee

At $K^1$, from the comment we gave before, the acting pair must be horizontal. However one horizontal pair can not make even number matter fields on each site, so  there are no contribution at this order. 

At $K^2$ order, next, we can consider two vertical pairs or two horizontal pairs. In both cases two pairs must {share the same link as}
\begin{align}
\frac{\braket{\Psi^A\vert \hat{T}\vert n,n}\vert_{K^2}}{K^2}=&\int dV T_G(U,V)d\phi A\left[
\underset{n}{
\begin{tikzpicture}
[q/.style={circle,draw,scale=0.5},
aq/.style={circle,draw,scale=0.5,fill}]
\node at (-0.2,1)[scale=0.7]{$a$};
\node at (-0.2,0.2)[scale=0.7]{$a$};
\node at (-0.2,0)[scale=0.7]{$b$};
\node at (0.4,1)[scale=0.7]{$c$};
\node at (0.4,0.2)[scale=0.7]{$c$};
\node at (0.4,0)[scale=0.7]{$b$};
\node at ( 0,0) [q] {};
\node at ( 0.2,0) [aq] {};
\node at ( 0,0.2) [aq] {};
\draw (0,0.3) -- (0,0.9);
\node at ( 0,1) [q] {};
\node at ( 0.2,0.2) [q] {};
\draw (0.2,0.3) -- (0.2,0.9);
\node at ( 0.2,1) [aq] {};
\end{tikzpicture}
}
+\sum_{m\not=n}\left(\underset{n}{\begin{tikzpicture}
[q/.style={circle,draw,scale=0.5},
aq/.style={circle,draw,scale=0.5,fill}]
\node at (-0.2,0)[scale=0.7]{$a$};
\node at (0.4,0)[scale=0.7]{$a$};
\node at ( 0,0) [q] {};
\node at ( 0.2,0) [aq] {};
\end{tikzpicture}
}\dots
\underset{m}{\begin{tikzpicture}
[q/.style={circle,draw,scale=0.5},
aq/.style={circle,draw,scale=0.5,fill}]
\node at (-0.2,1)[scale=0.7]{$b$};
\node at (-0.2,0)[scale=0.7]{$b$};
\node at (0.4,1)[scale=0.7]{$c$};
\node at (0.4,0)[scale=0.7]{$c$};
\node at ( 0,0) [aq] {};
\node at ( 0.2,0) [q] {};
\draw (0,0.1) -- (0,0.9);
\node at ( 0,1) [q] {};
\draw (0.2,0.1) -- (0.2,0.9);
\node at ( 0.2,1) [aq] {};
\end{tikzpicture}
}\right)\right.\nn\\
&+\left.
\underset{\;\;n\raise1ex\hbox{\scriptsize{$e$}}\;\;\;\;\;n+1}{
\begin{tikzpicture}
[q/.style={circle,draw,scale=0.5},
aq/.style={circle,draw,scale=0.5,fill}]
\node at (-0.2,0.2)[scale=0.7]{$a$};
\node at (-0.2,0)[scale=0.7]{$a$};
\node at (0.2,0.5)[scale=0.7]{$b$};
\node at (1.2,0.2)[scale=0.7]{$c$};
\node at (1.2,0)[scale=0.7]{$d$};
\node at ( 0,0) [q] {};
\node at ( 0,0.2) [aq] {};
\node at ( 0.2,0) [aq] {};
\draw (0.3,0) -- (0.5,0);
\draw [<-](0.5,0) -- (0.9,0);
\node at ( 1,0) [q] {};
\node at ( 0.2,0.2) [q] {};
\draw[->] (0.3,0.2) -- (0.5,0.2);
\draw (0.5,0.2) -- (0.9,0.2);
\node at ( 1,0.2) [aq] {};
\end{tikzpicture}
}+
\underset{\;\;n-1\;\;\;\raise1ex\hbox{\scriptsize{$d$}}\;\;n}{
\begin{tikzpicture}
[aq/.style={circle,draw,scale=0.5},
q/.style={circle,draw,scale=0.5,fill}]
\node at (0.2,0.2)[scale=0.7]{$c$};
\node at (0.2,0)[scale=0.7]{$c$};
\node at (-0.2,0.5)[scale=0.7]{$b$};
\node at (-1.2,0.2)[scale=0.7]{$a$};
\node at (-1.2,0)[scale=0.7]{$e$};
\node at ( 0,0) [q] {};
\node at ( 0,0.2) [aq] {};
\node at ( -0.2,0) [aq] {};
\draw [->](-0.3,0) -- (-0.5,0);
\draw (-0.5,0) -- (-0.9,0);
\node at ( -1,0) [q] {};
\node at ( -0.2,0.2) [q] {};
\draw (-0.3,0.2) -- (-0.5,0.2);
\draw [<-](-0.5,0.2) -- (-0.9,0.2);
\node at ( -1,0.2) [aq] {};
\end{tikzpicture}
}+
\sum_{m\not= n-1,n}\left(
\underset{n}{
\begin{tikzpicture}
[q/.style={circle,draw,scale=0.5},
aq/.style={circle,draw,scale=0.5,fill}]
\node at (-0.2,0.2)[scale=0.7]{$a$};
\node at (-0.2,0)[scale=0.7]{$a$};
\node at ( 0,0) [q] {};
\node at ( 0,0.2) [aq] {};
\end{tikzpicture}
}\dots
\underset{\;\;m\;\;\;\;m+1}{
\begin{tikzpicture}
[q/.style={circle,draw,scale=0.5},
aq/.style={circle,draw,scale=0.5,fill}]
\node at (0,0.2)[scale=0.7]{$b$};
\node at (1.2,0.2)[scale=0.7]{$c$};
\node at (1.2,0)[scale=0.7]{$d$};
\node at (0,0)[scale=0.7]{$e$};
\node at ( 0.2,0) [aq] {};
\draw (0.3,0) -- (0.5,0);
\draw [<-](0.5,0) -- (0.9,0);
\node at ( 1,0) [q] {};
\node at ( 0.2,0.2) [q] {};
\draw[->] (0.3,0.2) -- (0.5,0.2);
\draw (0.5,0.2) -- (0.9,0.2);
\node at ( 1,0.2) [aq] {};
\end{tikzpicture}
}
\right)\right]\nn\\\nn\\
=&\int dV T_G(U,V)\left[(\delta^b{}_a\delta^c{}_{b}+\delta^b{}_b\delta^c{}_{a})
\underset{n}{
\begin{tikzpicture}
[q/.style={circle,draw,scale=0.5},
aq/.style={circle,draw,scale=0.5,fill}]
\node at ( 0,0) [circle, draw,scale=0]{};
\node at (-0.2,1)[scale=0.7]{$a$};
\node at (0.4,1)[scale=0.7]{$c$};
\node at ( 0,1) [q] {};
\node at ( 0.2,1) [aq] {};
\end{tikzpicture}
}+\sum_{m\not= n}\delta^a{}_{a}\delta^{b}{}_{c}
\underset{m}{
\begin{tikzpicture}
[q/.style={circle,draw,scale=0.5},
aq/.style={circle,draw,scale=0.5,fill}]
\node at ( 0,0) [circle, draw,scale=0]{};
\node at (-0.2,1)[scale=0.7]{$b$};
\node at (0.4,1)[scale=0.7]{$c$};
\node at ( 0,1) [q] {};
\node at ( 0.2,1) [aq] {};
\end{tikzpicture}
}\right.\nn\\\nn\\\nn\\
&+(\delta^a{}_a\delta^b{}_{e}+\delta^a{}_e\delta^b{}_{a})\delta^c{}_d
\underset{\;\;n\;\;\;\;n+1}{
\begin{tikzpicture}
[q/.style={circle,draw,scale=0.5},
aq/.style={circle,draw,scale=0.5,fill}]
\node at (-0.1,-0.1)[scale=0.7]{$e$};
\node at (-0.1,0.2)[scale=0.7]{$b$};
\node at (1.1,-0.1)[scale=0.7]{$d$};
\node at (1.1,0.2)[scale=0.7]{$c$};
\draw[->] (0,-0.1) -- (0.5,-0.1);
\draw (0.5,-0.1) -- (1,-0.1);
\draw (0,0.2) -- (0.5,0.2);
\draw[<-] (0.5,0.2) -- (1,0.2);
\end{tikzpicture}
}
+\delta^a{}_e(\delta^c{}_c\delta^d{}_{b}+\delta^c{}_b\delta^d{}_{c})
\underset{\;\;n-1\;\;\;\;n}{
\begin{tikzpicture}
[q/.style={circle,draw,scale=0.5},
aq/.style={circle,draw,scale=0.5,fill}]
\node at (-0.1,-0.1)[scale=0.7]{$e$};
\node at (-0.1,0.2)[scale=0.7]{$a$};
\node at (1.1,-0.1)[scale=0.7]{$d$};
\node at (1.1,0.2)[scale=0.7]{$b$};
\draw[->] (0,-0.1) -- (0.5,-0.1);
\draw (0.5,-0.1) -- (1,-0.1);
\draw (0,0.2) -- (0.5,0.2);
\draw[<-] (0.5,0.2) -- (1,0.2);
\end{tikzpicture}
}\nn\\\nn\\
&\left.+\sum_{m\not= n-1,n}\delta^a{}_a\delta^b{}_e\delta^c{}_d
\underset{\;\;m\;\;\;\;m+1}{
\begin{tikzpicture}
[q/.style={circle,draw,scale=0.5},
aq/.style={circle,draw,scale=0.5,fill}]
\node at (-0.1,-0.1)[scale=0.7]{$e$};
\node at (-0.1,0.2)[scale=0.7]{$b$};
\node at (1.1,-0.1)[scale=0.7]{$d$};
\node at (1.1,0.2)[scale=0.7]{$c$};
\draw[->] (0,-0.1) -- (0.5,-0.1);
\draw (0.5,-0.1) -- (1,-0.1);
\draw (0,0.2) -- (0.5,0.2);
\draw[<-] (0.5,0.2) -- (1,0.2);
\end{tikzpicture}
}\right]
\nn\\\nn\\
=&\, (1+N) \,
\underset{n}{
\begin{tikzpicture}
[q/.style={circle,draw,scale=0.5},
aq/.style={circle,draw,scale=0.5,fill}]
\node at ( 0,0) [circle, draw,scale=0]{};
\node at ( 0,1) [q] {};
\node at ( 0.2,1) [aq] {};
\end{tikzpicture}
}+N \sum_{m\not= n} \,
\underset{m}{
\begin{tikzpicture}
[q/.style={circle,draw,scale=0.5},
aq/.style={circle,draw,scale=0.5,fill}]
\node at ( 0,0) [circle, draw,scale=0]{};
\node at ( 0,1) [q] {};
\node at ( 0.2,1) [aq] {};
\end{tikzpicture}
}+2N(1+N)+N^2(N_l-2) \,.
\end{align}
We finally obtain
\be
\hat{T}\ket{n,n}\vert_{K^2}=K^2N(NN_l+2)\ket{0}+K^2\ket{n,n}+K^2N\sum_m\ket{m,m}.
\ee

At $\mathcal{O}(K^3)$ order, there are three horizontal or vertical pairs. 
Only the  ``{\it U}'' shape diagram, consisting of two vertical and one horizontal pairs, 
are allowed, since other cases lead to an odd number of scalar fields on some site. 
{Employing}  rules (\ref{rule1}) and (\ref{rule2}) and taking care for the direction, 
we have 
\begin{align}
\frac{\braket{\Psi^A\vert \hat{T}\vert n,n}\vert_{K^3}}{K^3}
=\int dV T_G(U,V)d\phi&A\left[
\underset{\;\;n\;\;\;\;\;\;n+1}{
\begin{tikzpicture}
[q/.style={circle,draw,scale=0.5},
aq/.style={circle,draw,scale=0.5,fill}]
\node at ( 0,0.2) [q] {};
\node at ( 0,0) [aq] {};
\node at ( 0.2,1) [q] {};
\draw (0.2,0.9) -- (0.2,0.3);
\node at ( 0.2,0.2) [aq] {};
\node at ( 0.2,0) [q] {};
\draw [->](0.3,0) -- (0.6,0);
\draw (0.5,0) -- (0.9,0);
\node at ( 1,0) [aq] {};
\node at ( 1,0.2) [q] {};
\draw (1,0.3) -- (1,0.9);
\node at ( 1,1) [aq] {};
\end{tikzpicture}
}+
\underset{\;\;n\;\;\;\;\;\;n+1}{
\begin{tikzpicture}
[aq/.style={circle,draw,scale=0.5},
q/.style={circle,draw,scale=0.5,fill}]
\node at ( 0,0.2) [q] {};
\node at ( 0,0) [aq] {};
\node at ( 0.2,1) [q] {};
\draw (0.2,0.9) -- (0.2,0.3);
\node at ( 0.2,0.2) [aq] {};
\node at ( 0.2,0) [q] {};
\draw (0.3,0) -- (0.6,0);
\draw [<-](0.5,0) -- (0.9,0);
\node at ( 1,0) [aq] {};
\node at ( 1,0.2) [q] {};
\draw (1,0.3) -- (1,0.9);
\node at ( 1,1) [aq] {};
\end{tikzpicture}
}
\right.\nn\\\nn\\
&+
\underset{\;\;n-1\;\;\;\;\;\;n}{
\begin{tikzpicture}
[q/.style={circle,draw,scale=0.5},
aq/.style={circle,draw,scale=0.5,fill}]
\node at ( 0,0.2) [q] {};
\node at ( 0,0) [aq] {};
\node at ( -0.2,1) [q] {};
\draw (-0.2,0.9) -- (-0.2,0.3);
\node at ( -0.2,0.2) [aq] {};
\node at ( -0.2,0) [q] {};
\draw [->](-0.3,0) -- (-0.6,0);
\draw (-0.5,0) -- (-0.9,0);
\node at ( -1,0) [aq] {};
\node at ( -1,0.2) [q] {};
\draw (-1,0.3) -- (-1,0.9);
\node at ( -1,1) [aq] {};
\end{tikzpicture}
}+
\underset{\;\;n-1\;\;\;\;\;\;n}{
\begin{tikzpicture}
[aq/.style={circle,draw,scale=0.5},
q/.style={circle,draw,scale=0.5,fill}]
\node at ( 0,0.2) [q] {};
\node at ( 0,0) [aq] {};
\node at ( -0.2,1) [q] {};
\draw (-0.2,0.9) -- (-0.2,0.3);
\node at ( -0.2,0.2) [aq] {};
\node at ( -0.2,0) [q] {};
\draw (-0.3,0) -- (-0.6,0);
\draw [<-](-0.5,0) -- (-0.9,0);
\node at ( -1,0) [aq] {};
\node at ( -1,0.2) [q] {};
\draw (-1,0.3) -- (-1,0.9);
\node at ( -1,1) [aq] {};
\end{tikzpicture}
}
\nn\\\nn\\
&\left.+\sum_{m\not=n,n-1}\left(
\underset{n}{
\begin{tikzpicture}
[q/.style={circle,draw,scale=0.5},
aq/.style={circle,draw,scale=0.5,fill}]
\node at ( 0,0.2) [q] {};
\node at ( 0,0) [aq] {};
\end{tikzpicture}
}\dots
\underset{\;\;m\;\;\;\;\;\;m+1}{
\begin{tikzpicture}
[q/.style={circle,draw,scale=0.5},
aq/.style={circle,draw,scale=0.5,fill}]
\node at ( 0.2,1) [q] {};
\draw (0.2,0.9) -- (0.2,0.3);
\node at ( 0.2,0.2) [aq] {};
\node at ( 0.2,0) [q] {};
\draw [->](0.3,0) -- (0.6,0);
\draw (0.5,0) -- (0.9,0);
\node at ( 1,0) [aq] {};
\node at ( 1,0.2) [q] {};
\draw (1,0.3) -- (1,0.9);
\node at ( 1,1) [aq] {};
\end{tikzpicture}
}+
\underset{n}{
\begin{tikzpicture}
[q/.style={circle,draw,scale=0.5},
aq/.style={circle,draw,scale=0.5,fill}]
\node at ( 0,0.2) [q] {};
\node at ( 0,0) [aq] {};
\end{tikzpicture}
}\dots
\underset{\;\;m\;\;\;\;\;\;m+1}{
\begin{tikzpicture}
[aq/.style={circle,draw,scale=0.5},
q/.style={circle,draw,scale=0.5,fill}]
\node at ( 0.2,1) [q] {};
\draw (0.2,0.9) -- (0.2,0.3);
\node at ( 0.2,0.2) [aq] {};
\node at ( 0.2,0) [q] {};
\draw (0.3,0) -- (0.6,0);
\draw [<-](0.5,0) -- (0.9,0);
\node at ( 1,0) [aq] {};
\node at ( 1,0.2) [q] {};
\draw (1,0.3) -- (1,0.9);
\node at ( 1,1) [aq] {};
\end{tikzpicture}
}\right)\right]\nn\\\nn\\
=\int dVT_G(U,V)&\left[(N+1)\sum_{m=n-1,n}\left(
\underset{\;\;m\;\;\;\;\;\;m+1}{
\begin{tikzpicture}
[q/.style={circle,draw,scale=0.5},
aq/.style={circle,draw,scale=0.5,fill}]
\node at ( 0,1) [q] {};
\draw (0,0.9) -- (0,0);
\draw [->](0,0) -- (0.5,0);
\draw (0.5,0) -- (1,0);
\draw (1,0) -- (1,0.9);
\node at ( 1,1) [aq] {};
\end{tikzpicture}
}
+
\underset{\;\;m\;\;\;\;\;\;m+1}{
\begin{tikzpicture}
[aq/.style={circle,draw,scale=0.5},
q/.style={circle,draw,scale=0.5,fill}]
\node at ( 0,1) [q] {};
\draw (0,0.9) -- (0,0);
\draw (0,0) -- (0.5,0);
\draw [<-](0.5,0) -- (1,0);
\draw (1,0) -- (1,0.9);
\node at ( 1,1) [aq] {};
\end{tikzpicture}
}
\right)\right.\nn\\\nn\\
&+
\left.N\sum_{m\not= n-1,n}\left(
\underset{\;\;m\;\;\;\;\;\;m+1}{
\begin{tikzpicture}
[q/.style={circle,draw,scale=0.5},
aq/.style={circle,draw,scale=0.5,fill}]
\node at ( 0,1) [q] {};
\draw (0,0.9) -- (0,0);
\draw [->](0,0) -- (0.5,0);
\draw (0.5,0) -- (1,0);
\draw (1,0) -- (1,0.9);
\node at ( 1,1) [aq] {};
\end{tikzpicture}
}
+
\underset{\;\;m\;\;\;\;\;\;m+1}{
\begin{tikzpicture}
[aq/.style={circle,draw,scale=0.5},
q/.style={circle,draw,scale=0.5,fill}]
\node at ( 0,1) [q] {};
\draw (0,0.9) -- (0,0);
\draw (0,0) -- (0.5,0);
\draw [<-](0.5,0) -- (1,0);
\draw (1,0) -- (1,0.9);
\node at ( 1,1) [aq] {};
\end{tikzpicture}
}
\right)\right]\nn\\\nn\\
= \left(\frac{\lambda_{\bf F}}{\lambda_{\bf 1}}\right)\sum_m( N + & \delta_{m,n-1} +\delta_{m,n}) \left(
\underset{\;\;m\;\;\;\;\;\;m+1}{
\begin{tikzpicture}
[q/.style={circle,draw,scale=0.5},
aq/.style={circle,draw,scale=0.5,fill}]
\node at ( 0,0) [circle, draw,scale=0]{};
\node at ( 0,1) [q] {};
\draw [->](0.1,1) -- (0.5,1);
\draw (0.4,1) -- (0.9,1);
\node at ( 1,1) [aq] {};
\end{tikzpicture}
}
+
\underset{\;\;m\;\;\;\;\;\;m+1}{
\begin{tikzpicture}
[aq/.style={circle,draw,scale=0.5},
q/.style={circle,draw,scale=0.5,fill}]
\node at ( 0,0) [circle, draw,scale=0]{};
\node at ( 0,1) [q] {};
\draw (0.1,1) -- (0.5,1);
\draw [<-](0.5,1) -- (0.9,1);
\node at ( 1,1) [aq] {};
\end{tikzpicture}
}
\right).
\end{align}
As a result, we obtain 
\bea
\hat{T}\ket{n,n}\vert_{K^3}&=&
K^3\left(\frac{\lambda_{\bf F}}{\lambda_{\bf 1}}\right)(\ket{n,n+1}+\ket{n+1,n}+\ket{n-1,n}+\ket{n,n-1})\nn\\
\, &&+K^3N\left(\frac{\lambda_{\bf F}}{\lambda_{\bf 1}}\right)\sum_m\left(\ket{m,m+1}+\ket{m,m-1}\right) \,. 
\eea
\subsubsection{The detail for the calculation of \eqref{eigenlhs}}
\label{ii+1}
Here we show the derivation of \eqref{eigenlhs}, coefficient of $\ket{i,i}\ket{i+1,i+1}$ term at $K^6$ order. All we have to consider is $\hat{T}_6\ket{G^+_0}=\hat{T}_6\ket{0}$ and $\hat{T}_4\ket{G^+_2}=\sum_n\hat{T}_4\ket{n,n}$.

First let us consider $\hat{T_6}\ket{0}$. We have six meson-like pairs in $\hat{T}$, which act on $\ket{0}$. The four of them must be devoted to construct the future state $\ket{i,i}\ket{i+1,i+1}$ and the other two must be conjugated with each other in the horizontal direction. So we have following patterns of configurations to integrate:
\be
\underset{\;\;i\;\;\;\;\;\;i+1}{
\begin{tikzpicture}
[q/.style={circle,draw,scale=0.5},
aq/.style={circle,draw,scale=0.5,fill}]
\node at ( 0,1) [q] {};
\draw (0,0.9) -- (0,0.1);
\node at ( 0,0) [aq] {};
\node at ( 1,0) [q] {};
\draw (1,0.1) -- (1,0.9);
\node at ( 1,1) [aq] {};
\node at ( 0.2,1) [aq] {};
\draw (0.2,0.9) -- (0.2,0.1);
\node at ( 0.2,0) [q] {};
\node at ( 0.8,0) [aq] {};
\draw (0.8,0.1) -- (0.8,0.9);
\node at ( 0.8,1) [q] {};
\node at ( 0,-0.2) [q] {};
\draw [->](0.1,-0.2) -- (0.5,-0.2);
\draw (0.5,-0.2) -- (0.9,-0.2);
\node at ( 1,-0.2) [aq] {};
\node at ( 0,-0.4) [aq] {};
\draw (0.1,-0.4) -- (0.5,-0.4);
\draw [<-](0.5,-0.4) -- (0.9,-0.4);
\node at ( 1,-0.4) [q] {};
\end{tikzpicture}
},\;\;\;
\underset{i-1\;\;\;\;\;\;\;\;i\;\;\;\;\;\;\;\;i+1}{
\begin{tikzpicture}
[q/.style={circle,draw,scale=0.5},
aq/.style={circle,draw,scale=0.5,fill}]
\node at ( 0,1) [q] {};
\draw (0,0.9) -- (0,0.1);
\node at ( 0,0) [aq] {};
\node at ( 1,0) [q] {};
\draw (1,0.1) -- (1,0.9);
\node at ( 1,1) [aq] {};
\node at ( 0.2,1) [aq] {};
\draw (0.2,0.9) -- (0.2,0.1);
\node at ( 0.2,0) [q] {};
\node at ( 0.8,0) [aq] {};
\draw (0.8,0.1) -- (0.8,0.9);
\node at ( 0.8,1) [q] {};
\node at ( -1.2,0.2) [q] {};
\draw [->](-1.1,0.2) -- (-0.7,0.2);
\draw (-0.7,0.2) -- (-0.3,0.2);
\node at ( -0.2,0.2) [aq] {};
\node at ( -1.2,0) [aq] {};
\draw (-1.1,0) -- (-0.7,0);
\draw [<-](-0.7,0.) -- (-0.3,0.);
\node at ( -0.2,0) [q] {};
\end{tikzpicture}
},\;\;\;
\underset{i\;\;\;\;\;\;\;\;i+1\;\;\;\;\;\;\;\;i+2}{
\begin{tikzpicture}
[q/.style={circle,draw,scale=0.5},
aq/.style={circle,draw,scale=0.5,fill}]
\node at ( 0,1) [q] {};
\draw (0,0.9) -- (0,0.1);
\node at ( 0,0) [aq] {};
\node at ( 1,0) [q] {};
\draw (1,0.1) -- (1,0.9);
\node at ( 1,1) [aq] {};
\node at ( 0.2,1) [aq] {};
\draw (0.2,0.9) -- (0.2,0.1);
\node at ( 0.2,0) [q] {};
\node at ( 0.8,0) [aq] {};
\draw (0.8,0.1) -- (0.8,0.9);
\node at ( 0.8,1) [q] {};
\node at ( 1.2,0.2) [q] {};
\draw [->](1.3,0.2) -- (1.7,0.2);
\draw (1.7,0.2) -- (2.1,0.2);
\node at ( 2.2,0.2) [aq] {};
\node at ( 1.2,0) [aq] {};
\draw (1.3,0) -- (1.7,0);
\draw [<-](1.7,0.) -- (2.1,0.);
\node at ( 2.2,0) [q] {};
\end{tikzpicture}
},\;\;\;
\underset{i\;\;\;\;\;\;i+1}{
\begin{tikzpicture}
[q/.style={circle,draw,scale=0.5},
aq/.style={circle,draw,scale=0.5,fill}]
\node at ( 0,1) [q] {};
\draw (0,0.9) -- (0,0.1);
\node at ( 0,0) [aq] {};
\node at ( 1,0) [q] {};
\draw (1,0.1) -- (1,0.9);
\node at ( 1,1) [aq] {};
\node at ( 0.2,1) [aq] {};
\draw (0.2,0.9) -- (0.2,0.1);
\node at ( 0.2,0) [q] {};
\node at ( 0.8,0) [aq] {};
\draw (0.8,0.1) -- (0.8,0.9);
\node at ( 0.8,1) [q] {};
\end{tikzpicture}
}\dots
\underset{j\;\;\;\;\;\;j+1}{
\begin{tikzpicture}
[q/.style={circle,draw,scale=0.5},
aq/.style={circle,draw,scale=0.5,fill}]
\node at ( 0,-0.2) [q] {};
\draw [->](0.1,-0.2) -- (0.5,-0.2);
\draw (0.5,-0.2) -- (0.9,-0.2);
\node at ( 1,-0.2) [aq] {};
\node at ( 0,-0.4) [aq] {};
\draw (0.1,-0.4) -- (0.5,-0.4);
\draw [<-](0.5,-0.4) -- (0.9,-0.4);
\node at ( 1,-0.4) [q] {};
\end{tikzpicture}
},
\ee
where $j\not= i-1,i,i+1$. For the first configuration, the integration {can be done as}
\begin{align}
\int dV T_G(U,V)d\phi A
\begin{tikzpicture}
[q/.style={circle,draw,scale=0.5},
aq/.style={circle,draw,scale=0.5,fill}]
\node at ( 0,1) [q] {};
\draw (0,0.9) -- (0,0.1);
\node at ( 0,0) [aq] {};
\node at ( 1,0) [q] {};
\draw (1,0.1) -- (1,0.9);
\node at ( 1,1) [aq] {};
\node at ( 0.2,1) [aq] {};
\draw (0.2,0.9) -- (0.2,0.1);
\node at ( 0.2,0) [q] {};
\node at ( 0.8,0) [aq] {};
\draw (0.8,0.1) -- (0.8,0.9);
\node at ( 0.8,1) [q] {};
\node at ( 0,-0.2) [q] {};
\draw [->](0.1,-0.2) -- (0.5,-0.2);
\draw (0.5,-0.2) -- (0.9,-0.2);
\node at ( 1,-0.2) [aq] {};
\node at ( 0,-0.4) [aq] {};
\draw (0.1,-0.4) -- (0.5,-0.4);
\draw [<-](0.5,-0.4) -- (0.9,-0.4);
\node at ( 1,-0.4) [q] {};
\end{tikzpicture}
=&\int dV T_G(U,V)\left[(N+2)\begin{tikzpicture}
[q/.style={circle,draw,scale=0.5},
aq/.style={circle,draw,scale=0.5,fill}]
\node at ( 0,0) [circle, draw,scale=0]{};
\node at ( 0,1) [q] {};
\node at ( 1,1) [aq] {};
\node at ( 0.2,1) [aq] {};
\node at ( 0.8,1) [q] {};
\node at ( 1,1) [aq] {};
\end{tikzpicture}+
\begin{tikzpicture}
[q/.style={circle,draw,scale=0.5},
aq/.style={circle,draw,scale=0.5,fill}]
\node at ( 0,1) [q] {};
\draw (0,0.9) -- (0,0);
\draw [->](0,0) -- (0.5,0);
\draw (0.5,0) -- (1,0);
\draw (1,0) -- (1,0.9);
\node at ( 1,1) [aq] {};
\node at ( 0.2,1) [aq] {};
\draw (0.2,0.9) -- (0.2,0.1);
\draw (0.2,0.1) -- (0.5,0.1);
\draw [<-](0.5,0.1) -- (0.8,0.1);
\draw (0.8,0.1) -- (0.8,0.9);
\node at ( 0.8,1) [q] {};
\node at ( 1,1) [aq] {};
\end{tikzpicture}
\right]\nn\\
=&\int dV T_G(U,V)\left[(N+2+\frac{1}{N})
\begin{tikzpicture}
[q/.style={circle,draw,scale=0.5},
aq/.style={circle,draw,scale=0.5,fill}]
\node at ( 0,0) [circle, draw,scale=0]{};
\node at ( 0,1) [q] {};
\node at ( 1,1) [aq] {};
\node at ( 0.2,1) [aq] {};
\node at ( 0.8,1) [q] {};
\node at ( 1,1) [aq] {};
\end{tikzpicture}+
\begin{tikzpicture}
[a/.style={rectangle,draw,scale=0.7}]
\node at ( 0,1) [a] {};
\draw [double](0,0.9) -- (0,0);
\draw [double](0,0) -- (1,0);
\draw [double](1,0) -- (1,0.9);
\node at ( 1,1) [a] {};
\end{tikzpicture}
\right]\nn\\
=&(N+2+\dfrac{1}{N})
\begin{tikzpicture}
[q/.style={circle,draw,scale=0.5},
aq/.style={circle,draw,scale=0.5,fill}]
\node at ( 0,0) [circle, draw,scale=0]{};
\node at ( 0,1) [q] {};
\node at ( 1,1) [aq] {};
\node at ( 0.2,1) [aq] {};
\node at ( 0.8,1) [q] {};
\node at ( 1,1) [aq] {};
\end{tikzpicture}
+\dfrac{\lambda_{adj}}{\lambda_{\bf 1}}
\begin{tikzpicture}
[a/.style={rectangle,draw,scale=0.7}]
\node at ( 0,0) [circle, draw,scale=0]{};
\node at ( 0,1) [a] {};
\draw [double](0.1,1) -- (0.9,1);
\node at ( 1,1) [a] {};
\end{tikzpicture}
\;,
\end{align}
where we use (\ref{formuladec}). For the other configurations, we have
\begin{align}
\int dV T_G(U,V)d\phi A&\left[
\underset{i-1\;\;\;\;\;\;\;\;i\;\;\;\;\;\;\;\;i+1}{
\begin{tikzpicture}
[q/.style={circle,draw,scale=0.5},
aq/.style={circle,draw,scale=0.5,fill}]
\node at ( 0,1) [q] {};
\draw (0,0.9) -- (0,0.1);
\node at ( 0,0) [aq] {};
\node at ( 1,0) [q] {};
\draw (1,0.1) -- (1,0.9);
\node at ( 1,1) [aq] {};
\node at ( 0.2,1) [aq] {};
\draw (0.2,0.9) -- (0.2,0.1);
\node at ( 0.2,0) [q] {};
\node at ( 0.8,0) [aq] {};
\draw (0.8,0.1) -- (0.8,0.9);
\node at ( 0.8,1) [q] {};
\node at ( -1.2,0.2) [q] {};
\draw [->](-1.1,0.2) -- (-0.7,0.2);
\draw (-0.7,0.2) -- (-0.3,0.2);
\node at ( -0.2,0.2) [aq] {};
\node at ( -1.2,0) [aq] {};
\draw (-1.1,0) -- (-0.7,0);
\draw [<-](-0.7,0.) -- (-0.3,0.);
\node at ( -0.2,0) [q] {};
\end{tikzpicture}
}\;\;+\;\;
\underset{i\;\;\;\;\;\;\;\;i+1\;\;\;\;\;\;\;\;i+2}{
\begin{tikzpicture}
[q/.style={circle,draw,scale=0.5},
aq/.style={circle,draw,scale=0.5,fill}]
\node at ( 0,1) [q] {};
\draw (0,0.9) -- (0,0.1);
\node at ( 0,0) [aq] {};
\node at ( 1,0) [q] {};
\draw (1,0.1) -- (1,0.9);
\node at ( 1,1) [aq] {};
\node at ( 0.2,1) [aq] {};
\draw (0.2,0.9) -- (0.2,0.1);
\node at ( 0.2,0) [q] {};
\node at ( 0.8,0) [aq] {};
\draw (0.8,0.1) -- (0.8,0.9);
\node at ( 0.8,1) [q] {};
\node at ( 1.2,0.2) [q] {};
\draw [->](1.3,0.2) -- (1.7,0.2);
\draw (1.7,0.2) -- (2.1,0.2);
\node at ( 2.2,0.2) [aq] {};
\node at ( 1.2,0) [aq] {};
\draw (1.3,0) -- (1.7,0);
\draw [<-](1.7,0.) -- (2.1,0.);
\node at ( 2.2,0) [q] {};
\end{tikzpicture}
}\right.\nn\\
&+\sum_{j\not= j-1,j,j+1}\left.
\underset{i\;\;\;\;\;\;i+1}{
\begin{tikzpicture}
[q/.style={circle,draw,scale=0.5},
aq/.style={circle,draw,scale=0.5,fill}]
\node at ( 0,1) [q] {};
\draw (0,0.9) -- (0,0.1);
\node at ( 0,0) [aq] {};
\node at ( 1,0) [q] {};
\draw (1,0.1) -- (1,0.9);
\node at ( 1,1) [aq] {};
\node at ( 0.2,1) [aq] {};
\draw (0.2,0.9) -- (0.2,0.1);
\node at ( 0.2,0) [q] {};
\node at ( 0.8,0) [aq] {};
\draw (0.8,0.1) -- (0.8,0.9);
\node at ( 0.8,1) [q] {};
\end{tikzpicture}
}\dots
\underset{j\;\;\;\;\;\;j+1}{
\begin{tikzpicture}
[q/.style={circle,draw,scale=0.5},
aq/.style={circle,draw,scale=0.5,fill}]
\node at ( 0,-0.2) [q] {};
\draw [->](0.1,-0.2) -- (0.5,-0.2);
\draw (0.5,-0.2) -- (0.9,-0.2);
\node at ( 1,-0.2) [aq] {};
\node at ( 0,-0.4) [aq] {};
\draw (0.1,-0.4) -- (0.5,-0.4);
\draw [<-](0.5,-0.4) -- (0.9,-0.4);
\node at ( 1,-0.4) [q] {};
\end{tikzpicture}
},\right]\nn\\
=&(2+NN_l-N)
\begin{tikzpicture}
[q/.style={circle,draw,scale=0.5},
aq/.style={circle,draw,scale=0.5,fill}]
\node at ( 0,0) [circle, draw,scale=0]{};
\node at ( 0,1) [q] {};
\node at ( 1,1) [aq] {};
\node at ( 0.2,1) [aq] {};
\node at ( 0.8,1) [q] {};
\node at ( 1,1) [aq] {};
\end{tikzpicture}
.
\end{align}

For $\hat{T}_4\sum_n\ket{n,n}$, all of pairs in $\hat{T}$ should be used to make $\ket{i,i}\ket{i+1}\ket{i+1}$. So we have
\begin{align}
\int dV T_G(U,V)d\phi A&\left[
\underset{\;\;i\;\;\;\;\;\;i+1}{
\begin{tikzpicture}
[q/.style={circle,draw,scale=0.5},
aq/.style={circle,draw,scale=0.5,fill}]
\node at ( 0,1) [q] {};
\draw (0,0.9) -- (0,0.1);
\node at ( 0,0) [aq] {};
\node at ( 1,0) [q] {};
\draw (1,0.1) -- (1,0.9);
\node at ( 1,1) [aq] {};
\node at ( 0.2,1) [aq] {};
\draw (0.2,0.9) -- (0.2,0.1);
\node at ( 0.2,0) [q] {};
\node at ( 0.8,0) [aq] {};
\draw (0.8,0.1) -- (0.8,0.9);
\node at ( 0.8,1) [q] {};
\node at ( 0,-0.2) [q] {};
\node at ( 0.2,-0.2) [aq] {};
\end{tikzpicture}
}\;\;+\;\;
\underset{\;\;i\;\;\;\;\;\;i+1}{
\begin{tikzpicture}
[q/.style={circle,draw,scale=0.5},
aq/.style={circle,draw,scale=0.5,fill}]
\node at ( 0,1) [q] {};
\draw (0,0.9) -- (0,0.1);
\node at ( 0,0) [aq] {};
\node at ( 1,0) [q] {};
\draw (1,0.1) -- (1,0.9);
\node at ( 1,1) [aq] {};
\node at ( 0.2,1) [aq] {};
\draw (0.2,0.9) -- (0.2,0.1);
\node at ( 0.2,0) [q] {};
\node at ( 0.8,0) [aq] {};
\draw (0.8,0.1) -- (0.8,0.9);
\node at ( 0.8,1) [q] {};
\node at ( 0.8,-0.2) [q] {};
\node at ( 1,-0.2) [aq] {};
\end{tikzpicture}
}\;\;+\;\;\sum_{j\not=i,i+1}
\underset{\;\;i\;\;\;\;\;\;i+1}{
\begin{tikzpicture}
[q/.style={circle,draw,scale=0.5},
aq/.style={circle,draw,scale=0.5,fill}]
\node at ( 0,1) [q] {};
\draw (0,0.9) -- (0,0.1);
\node at ( 0,0) [aq] {};
\node at ( 1,0) [q] {};
\draw (1,0.1) -- (1,0.9);
\node at ( 1,1) [aq] {};
\node at ( 0.2,1) [aq] {};
\draw (0.2,0.9) -- (0.2,0.1);
\node at ( 0.2,0) [q] {};
\node at ( 0.8,0) [aq] {};
\draw (0.8,0.1) -- (0.8,0.9);
\node at ( 0.8,1) [q] {};
\end{tikzpicture}
}\dots
\underset{j}{
\begin{tikzpicture}
[q/.style={circle,draw,scale=0.5},
aq/.style={circle,draw,scale=0.5,fill}]
\node at ( 0,0) [q] {};
\node at ( 0.2,0) [aq] {};
\end{tikzpicture}
}
\right]\nn\\
&=(NN_l+2)
\begin{tikzpicture}
[q/.style={circle,draw,scale=0.5},
aq/.style={circle,draw,scale=0.5,fill}]
\node at ( 0,0) [circle, draw,scale=0]{};
\node at ( 0,1) [q] {};
\node at ( 1,1) [aq] {};
\node at ( 0.2,1) [aq] {};
\node at ( 0.8,1) [q] {};
\node at ( 1,1) [aq] {};
\end{tikzpicture}
. 
\end{align}
Combining all  results, the coefficient of $\ket{i,i}\ket{i+1,i+1}$ becomes $2NN_l+6+\dfrac{1}{N}$.



\section{$\mathcal{O}(K^2)$ eigenstates and eigenvalues of $\hat{T}$}
\label{eigenstatesfortransfer}
In this appendix, we derive eigenvalues and their eigenfunctions of the transfer matrix $\hat{T}$ at $O(K^2)$,
where $\ket{0}$ and  $\ket{n, m}$ with $\vert n - m\vert \le 2$ mix with each other. 
First, we classify these eigenstates depending on the value of $f_0$ (zero or nonzero) as  
\begin{align}
\ket{G}_K\equiv &\, f_0 \ket{0}+\sum_{n}a_n\ket{n,n}+\sum_{n}b_n\ket{n,n+1}+\sum_{n}c_n\ket{n,n-1} \nn \\
&\qquad +\sum_{n}d_n\ket{n,n+2}+\sum_{n}e_n\ket{n,n-2} \,, \nn\\
\ket{E}_K \equiv &
\sum_{n}a_n\ket{n,n}+\sum_{n}b_n\ket{n,n+1}+\sum_{n}c_n\ket{n,n-1} \nn \\
& \qquad+\sum_{n}d_n\ket{n,n+2}+\sum_{n}e_n\ket{n,n-2}, \label{eq:XYdeno}
\end{align}
which correspond to $f_0 \ne 0$ case and $f_0=0$ case, respectively.
Here $\ket{G}_K$'s should include $\ket{0}$ while
$\ket{E}_K$'s denote the complement of $\ket{G}_K$'s.\footnote{$G$ for $\ket{G}_K$ means that it contains the strong coupling ground state $\ket{0}$, while $E$ for $\ket{E}_K$ represents the lattice excited states. Their subscript $K$ denotes that the state depends on $K$.} 

All relevant eigenvalues and eigenfunctions are obtained as follows.  
\begin{itemize}
\item States $\ket{G^\pm}_K$ with  eigenvalues $G^\pm_K$ are given by
\begin{align}
\ket{G^\pm}_K&:=\ket{0}+\sum_na_n^\pm\ket{n,n} ,\quad \mbox{where}  \quad
a^\pm_n =\dfrac{K^2}{G^\pm_K-(1+NN_\ell)K^2} \,, \\
G^\pm_K&=\dfrac{1}{2}\{1+K^2(1+2NN_\ell)\}\nn\\
&\quad \pm\dfrac{1}{2}\sqrt{1-2(1-2NN_{\ell})K^2+\{1+4N(NN_\ell+2)N_{\ell}\}K^4}.
\end{align}
\item State $\ket{G^{bc}}_K$ with the eigenvalue $G^{bc}_K$ is given by 
\begin{align}
\ket{G^{bc}}_K&:=K\ket{0}+\dfrac{K}{\frac{\lambda_{\bf F}}{\lambda_{\bf 1}}-(1+NN_\ell)}\sum_n\ket{n,n} \nn \\
& \qquad \qquad +\sum_{n}(b^G_n\ket{n,n+1}+c^G_n\ket{n,n-1}) \,, \label{eq:Xbc}\\
G^{bc}_K&=K^2\left(\dfrac{\lambda_{\bf F}}{\lambda_{\bf 1}}\right) \,, 
\end{align}
where coefficients $b^G_n$ and $c^G_n$ must satisfy 
\begin{align}
\sum_n(b^G_n+c^G_n)&=-\dfrac{1}{N}-\dfrac{N_\ell}{\frac{\lambda_{\bf F}}{\lambda_{\bf 1}}-(1+NN_\ell)} \nn \\
&\quad +K^2\left[\dfrac{1}{N}\left(\dfrac{\lambda_{\bf F}}{\lambda_{\bf 1}}\right)-N_\ell-(NN_\ell+2)\dfrac{N_\ell}{\frac{\lambda_{\bf F}}{\lambda_{\bf 1}}-(1+NN_\ell)}\right] \,. 
\end{align}
\item State $\ket{G^{de}}_K$ with the eigenvalue $G^{de}_K$ is given by
\begin{align}
\ket{G^{de}}_K&:=K^2\ket{0}+\dfrac{K^2}{\left(\frac{\lambda_{\bf F}}{\lambda_{\bf 1}}\right)^2-(1+NN_\ell)}\sum_n\ket{n,n} \nn \\
& \quad +\sum_{n}(d^G_n\ket{n,n+2}+e^G_n\ket{n,n-2})\\
G^{de}_K&=K^2\left(\dfrac{\lambda_{\bf F}}{\lambda_{\bf 1}}\right)^2 \,.
\end{align}
where coefficients $d^G_n$ and $e^G_n$ must satisfy 
\begin{align}
\sum_n(d^G_n+e^G_n)&=-\dfrac{1}{N}-\dfrac{N_\ell}{\left(\frac{\lambda_{\bf F}}{\lambda_{\bf 1}}\right)^2-(1+NN_\ell)} \nn \\
&\quad +K^2\left[\dfrac{1}{N}\left(\dfrac{\lambda_{\bf F}}{\lambda_{\bf 1}}\right)^2-N_\ell-(NN_\ell+2)\dfrac{N_\ell}{\left(\frac{\lambda_{\bf F}}{\lambda_{\bf 1}}\right)^2-(1+NN_\ell)}\right] \,.
\end{align}
\item State $\ket{E^{a}}_K$ with the eigenvalue $E^{a}_K$ is given by  
\begin{align}
\ket{E^a}_K&:=\sum_na^E_n\ket{n,n}\quad \mbox{where} \quad \sum_na^E_n=0 \,, \\
E^{a}_K&=K^2 \,, 
\end{align}
\item State $\ket{E^{bc}}_K$, which gives the eigenvalues $E^{bc}_K$, defined as 
\begin{align}
\ket{E^{bc}}_K&:=\sum_n(b^E_n\ket{n,n+1}+c^E_n\ket{n,n-1}) \,, 
\, \mbox{where} \,\, \sum_n(b^E_n+c^E_n)=0  \,,\\
E^{bc}_K&=K^2\left(\dfrac{\lambda_{\bf F}}{\lambda_{\bf 1}}\right) \,. 
\end{align}
\item State $\ket{E^{de}}_K$ with the eigenvalue $E^{de}_K$ is given by
\begin{align}
\ket{E^{de}}_K&:=\sum_n(d^Y_n\ket{n,n+2}+e^Y_n\ket{n,n-2}) \,, \, \mbox{where} \,\,  \sum_n(d^Y_n+e^Y_n)=0 \,, \\
E^{de}_K&=K^2\left(\dfrac{\lambda_{\bf F}}{\lambda_{\bf 1}}\right)^2  \,.
\end{align}
\end{itemize}




\end{document}